\documentclass[a4paper,11pt]{article}
\pdfoutput=1 

\usepackage{jinstpub} 
\usepackage{lineno}
\usepackage{doi}
\usepackage{siunitx}
\usepackage[margin=1cm]{caption} 
\usepackage{placeins}  

\usepackage{nsw_electronics-defs}
\usepackage{nsw_electronics-cmds}
\usepackage{lipsum}
\usepackage{enumitem}
\usepackage{comment}
\usepackage{multirow}

\usepackage{multicol}
\usepackage[toc,xindy,acronym,nonumberlist]{glossaries}
\renewcommand{\glossarysection}[2][]{}

\newglossaryentry{MSB}{name={MSB},
    description={Most Significant Bit},nonumberlist}

\newglossaryentry{SCA}{name={SCA},
    description={Slow Control Adapter},nonumberlist}

\newglossaryentry{HDLC}{name={HDLC},
    description={High-Level Data Link Control},nonumberlist}

\newglossaryentry{SPI}{name={SPI},
    description={Serial Peripheral Interface},nonumberlist}

\newglossaryentry{ROC}{name={ROC},
    description={Read Out Controller},nonumberlist}

\newglossaryentry{MO}{name={MO},
    description={Monitoring Output of the VMM ASIC},nonumberlist}

\newglossaryentry{SEU}{name={SEU},
    description={Single Event Upset},nonumberlist}

\newglossaryentry{TMR}{name={TMR},
    description={Triple Modular Redundancy},nonumberlist}

\newglossaryentry{TID}{name={TID},
    description={Total Ionizing Dose},nonumberlist}

\newglossaryentry{DICE}{name={DICE},
    description={Dual Interlocked Cells},nonumberlist}

\newglossaryentry{Micromegas}{name={Micromegas},
    description={Micro Mesh Gaseous Structure},nonumberlist}

\newglossaryentry{BGA}{name={BGA},
    description={Ball Grid Array},nonumberlist}

\newglossaryentry{ESD}{name={ESD},
    description={Electrostatic Discharge},nonumberlist}

\newglossaryentry{SLVS}{name={SLVS},
    description={Scalable Low Voltage Signaling},nonumberlist}

\newglossaryentry{EAR}{name={EAR},
    description={Export Administration Regulations},nonumberlist}

\newglossaryentry{DDR}{name={DDR},
    description={Double Data Rate},nonumberlist}

\newglossaryentry{FIFO}{name={FIFO},
    description={First In First Out},nonumberlist}

\newglossaryentry{LSB}{name={LSB},
    description={Least Significant Bit},nonumberlist}

\newglossaryentry{ToT}{name={ToT},
    description={Time over Threshold},nonumberlist}

\newglossaryentry{TtP}{name={TtP},
    description={Threshold to Peak},nonumberlist}

\newglossaryentry{PtP}{name={PtP},
    description={Pulse at Peak},nonumberlist}

\newglossaryentry{PtT}{name={PtT},
    description={Peak to Threshold},nonumberlist}

\newglossaryentry{DDF}{name={DDF},
    description={Delayed Dissipative Feedback},nonumberlist}

\newglossaryentry{MOSFET}{name={MOSFET},
    description={Metal Oxide Semiconductor Field Effect Transistor},nonumberlist}

\newglossaryentry{CMOS}{name={CMOS},
    description={Complementary Metal Oxide Semiconductor},nonumberlist}

\newglossaryentry{GUI}{name={GUI},
    description={Graphical User Interface},nonumberlist}

\newglossaryentry{DAC}{name={DAC},
    description={Digital to Analog Converter},nonumberlist}

\newglossaryentry{DAQ}{name={DAQ},
    description={Data Acquisition},nonumberlist}

\newglossaryentry{DCS}{name={DCS},
    description={Detector Control System},nonumberlist}

\newglossaryentry{sTGC}{name={sTGC},
    description={small strip Thin Gap Chambers},nonumberlist}

\newglossaryentry{ART}{name={ART},
    description={Address in Real Time},nonumberlist}

\newglossaryentry{ASIC}{name={ASIC},
    description={Application-Specific Integrated Circuit},nonumberlist}

\newglossaryentry{FPGA}{name={FPGA},
    description={Field Programmable Gate Array},nonumberlist}

\newglossaryentry{TAC}{name={TAC},
    description={Time to Amplitude Converter},nonumberlist}

\newglossaryentry{LVDS}{name={LVDS},
    description={Low Voltage Differential Signal},nonumberlist}

\newglossaryentry{ADC}{name={ADC},
    description={Analog to Digital Converter},nonumberlist}

\newglossaryentry{PDO}{name={PDO},
    description={Peak Detector Output of the VMM ASIC},nonumberlist}

\newglossaryentry{TDO}{name={TDO},
    description={Time Detector Output of the VMM ASIC},nonumberlist}

\newglossaryentry{CA}{name={CA},
    description={Charge Amplifier},nonumberlist}

\newglossaryentry{PD}{name={PD},
    description={Peak Detector},nonumberlist}

\newglossaryentry{TD}{name={TD},
    description={Time Detector},nonumberlist}

\newglossaryentry{NSW}{name={NSW},
    description={New Small Wheel},nonumberlist}

\newglossaryentry{ENC}{name={ENC},
    description={Equivalent Noise Charge},nonumberlist}

\newglossaryentry{TCP}{name={TCP},
    description={Transmission Control Protocol},nonumberlist}

\newglossaryentry{UDP}{name={UDP},
    description={User Datagram Protocol},nonumberlist}

\newglossaryentry{MPV}{name={MPV},
    description={Most Probable Value},nonumberlist}

\newglossaryentry{RT}{name={RT},
    description={Radon Transform},nonumberlist}

\newglossaryentry{TDS}{name={TDS},
    description={Trigger Data Serializer},nonumberlist}

\newglossaryentry{ADDC}{name={ADDC},
    description={ART Data Driver Card},nonumberlist}

\newglossaryentry{ALTI}{name={ALTI},
    description={ATLAS Local Trigger Interface},nonumberlist}

\newglossaryentry{ART ASIC}{name={ART ASIC},
    description={Address in Real Time ASIC},nonumberlist}

\newglossaryentry{ATCA}{name={ATCA},
    description={Advanced Telecommunications Computing Architecture},nonumberlist}

\newglossaryentry{BC}{name={BC},
    description={Bunch Crossing},nonumberlist}

\newglossaryentry{BCID}{name={BCID},
    description={Bunch Crossing IDentification},nonumberlist}

\newglossaryentry{BCR}{name={BCR},
    description={Bunch Crossing Reset},nonumberlist}

\newglossaryentry{BER}{name={BER},
    description={Bit Error Rate},nonumberlist}

\newglossaryentry{Big Wheel}{name={Big Wheel},
    description={The large wheels of the End cap MDT and TGC Muon Spectrometer},nonumberlist}

\newglossaryentry{CDR}{name={CDR},
    description={Clock and Data Recover},nonumberlist}

\newglossaryentry{CRC}{name={CRC},
    description={Cyclic Redundancy Check},nonumberlist}

\newglossaryentry{CTP}{name={CTP},
    description={Central Trigger Processor},nonumberlist}

\newglossaryentry{DDO}{name={DDO},
    description={Direct Data Output},nonumberlist}

\newglossaryentry{ECC}{name={ECC},
    description={Error Correcting Code},nonumberlist}

\newglossaryentry{E-links}{name={E-links},
    description={Serial I/O Electrical Links of the GBTx ASIC},nonumberlist}

\newglossaryentry{FEAST ASIC}{name={FEAST ASIC},
    description={A CERN produced radiation hard DC-DC point-of-load voltage converter},nonumberlist}

\newglossaryentry{FEASTMP}{name={FEASTMP},
    description={A pluggable module base on the FEAST ASIC},nonumberlist}

\newglossaryentry{FELIX}{name={FELIX},
    description={Front-End LInk eXchange},nonumberlist}

\newglossaryentry{Flexible chains}{name={Flexible chains},
    description={dddd},nonumberlist}

\newglossaryentry{GBT-FPGA}{name={GBT-FPGA},
    description={FPGA firmware that emulates the GBTx ASIC},nonumberlist}

\newglossaryentry{GBTx ASIC}{name={GBTx ASIC},
    description={Gigabit Transfer ASIC},nonumberlist}

\newglossaryentry{GPIO}{name={GPIO},
    description={General Purpose Input Output},nonumberlist}

\newglossaryentry{GTH}{name={GTH},
    description={Gigabit Transceiver, generation H},nonumberlist}

\newglossaryentry{HL-LHC}{name={HL-LHC},
    description={High Luminosity LHC},nonumberlist}

\newglossaryentry{I2C}{name={I2C},
    description={Inter-Integrated Circuit: A synchronous, multi-controller/multi-target (master/slave), packet switched, single-ended, serial communication bus},nonumberlist}

\newglossaryentry{ICS}{name={ICS},
    description={Intermediate Conversion Stage: The voltage converters on the NSW Rim supplying $\sim$11\,V to the on-detector electronics from 280\,V},nonumberlist}

\newglossaryentry{IP}{name={IP},
    description={Interaction Point},nonumberlist}

\newglossaryentry{IPMC}{name={IPMC},
    description={Intelligent Platform Management Controller: A autonomous controller that provides management and monitoring independently of the host system's CPU},nonumberlist}

\newglossaryentry{JTAG}{name={JTAG},
    description={A serial protocol used to configure FPGAs},nonumberlist}

\newglossaryentry{L0A}{name={L0A},
    description={Level-0 Trigger Accept },nonumberlist}

\newglossaryentry{L1A}{name={L1A},
    description={Level-1 Trigger Accept},nonumberlist}

\newglossaryentry{L1DDC}{name={L1DDC},
    description={Level-1 Data Driver Card},nonumberlist}

\newglossaryentry{Large Box}{name={Large Box},
    description={dddd},nonumberlist}

\newglossaryentry{LC}{name={LC},
    description={A type of optical fibre connector},nonumberlist}

\newglossaryentry{LHC}{name={LHC},
    description={Large Hadron Collider},nonumberlist}

\newglossaryentry{LUT}{name={LUT},
    description={Look-Up Table},nonumberlist}

\newglossaryentry{MiniSAS}{name={MiniSAS},
    description={A high speed, multi-lane twin-ax ribbon cable},nonumberlist}

\newglossaryentry{MMC}{name={MMC},
    description={Module Management Controller},nonumberlist}

\newglossaryentry{MMFE8}{name=MMFE8{},
    description={Front-end board with 8 VMM's for the Micromegas},nonumberlist}

\newglossaryentry{MTP}{name={MTP},
    description={Multi-fibre Termination Push-on, an multi-fibre optical connector},nonumberlist}

\newglossaryentry{MTx}{name={MTx},
    description={Miniature Transmitter},nonumberlist}

\newglossaryentry{MUCTPI}{name={MUCTPI},
    description={MUon to Central Trigger Processor Interface},nonumberlist}

\newglossaryentry{NIC}{name={NIC},
    description={Network Interconnect Card},nonumberlist}

\newglossaryentry{OM3}{name={OM3},
    description={A multi-mode glass fibre for data transmission of up to 300\,meters at a data rate of 10\,Gb/s},nonumberlist}

\newglossaryentry{OPC UA}{name={OPC UA},
    description={OPC Unified Architecture is a machine-to-machine communication protocol used for industrial automation},nonumberlist}

\newglossaryentry{pad-FEB}{name={pad-FEB},
    description={Front-end board for sTGC pads and wires},nonumberlist}

\newglossaryentry{PLL}{name={PLL},
    description={Phase-Locked loop},nonumberlist}

\newglossaryentry{PRBS}{name={PRBS},
    description={Pseudo-Random Bit Sequence},nonumberlist}

\newglossaryentry{RTM}{name={RTM},
    description={Rear Transition Module},nonumberlist}

\newglossaryentry{SCA ASIC}{name={SCA ASIC},
    description={Slow Control Adaptor ASIC},nonumberlist}

\newglossaryentry{SCAx}{name={SCAx},
    description={FPGA emulation of the Slow Control Adaptor ASIC},nonumberlist}

\newglossaryentry{Sector Logic}{name={Sector Logic},
    description={The electronics modules that generate muon candidates by matching track segments found in the NSW Micromegas and sTGC detectors with those found in the Big Wheel TGC detectors},nonumberlist}

\newglossaryentry{sFEB}{name={sFEB},
    description={Front-end board for sTGC strips},nonumberlist}

\newglossaryentry{SFP}{name={SFP},
    description={Small Form-factor Pluggable is a compact, hot-pluggable network interface module format used for data communication},nonumberlist}

\newglossaryentry{Small Box}{name={Small Box},
    description={dddd},nonumberlist}

\newglossaryentry{sROC}{name={sROC},
    description={dddd},nonumberlist}

\newglossaryentry{swROD}{name={swROD},
    description={Software ROD is the software infrastructure for event fragment building, buffering and detector-specific processing of ATLAS event data},nonumberlist}

\newglossaryentry{TTC}{name={TTC},
    description={Trigger, Timing and Control},nonumberlist}

\newglossaryentry{TVS}{name={TVS},
    description={Transient Voltage Suppressor},nonumberlist}

\newglossaryentry{Twinax}{name={Twinax},
    description={A type of cable similar to coaxial cable, but with two inner conductors instead of one. Used especially for high-speed differential signaling},nonumberlist}

\newglossaryentry{USA15}{name=USA15{},
    description={The underground radiation protected rooms for ATLAS readout, trigger and DCS equipment},nonumberlist}

\newglossaryentry{VME}{name={VME},
    description={Versa Module Eurocard is a computer bus standard. It is physically based on Eurocard sizes, mechanics and connectors, but uses its own signalling system},nonumberlist}

\newglossaryentry{VTRx}{name={VTRx},
    description={Versatile Transmitter plus Receiver optical interaface},nonumberlist}

\newglossaryentry{VTTx}{name={VTTx},
    description={Versatile Twin-Transmitter optical interaface},nonumberlist}

\newglossaryentry{ZEBRA}{name={ZEBRA},
    description={A Zebra strip connector is comprised of a malleable insulation strip with alternating conducting lines},nonumberlist}

\newglossaryentry{VMM}{name={VMM},
   description={The custom front-end ASIC for the Muon NSW},nonumberlist}

\makeglossaries

\title{\boldmath The New Small Wheel electronics}

 \author[a,1]{G.~Iakovidis\note{Corresponding author},}
 \author[b,1]{L.~Levinson,}

\author[t,8]{Y.~Afik,}
\author[f]{C.~Alexa,}
\author[d]{T.~Alexopoulos,}
\author[p]{J.~Ameel,}
\author[p]{D.~Amidei,}
\author[v,2]{D.~Antrim\note{Now at Google, CA; United States of America},}
\author[j]{A.~Badea,}
\author[a,d,3]{C.~Bakalis\note{Now at SLAC National Accelerator Laboratory, Stanford CA; United States of America},}
% ...

\affiliation[a]{Brookhaven National Laboratory, Upton NY; United States of America}
\affiliation[b]{Weizmann Institute of Science, Rehovot; Israel}
% ...
\affiliation[d]{National Technical University of Athens, Zografou; Greece}
% ...
\affiliation[f]{Horia Hulubei National Institute of Physics and Nuclear Engineering, Bucharest; Romania}
% ...
\affiliation[j]{Harvard University, Cambridge MA; United States of America}
% ...
\affiliation[p]{University of Michigan, Ann Arbor MI; United States of America}
% ...
\affiliation[t]{Technion, Israel Institute of Technology, Haifa; Israel}
% ...
\affiliation[v]{University of California Irvine, Irvine CA; United States of America}
% ...

\emailAdd{giakovidis@bnl.gov}
\emailAdd{lorne.levinson@weizmann.ac.il}

\abstract{
The increase in luminosity, and consequent higher backgrounds, of the LHC upgrades require improved rejection of fake tracks in the forward region of the ATLAS Muon Spectrometer.
The New Small Wheel upgrade of the Muon Spectrometer aims to reduce the large background of fake triggers from track segments that don't originate from the interaction point.
The New Small Wheel employs two detector technologies, the resistive strip \MM detectors and the ``small'' Thin Gap Chambers, with a total of 2.45\,million electrodes to be sensed.
The two technologies require the design of a complex electronics system given that it consists of two different detector technologies and is required to provide both precision readout and a fast trigger.
It will operate in a high background radiation region up to about 20\,kHz/cm$^{2}$ at the expected HL-LHC luminosity of \lumihllhchigh.
The architecture of the system is strongly defined by the GBTx data aggregation ASIC, the newly-introduced FELIX data router and the software based data handler of the ATLAS detector.
The electronics complex of this new detector was designed and developed in the last ten years and consists of multiple radiation tolerant Application Specific Integrated Circuits, multiple front-end boards, dense boards with FPGA's and purpose-built Trigger Processor boards within the ATCA standard.
The New Small Wheel has been installed in 2021 and is undergoing integration within ATLAS for LHC Run\,3.
It should operate through the end of Run\,4 (December 2032).
In this manuscript, the overall design of the New Small Wheel electronics is presented.}

\keywords{LHC, HL-LHC, ATLAS, NSW, GBTx, sTGC, MicroMegas, Application Specific Integrated Circuits, Data acquisition circuits, Data acquisition concepts, Digital electronic circuits,
          Front-end electronics for detector readout, Trigger concepts and systems (hardware and software),
          VLSI circuits, Trigger algorithms}
\arxivnumber{2303.12571} 

\dedicated{In memory of \\
Dr. Stephanie Ulrike Zimmermann \\
(1973 -- 2020)}

\begin{document}

\maketitle

%
\clearpage  

\section{Introduction}
\label{sec:intro}

The New Small Wheel (NSW)\,\cite{nswTDR} is an upgrade of the innermost forward station of the Muon Spectrometer at the ATLAS experiment\,\cite{ATLAS_2008} at CERN.
The High Luminosity LHC (HL-LHC) will provide substantially increased luminosity; therefore, higher background rates are expected.
The upgrades to ATLAS for HL-LHC are done in two phases: Phase\,1 for Run\,3, which began in mid-2022 and Phase\,2 for Run\,4, which is expected to begin in mid-2029.
The initial configuration of the Small Wheels of the Muon Spectrometer does not allow the rejection of the fake triggers from the increased background,	and moreover, efficiency loss is expected at high particle rate.
The New Small Wheel will reduce the significant background of fake triggers from track segments that do not originate from the interaction point by providing a track segment to the ATLAS Level 1 trigger logic to match with hit coincidences in the Big Wheel\,\cite{l1_muon_twiki}.
Also, it will cope with a ten-fold increase in the ATLAS Level-1 trigger rate.
It was installed in the ATLAS cavern in 2021 and is undergoing  commissioning.
Operating in a high background radiation region (up to about 20\,kHz/cm$^{2}$ at the expected HL-LHC luminosity of \lumihllhchigh),
it reconstructs muon tracks with high precision and furnishes pointing track segments to the ATLAS \Lone trigger.
In this manuscript, the overall design of the New Small Wheel electronics is presented.

The NSW employs two gaseous detector technologies: sTGC\,\cite{ABUSLEME201685} and \MM\,\cite{ALEXOPOULOS2010161}.
The sTGC provides bunch crossing assignment with high radial resolution from strips and rough $\phi$ resolution from pads;
the \MM strips provide even better radial resolution, and a good $\phi$ coordinate due to its stereo strips layout.
Both technologies are used for triggering and for track reconstruction.
For triggering, the sTGC is expected to provide better timing resolution than Micromegas and a higher angular resolution due to its greater separation between its first four and last four layers.
This performance is still to be demonstrated during Run\,3 operation of the NSW.

The NSW envelope is a disk, $\sim$10\,m in diameter and $\sim$1\,m thick.
It lies between the Endcap Liquid Argon Calorimeter and the Endcap Toroid, centred at $z=7.3$\,m.
Each NSW comprises 16 sectors, 8 small and 8 large alternating each other.
Each sector consists of eight layers of each detector technology, arranged along the beam axis as follows: a 4-layer sTGC wedge, two 4-layer \MM wedges on either side of a support structure and another 4-layer sTGC wedge.
There are $\sim$2.1\,million \MM strips (4 of the 8 layers have stereo strip layout) $\sim$282\,k sTGC strip, $\sim$47\,k sTGC pad and $\sim$24\,k sTGC wire analog readouts
for a total of $\sim$2.45\,million channels.
The NSW electronics follows the NSW detector organization in a hierarchy from endcaps, sectors, layers, and finally, radial position.
A large and a small sector cover one octant.
The sectors overlap slightly, but are independent, with no communication between them.

The NSW performance criteria are demanding.
In particular, the precision reconstruction of track segments for the offline analysis requires a spatial resolution of $\sim$100\um per layer to provide good momentum resolution along the radial direction,
over a 4\,m active radius surface. The track segments for the \Lone trigger must be reconstructed online with a polar angular resolution of approximately 1\,mrad,
in order to match the 1\,mrad angular resolution of the middle and outer muon stations planned for the Phase\,2 upgrade for the HL- LHC.
The trigger requires that for a valid NSW segment, the angle between it and the infinite momentum track from the interaction point be less than $\pm$15\,mrad. (The angle is configurable up to\,$\pm$15\,mrad in steps of 1\,mrad.)
In Phase\,1, NSW segments found at every bunch crossing are extrapolated to match the \emph{hit} coincidences in the muon stations downstream from the endcap toroid magnet.
In Phase\,2, they are used with \emph{segments} in the downstream muon stations to measure a track's transverse momentum, $p_\mathrm{T}$, for the trigger algorithm.

\vspace{12pt}
\noindent Constraints strongly affecting the design of the electronics include:
\begin{itemize}[topsep=2pt, itemsep=0pt, parsep=1pt]
   \item Sophisticated Front-end readout is required for 2.45\,million analog channels with differing signal characteristics.
   \item Tightly limited latency for the Phase\,1 trigger path of $\sim$1.1\us including $\sim$500\ns of time-of-flight, fibre and cable propagation
   \item Radiation at the inner radius (ten-year total dose of 350\,kRad, simulated, with safety factors, requires radiation and Single Event Upset tolerant ASICs and testing of any commodity electronics used).
            FPGAs can be used only on the rim of the wheel and only if they use SEU mitigation techniques.
   \item Magnetic fields up to 5\,kG strongly limit the use of non-air-core inductors in filters and DC-DC converters.
   \item Limited space implies dense circuit boards, active cooling.
   \item Limited access after installation requires redundancy.
   \item Since the on-detector electronics cannot be replaced for Phase\,2, the Phase\,2 requirements must be met with Phase\,1 technology.
   \item The decision between a single level and a two-level hardware trigger could not be made before freezing the NSW ASIC requirements,
            thus requiring the readout ASIC to support both.
   \item Need to support two detector technologies with as many shared elements as possible.
   \item Need for state-of-the-art gigabit serial transceivers and interconnections
   \item Need of on-board voltage conversion requires very careful design of the power distribution to maintain a high signal-to-noise ratio.
\end{itemize}

\para{Actual clock frequencies}
All the NSW clocks are inherited from the LHC bunch crossing clock, which is 40.079\,MHz.
However, for convenience in this document, the bunch crossing (BC) clock and its integer multiples are referred to as 40\,MHz, 80\,MHz, 160\,MHz, etc.
For example, the actual frequency of the 160\,MHz clock is 160.316\,MHz.

\subsection{Radiation and magnetic field tolerance}
The radiation level diminishes as the radius increases; see Table\,\ref{tab:radEnv}.
It is high enough to force using  ASICs on the detectors, but FPGAs are manageable, with Single Event Upset (SEU) mitigation, on the rim of the wheel.
All on-detector components were tested to confirm that they withstand the expected Total Ionisation Dose (TID).
All on-detector non-ASIC components are stateless, so their SEUs are not fatal.
The modern jitter cleaners required by FPGA gigabit transceivers have both analog and digital logic and, unfortunately, were found to have fatal Single Event Effects even in the radiation environment on the Rim.
See\,\cite{Ameel, amideiDCDC, ATL-MUON-PUB-2022-001, NSWenv}.

The magnetic field, combined from the solenoid and endcap toroid, is neither uniform in radius, $R$, nor azimuth, $\phi$.
Its highest value lies exactly on the Rim and the edges of the detectors where the electronic boards are located.
Either air-core inductors, or inductors that could be shielded and oriented to minimize the effect of the magnetic field, were used\,\cite{NSWenv}.

\begin{table}[ht!]
\centering
   \caption{Simulated radiation loads and magnetic fields, from\,\cite{NSWenv}  for the NSW after 10\,years at high luminosity LHC, $\mathcal{L}=\rm 5{\times}10^{34} cm^{-2} s^{-1}$,
              for Total Ionization Dose (TID),
              Non-Ionizing Energy Loss (NIEL),
              Single Event Effect (SEE).
              Safety factors are not included.}
   \label{tab:radEnv}
   \begin{tabular}{l l l}
     \toprule
     \textbf{ } & \textbf{Inner radius (R=1\,m)} & \textbf{Outer Rim (R=5\,m)}\\
     \midrule
       TID ($\gamma$) & 780\,Gy  & 26\,Gy\\
       NIEL (fast neutrons) & $\rm 3.9{\times}10^{13} \        \emph{n}/cm^2$ &  $\rm 1.2{\times}10^{12} \   \emph{n}/cm^2$\\
       SEE (protons)        & $\rm 6.7{\times}10^{12} \        \emph{p}/cm^2$ &  $\rm 2.2{\times}10^{11} \   \emph{p}/cm^2$\\
       B\,field & $\le$\,1\,kG & max 5\,kG\\
    \bottomrule
    \end{tabular}
\end{table}



\section{Overall architecture of the electronics}
\label{sec:overall}

The overall architecture of the NSW electronics is shown in Figure\,\ref{fig:LL_NSW_ElxOvr}.
There are five data and signal paths:
\begin{itemize}[topsep=0pt, itemsep=0pt, parsep=0pt]  
   \item Synchronization, trigger, test pulse and reset signals and the LHC bunch-crossing clock
   \item Digitized readout data from the detector including monitoring and calibration data
   \item Configuration parameters and commands to the Front-ends and status indicators from the Front-ends
   \item Busy signal path
   \item Digitized data from the detector for the trigger path
\end{itemize}
These five paths are elaborated below.

\subsection{The GBTx ASIC and FELIX}
The use of the radiation-hard GBTx ASIC\,\cite{Moreira:2009pem, Wyllie:2012cua} to provide fibre connections between the collision cavern and the radiation-protected room, drastically impacts the architecture of the NSW electronics.
The first three paths are time-multiplexed together on fibres to and from a GBTx ASIC which interfaces to separate electrical ``E-links'' on twin-ax cables (also known as MiniSAS cables)\,\cite{8F36,twin-ax} to electronics boards on the detector.
These slow links carry Front-end readout, calibration and detector monitoring data from the Front-end, and configuration and control data to the Front-end.
They are aggregated to the 4.8\,Gb/s fibres by the GBTx ASICs on the Level-1 Data Driver Cards\,(L1DDC).
See Sections\,\ref{sec:GBTx} and\,\ref{sec:L1DDC}.
In the \MM trigger path, the GBTx is used as a point-to-point 4.8\,Gb/s link between the \gls{ADDC} and the Trigger Processor.
The data transported are not considered as E-links.
See Sections\,\ref{sec:GBTx} and\,\ref{sec:addc}.

\begin{figure}[ht]
\centering
\includegraphics[width=0.96\textwidth]{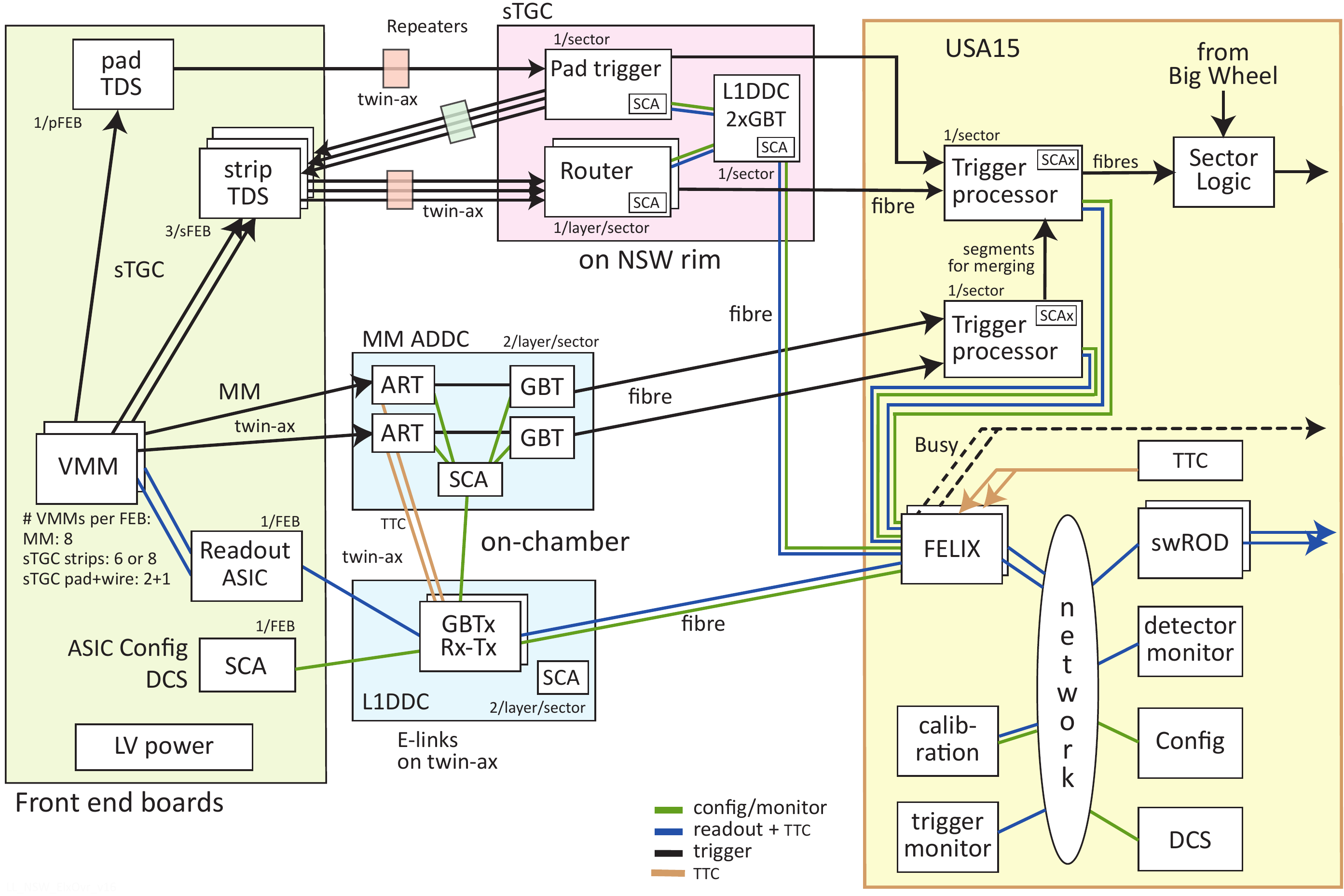}
\caption{Overview of the NSW electronics.
         There are electronics boards on the detector, on the rim of the wheel and in the radiation-protected room (USA15).
         The objects shown connected to the network, except FELIX, may be computers or processes.
         The \gls{Sector Logic}, \gls{FELIX} and \gls{swROD} are part of the ATLAS Trigger \& DAQ project.
         The quantities of each board and ASIC is given in Table\,\ref{tab:boardsASICs}.}
\label{fig:LL_NSW_ElxOvr}
\end{figure}

The fibres from the L1DDCs that carry the E-links connect to FELIX,
the Front End Link eXchange (FELIX)\,\cite{Panduro-Vasquez:2022oR, Levinson:2799865, Paramonov:2021jpz, PanduroVazquez:2020mnk, felixHW, Trovato:2019pui, FelixUserGuide}.
The use of FELIX also strongly impacts the architecture of the NSW Electronics.
FELIX routes E-links to and from endpoints on a standard Ethernet network.
It replaces the previous ATLAS front-end readout architecture.
FELIX is part of all but the trigger data paths.
FELIX separates data transport from data processing:
data are transported by FELIX, a detector-neutral custom hardware plus software device;
data are processed or sent by detector-specific software; for example, readout data is processed by the swROD\,\cite{9448326}.
FELIX is described in more detail in Section\,\ref{sec:felix}.



Following Figure\,\ref{fig:LL_NSW_ElxOvr}, the components and data flow for each of the five paths are summarized here, with details for all the components in Sections\,\ref{sec:asics}\,and\,\ref{sec:boards}.

\subsection{Synchronization, trigger, reset, test pulse signals and LHC clock path}
This path begins with the \gls{ALTI}\,\cite{ALTItwiki} Timing, Trigger \& Control (TTC) module,
which sends a serial stream containing the LHC Bunch Crossing (BC) clock and control signals via optical fan-outs and dedicated fibres to each FELIX FPGA.
The control signals can be sent synchronously to the BC clock with a defined offset in BC's from the beginning of the LHC orbit.
FELIX decodes the control signals, recovers and jitter-cleans the BC clock.
The control signals, Level-1 Trigger Accept (\gls{L1A}), Level-0 Trigger Accept (\gls{L0A}), Event Counter Reset (ECR), Bunch Crossing Counter Reset (BCR), Test-Pulse (TP), Soft-Reset (SR), Orbit Count Reset (OCR) and the BC clock
are sent to the Readout Controller (ROC), see Section\,\ref{sec:ROC}, on the Front-end boards, to the Rim boards and to the \MM trigger boards via the GBTx ASIC on the L1DDC boards.
The BC clock is recovered by the GBTx ASIC and is sent to all ASICs in the system.
The Trigger Processors receive the same signals via firmware\,\cite{GBT-FPGA, GBT_FPGA2} that emulates the GBTx.
The higher speed reference clocks of all serial links are multiples of the recovered BC clock.
The Readout Controller on each Front-end board forwards the needed control signals and bunch crossing clock, with configurable delays, to the various ASICs on the board.
This path has a fixed latency, reproducible across power cycles, from the Central Trigger Processor (\gls{CTP}) to ALTI to the Front-end ASICs.

\subsection{Digitized detector data path}
The digitized detector data path begins with the 64-channel \gls{VMM} Front-end ASIC, see \ref{sec:VMM}, which senses the detector readout electrodes.
Both Micromegas and sTGC strips as well as sTGC pad and wire signals are processed by digitizing their peak voltage after amplification and shaping.
The time of the peak or threshold crossing (configurable) relative to the BC clock edge is also digitized.
The digitized data is tagged by the VMM with the value, the ``BCID'', of a BC counter which remains attached to the data as it moves through the system.
The charge and time information are buffered in the VMM until the L0A (Level-0 Accept) trigger signal is received and sent to the ROC.
The data are  held in the buffer until they become older than the BC window of interest. The ROC buffers the data until the L1A (Level-1 Accept) trigger signal is received.\footnote{The L1A originates from the ATLAS Central Trigger. On receipt of L1A, FELIX sends out L0A; it then sends L1A after a configurable delay. This allows time for the transfer from VMM to ROC.}
The Read Out Controller aggregates up to eight VMMs and sends the data for those BCs selected by the L1A trigger to a GBTx ASIC on the L1DDC card via one or more copper twin-ax serial readout ``E-links''.
The GBTx aggregates several up to 320\,Mb/s readout E-links onto a 4.8\,Gb/s optical link.
FELIX then receives data from several optical links and routes the data from the readout E-links to the ``software Read Out Driver'' (swROD) which subscribes to and processes many readout E-links.
The swROD finally sends the Front-end data in an ATLAS standard format to the High Level Trigger (HLT).
The swROD provides a sampled data stream for monitoring user-defined parameters and data flow directly or by other separate monitoring processes.

In addition to the L1A event flow, calibration triggers can move Front-end data along the path,
either to a swROD for offline calibration or to a separate swROD that does not connect to the HLT, but instead acts as a dedicated calibration processor.
Additionally, the Trigger Processors send monitoring data for non-triggered bunch crossings to an instance of the swROD (without connection to HLT) that monitors the Trigger Processors.

\subsection{Configuration, command and status path}
The configurable operating parameters for all NSW electronics components are controlled via the SCA ASIC\,\cite{GBT-SCA} on every board,
or its emulator in FPGA firmware, SCAx\,\cite{SCAxIEEE} (as shown in Figure\,\ref{fig:LL_NSW_ElxOvr}).
An SCA server based on the \glslink{OPC UA}{OPC\,UA}\,\cite{OPC-UA} architecture is the software interface to the SCA and SCAx.
The communication is realized via FELIX and GBTx E-links.
Configuration and status reporting processes access the various SCA and SCAx configuration and status registers through several OPC\,UA clients.
Commands such as resets, are also sent on this path.
The same path is shared with the Detector Control System's (\gls{DCS}) OPC\,UA client for monitoring board and ASIC conditions, such as voltages and temperatures, that are sampled by the SCA.

\subsection{Busy path} Should a FELIX FPGA or server be near to overflowing its buffer space, it can assert a BUSY signal on a dedicated electrical line.
The ``OR'' of all such lines requests the Central Trigger to stop generating the Level-1 Accept triggers, thus preventing the Front-ends from (eventually) sending more data, thereby allowing FELIX to clear its buffers.
Busy can be asserted when FELIX suspends transmission of events in response to the swROD sending an XOFF via Ethernet.
The NSW Read Out Controller can send BUSY requests to FELIX, but this feature is not foreseen to be active during normal detector operation.
The Pad Triggers and Trigger Processors will assert BUSY should their buffers become full.

\subsection{Trigger data path}
\label{sec:trigpath}
The trigger paths contain hits from the VMM ``Time-over-Threshold' (ToT)' outputs for the sTGC pads, the VMM ``direct'' 6-bit charge data outputs for sTGC strips, and the VMM ``Address in Real Time' (ART)' output for the \MM.
The digitized data is tagged by the trigger data serializers (the ART and TDS ASICs) with the value, the ``BCID'', of a BC counter which remains attached to the data as it moves through the system.
All the trigger primitives are sent to the Trigger Processors, one per sector per detector type.
The sTGC and \MM Trigger Processors separately find track segments from hits in their respective detectors.
For sTGC, hits are the centroids of charges induced on strips.
For \MM, hits are the channel number of the first hit in the bunch crossing per Front-end VMM ASIC.
Hits are used to find track segments pointing to the Interaction Point.
On every bunch crossing, the sTGC Trigger Processor merges the sTGC and \MM segments and sends them to the Sector Logic.
The Trigger Processor is described in Section\,\ref{sec:trigproc}.

\begin{figure}[ht]
\centering
\includegraphics[width=0.9\textwidth]{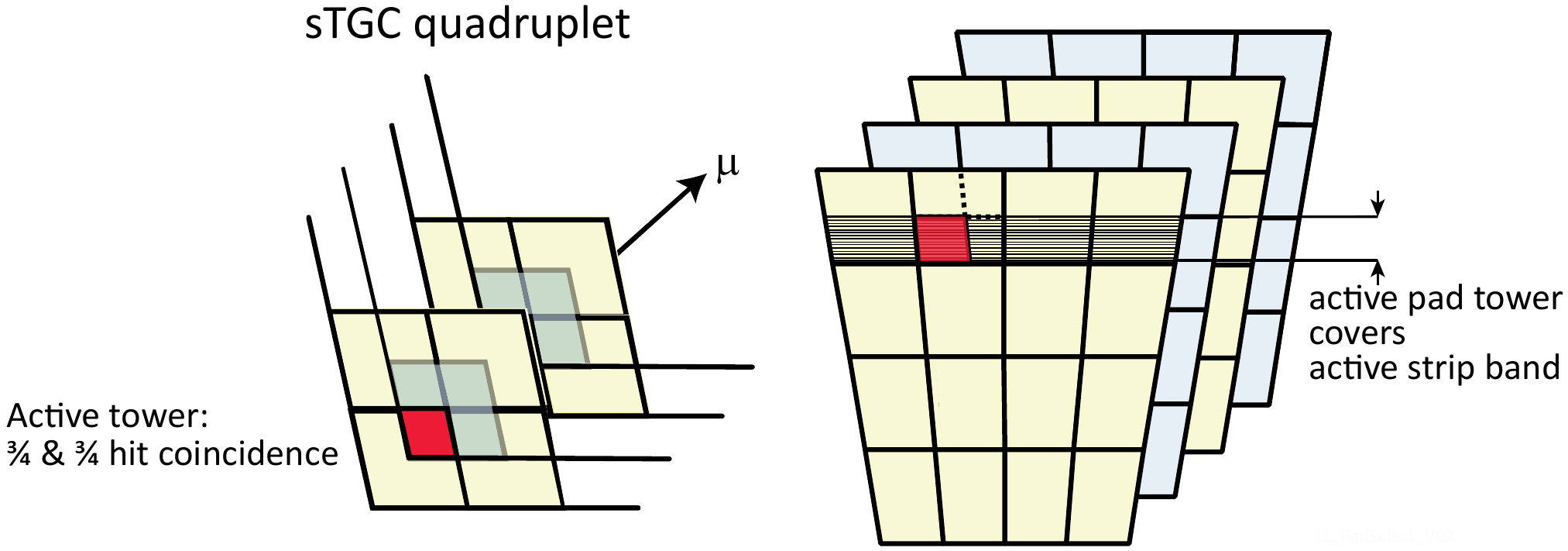}
\caption{A band of strips in each layer is selected by a particle making a 3-out-of-4 hit coincidence in a pointing tower of sTGC pads in each quadruplet.
The pads in half of the layers are shifted by half a pad in both directions to increase the resolution.
Eight-layer towers pointing to the interaction point are defined by the overlapping physical pads (shown in grey) which identify a logical pad (in red) in each layer.}
\label{fig:LL_PadSelect}
\end{figure}

\label{sec:sTGCtrigger}
\para{The sTGC trigger}
Ideally, one would read out all the strips in a sector directly to that sector's Trigger Processor.
However, reading out a 6-bit charge for each of 282,000 strips on every bunch crossing would require the huge bandwidth of almost 70\,Tb/s.
The power, cooling, and cost of current electronics are prohibitive.
In order to reduce the number of strips to be transferred to the Trigger Processor, the NSW uses eight-layer towers of sTGC pads pointing to the interaction point to provide a pre-trigger.
See Figure\,\ref{fig:LL_PadSelect}.

The pre-trigger per sector is formed by the Pad Trigger board (Section\,\ref{sec:pad_trigger}) on the rim of the NSW.
Coincidences between layers of the towers identify up to four bands of strips in each of the eight layers.
The Pad Trigger signals the TDS ASICs that contain those bands to transmit the strip charges to the Trigger Processor via the Router (Section\,\ref{sec:router}).
Note the zigzag path in Figure\,\ref{fig:LL_NSW_ElxOvr} and its consequent significant increase in trigger latency.
The Trigger Processor receives the strip charges and calculates a centroid for each layer.
These are used to calculate $r$ and $\Delta\theta$ of a track segment and to apply the $\Delta\theta$ cut mentioned above.

\begin{figure}[ht]
\centering
\includegraphics[width=0.7\textwidth]{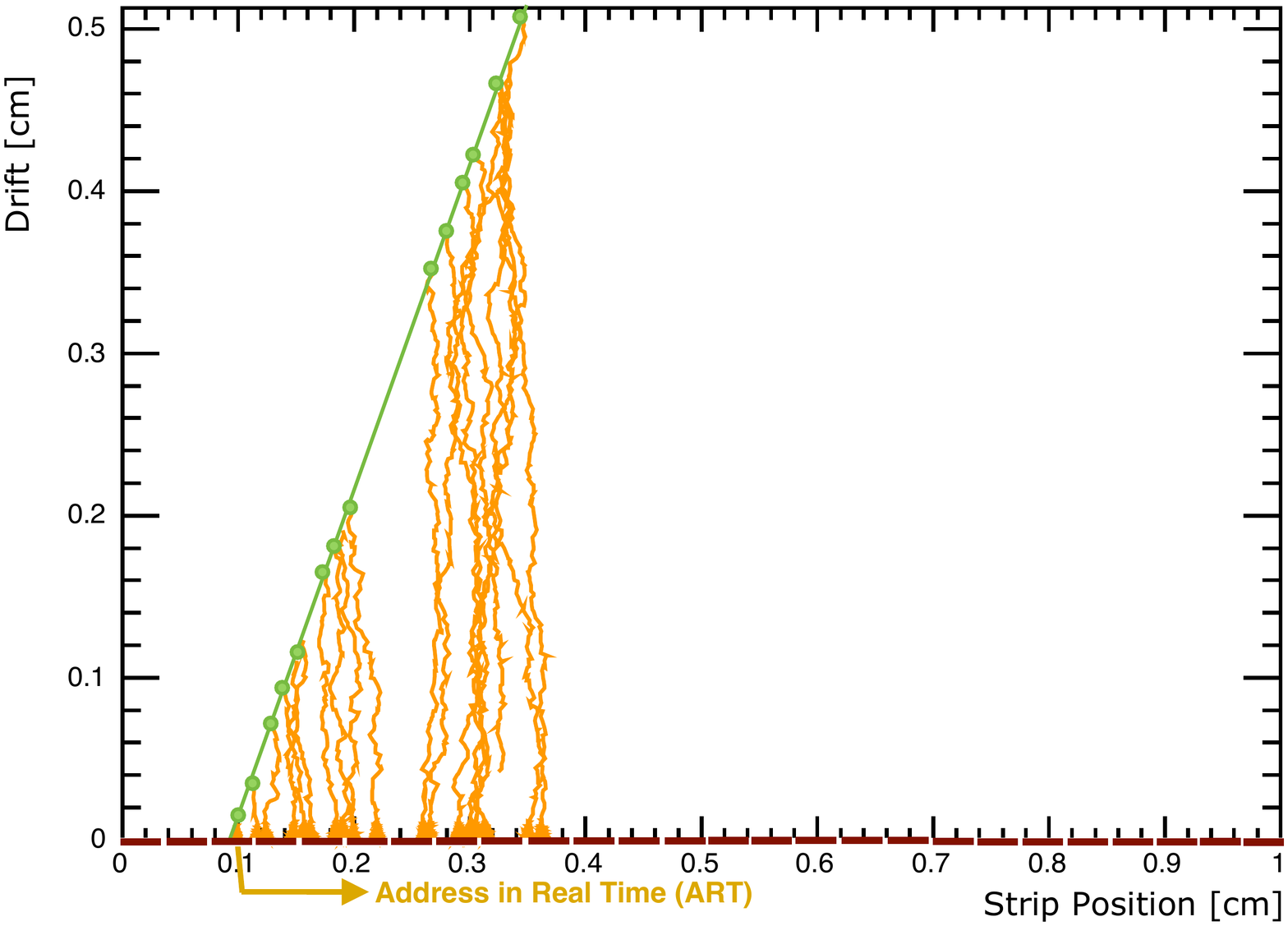}
\caption{A simulated event showing the ionisation from a particle crossing the \MM detector at an angle.
The address of the strip to which the charge arrives first is output as an ART signal\,\cite{georgePhd}.}
\label{fig:ART_concept}
\end{figure}

\label{sec:mmtrigger}
\para{The Micromegas trigger,} although being a 2.1\,M channel system, utilises the Address in Real Time (ART) output of the VMM to scale down the system to $\sim$262\,k channels for the trigger. The concept utilises the fine pitch of the \MM detectors and the spread of ionisation charge for particles crossing the detector at an angle\,\cite{georgePhd}. The VMM sends out the address of the channel that presents the earliest signal in every bunch crossing.
For the fine $\sim$0.45\,mm strip pitch, this is a good approximation of the coordinate perpendicular to the strips direction; see Figure\,\ref{fig:ART_concept}.
The address is sent to the ART ASIC on the ADDC board which receives the addresses of 32~VMMs and selects eight of them to be sent out to the Trigger Processor.
The 32~ART ASICs in a sector collect the addresses from all the eight layers and transmit them serially on 32~fibres to the Trigger Processor (one per sector) which forms track segments.
Since the drift time of the \MM detectors can extend up to 150\,ns, the ART ASIC has the option to mask the input of a VMM which has already provided the strip with the earliest time within the drift time of the \MM.
Therefore, signals originating from the same particle track, but from a different strip, can be discarded if desired.


\section{FELIX}
\label{sec:felix}

\begin{figure}[th]
\centering
\includegraphics[width=0.85\textwidth]{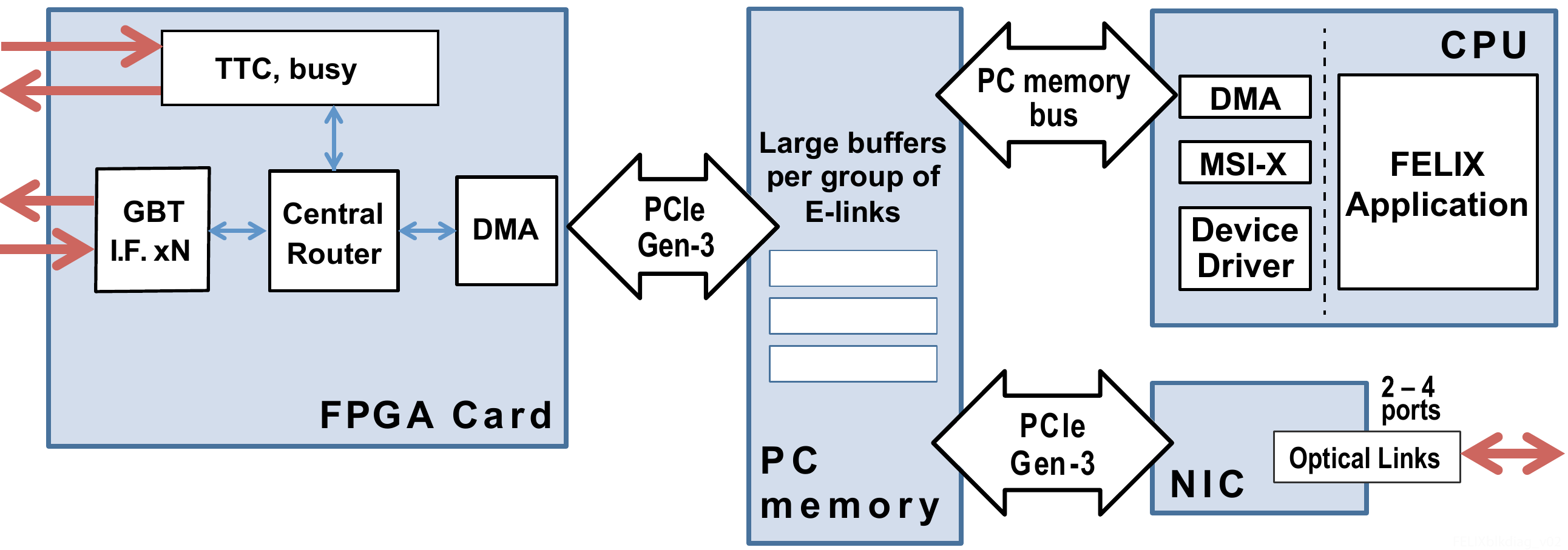}
\caption{Block diagram of the FELIX server. A server may contain up to two FPGA cards.
         The  Network Interface Card (NIC) for NSW has two 25\,Gb/s ports and supports RDMA\,\cite{RDMA}.} 
\label{fig:FELIXblkdiag}
\end{figure}

The Front End Link eXchange (FELIX), developed by the ATLAS Trigger and DAQ project\,\cite{Panduro-Vasquez:2022oR, Levinson:2799865, Paramonov:2021jpz, PanduroVazquez:2020mnk, felixHW, Trovato:2019pui, FelixUserGuide}, 
interfaces 4.8\,Gb/s optical links from GBTx ASICs that aggregate several slow serial copper ``E-links'', to an industry standard Ethernet network.
These slow links carry Front-end readout, configuration, calibration and detector monitoring data.
Acting similarly to a network switch, FELIX routes these slow links individually between Front-end electronics and software processes on the network, such as those shown in Figure\,\ref{fig:LL_NSW_ElxOvr}.
Furthermore, it distributes the TTC (Timing, Trigger and Control) signals, including the LHC Bunch Crossing clock, to all the NSW electronics.
FELIX is built from custom PCIe FPGA cards\,\cite{felixHW} hosted in commercial Linux servers, each equipped with a high-performance Ethernet interface card; see Figure\,\ref{fig:FELIXblkdiag}.
FELIX provides a common platform for some ATLAS Phase\,1 subsystems and will do so for all ATLAS subsystems in Phase\,2.
The use of commodity components and the sharing of a common platform reduces hardware, firmware and software effort.

The New Small Wheel uses the version of the FELIX FPGA card with 24 4.8\,Gb/s links in each direction in so-called GBT mode.
Each From-detector optical link carries data from between 9 and 21 slow, copper twin-ax serial links, ``E-links''; To-detector links carry between 7~and 18 serial E-links.
The slow serial E-links are aggregated to the 4.8\,Gb/s fibre by the GBTx ASICs on L1DDC boards or directly by the Trigger Processor Carrier FPGA.


FELIX receives TTC information from an ALTI\,\cite{ALTItwiki} Timing, Trigger \& Control (TTC) module via a fibre connection.
It also asserts BUSY on a dedicated electrical line should its FELIX FPGA or server buffers be near to overflowing.

For Phase\,1, there are 60 FELIX FPGA boards in 30 servers.
Phase\,2 requires 40~additional boards (based on 24~input links per FPGA).
Although the Phase\,2 fibres reach the radiation-protected underground room (USA15), the Phase\,2 FELIX boards and servers will not be installed until Phase\,2.

%

\subsection{Geographic names}
\label{sec:geonames}

There are over 22,000 E-links in the NSW.
For the software to have the ability to address them according to their data type and the exact region of the NSW to which they are connected,
each one is given a geographical or logical name, a so-called Detector Resource Name\,\cite{geonames}.
Referencing by geographic name is much clearer, less error prone and easier to maintain than by its physical FELIX connection.
Also switching to spare fibres or spare FELIX servers is transparent to the software.
The names are independent of specific connections of the multi-fibre bundles from the detector to FELIX boards and servers.
The mapping is done via ``FELIX-ID''s\,\cite{FELIXID}, with NSW experts maintaining the translation from fibres and E-links to FELIX-IDs and
FELIX experts maintaining the translation from FELIX-IDs to  FELIX servers.
An example geographical name is:
``{\small\textsf{MM-A/V0/L1A/strip/S4/L3/R11/E3}}''.
It refers to L1A data from Micromegas, Version\,0, Endcap\,A, strips, Sector\,4, Layer\,3, Radial position\,11, Readout controller E-link\,3.
Corresponding to the string is a 32-bit binary representation that can be used within data records to identify their data.


\section{Trigger Processor}
\label{sec:trigproc}

\begin{figure}[htb]
\centering
\includegraphics[width=0.95\textwidth]{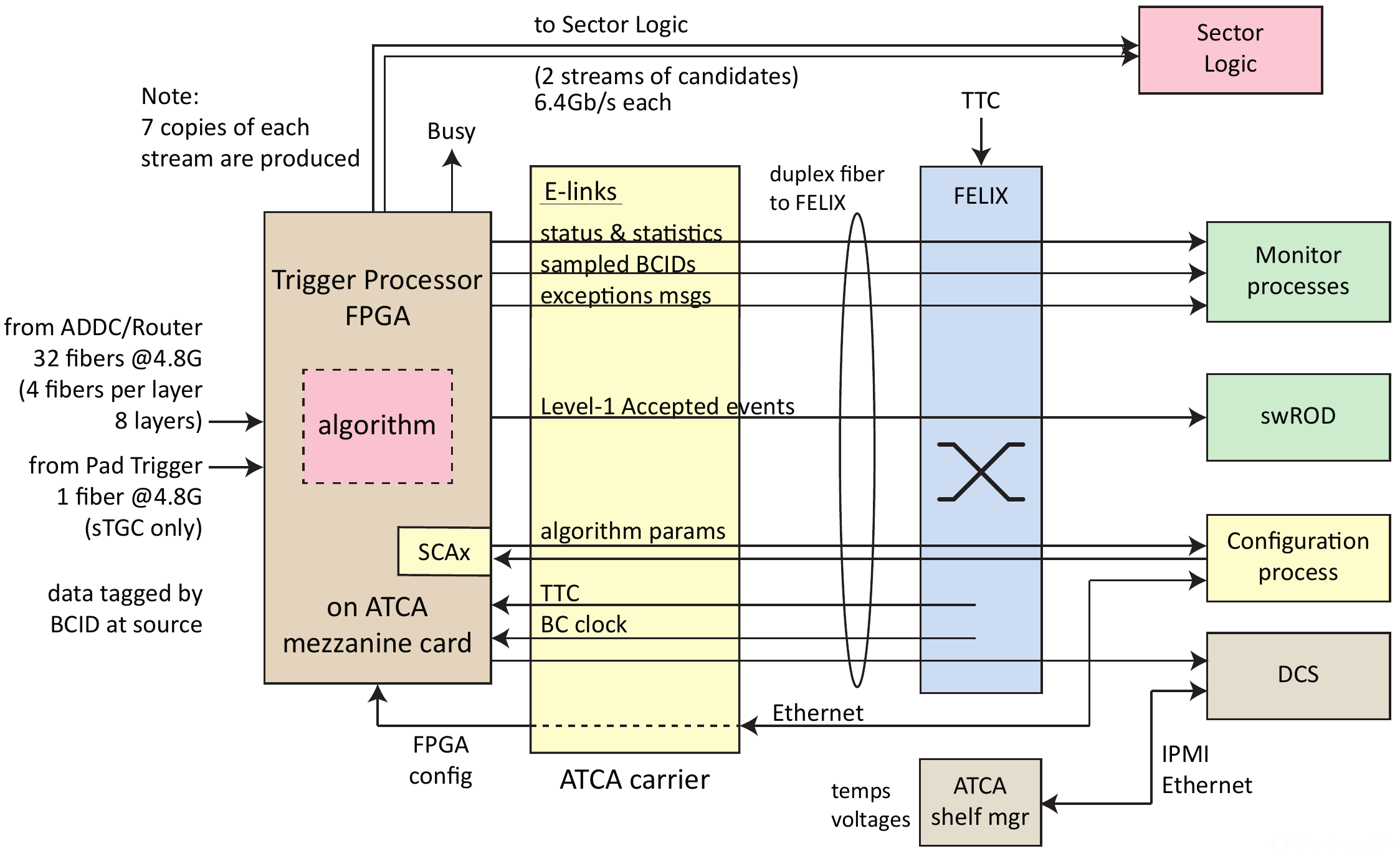}
\caption{Context diagram for the Trigger Processor }
\label{fig:TrigProcContext}
\end{figure}

The Trigger Processor for each sector is located in the radiation-protected room, USA15.
On every bunch crossing, each NSW Trigger Processor sends to the Sector Logic up to eight unique track segments that point to the \gls{Big Wheel}.
The segments must point within $\pm$15\,mrad of the infinite momentum track from the interaction point.
If there is a NSW track segment that matches a Big Wheel track segment, the Sector Logic sends the \pT information of the segment from the Big Wheel with a ``NSW'' flag to the Muon-to-Central Trigger Processor Interface (MUCTPI).
If there is no NSW track segment with a Big Wheel track segment, the Sector Logic sends the \pT information of the segment without the flag.
The MUCTPI/CTP can make a decision, according to the Level-1 trigger menu, based on the \pT and the flags.
Due to the mismatch in Big Wheel and New Small Wheel sector coverage, the Trigger Processor sends its segments to up to seven Sector Logic modules.
A context diagram for the Trigger Processor is shown in Figure\,\ref{fig:TrigProcContext}.

The Trigger Processors find track segments from the trigger primitives that they receive as described in Section\,\ref{sec:trigpath}.
The \MM ART ASICs collect hits from the \MM Front-end boards and serialize them onto a fibre link to the Trigger Processor.
For bands of strips, the sTGC strip-TDS ASICs send their strip charge information to the sTGC Trigger Processor via the Routers.
In both cases, 32~fibres at 4.8\,Gb/s connect to the Trigger Processor per sector.
The sTGC Trigger Processor also receives input from the Pad Trigger board of its sector.


\subsection{Hardware}
\begin{figure}[t]
\centering
\includegraphics[width=0.99\textwidth]{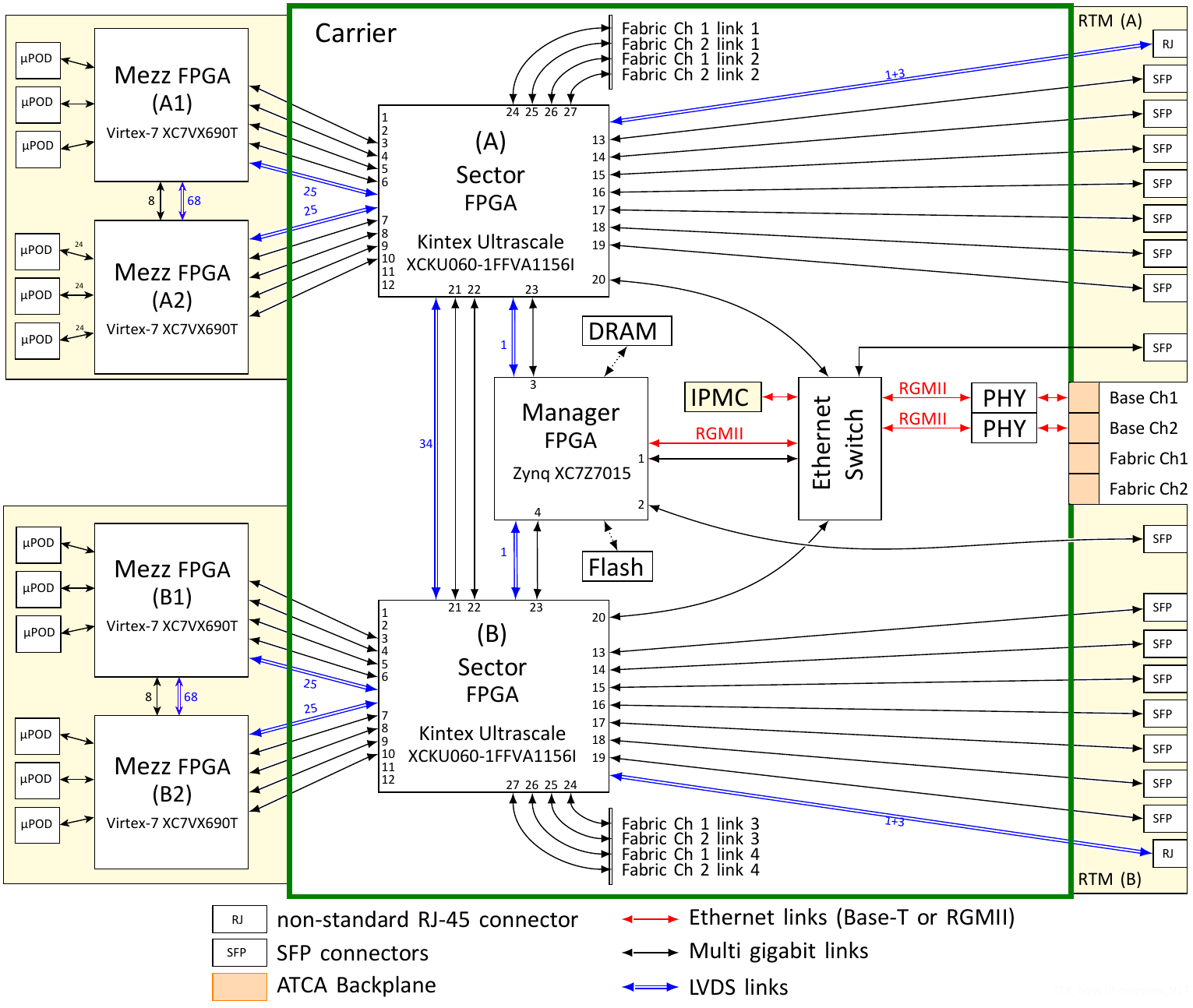}
\caption{Block diagram of the Trigger Processor ATCA card showing the Carrier, two mezzanines, the Rear Transition Module (RTM)
            and an Intelligent Platform Management Controller (\gls{IPMC}) card.
            Each mezzanine hosts an sTGC FPGA and a \MM FPGA for a sector.}
\label{fig:TCP_NSWTP-overview}
\vspace{-10pt}
\end{figure}

The NSW Trigger Processor platform is built according to the Advanced Telecommunications Computing Architecture (\gls{ATCA}) standard\,\cite{ATCA}.
A Trigger Processor ``blade'' consists of a carrier card, two mezzanine cards and a Rear Transition Module (RTM).
See Figure\,\ref{fig:TCP_NSWTP-overview}.
The sTGC and \MM use the same hardware for their trigger logic, but with different firmware.
A single mezzanine card with two Xilinx FPGAs\footnote{XC7VX690T}, one for \MM data and one for sTGC data, handles one sector.
The two FPGAs can communicate with each other via 68 fast (640\,Mb/s) low-latency \gls{LVDS} signals.
The ATCA blade supports two such mezzanines. The backplane signal connections are so far not used.
More information on the Trigger Processor hardware can be found in\,\cite{NswTpHW}.

\para{Carrier}
The Carrier conforms to the \gls{ATCA} standard.
Two Xilinx Ultrascale FPGAs\footnote{XCKU060-1FFVA1156I}, one for each sector, hold the FELIX interface.
Should it be required, buffering and preparing the Level-1 accepted data could be moved here from the mezzanines.
A Xilinx Zynq System-on-Chip FPGA\footnote{XC7Z7015\,\cite{zynq}}, running Linux CentOS\,7 on the Zynq's 32-bit ARM CPU, provides some board management functions.
An on-board Ethernet switch connects the Carrier's Sector FPGAs, the Zynq processor, the Zynq FPGA fabric and the CERN IPMC card\,\cite{CERN-IPMC} to the external network.
Board management functions include
configuring the Carrier and Mezzanine FPGAs, via the Xilinx Virtual Cable (XVC)\,\cite{XVC} over the Ethernet network,
configuring jitter cleaners, clocks and the Ethernet switch and reading the board ID.
XVC can also be used as a debug interface for the firmware of all the other FPGAs.
Other board management functions are provided by the IPMC card.
In addition, there is a dedicated high-speed serial link between the Zynq and each of the two Carrier FPGAs which may be used for debugging or monitoring purposes.

\para{Mezzanines\footnote{The custom mezzanines do not conform to the ATCA mezzanine standard.}}
In addition to the two FPGAs mentioned above, the mezzanine connects to 72 fibres via three 12-channel 10\,Gb/s microPOD\,\cite{microPOD} optical receivers, three 12-channel 10\,Gb/s microPOD optical transmitters per FPGA
and jitter cleaners for the required design clocks and FPGA transceiver reference clocks.
A Module Management Controller (\gls{MMC}) from Samway Electronic SRL\,\cite{Samway,MMC} on each mezzanine communicates with the IPMC management card on the Carrier.

\para{Rear Transition Module (RTM)}
The RTM provides several \gls{SFP} cages for serial transceivers: one connected to the on-board switch for an Ethernet transceiver, another to the Zynq,
and several connected to the Sector FPGAs for the fibre transceivers to FELIX.
The RJ-45 connectors provide four LVDS lines to each of the Sector FPGAs.
The RTM provides also a clock input which may alternatively supply a clock to the entire Carrier system.
For self-triggering of the NSW, an LVDS trigger output signal is available from one of the RJ-45 connectors.
There is also an MMC on the RTM.


\subsection{FPGA firmware}
\begin{figure}[ht]
\centering
\includegraphics[width=0.99\textwidth]{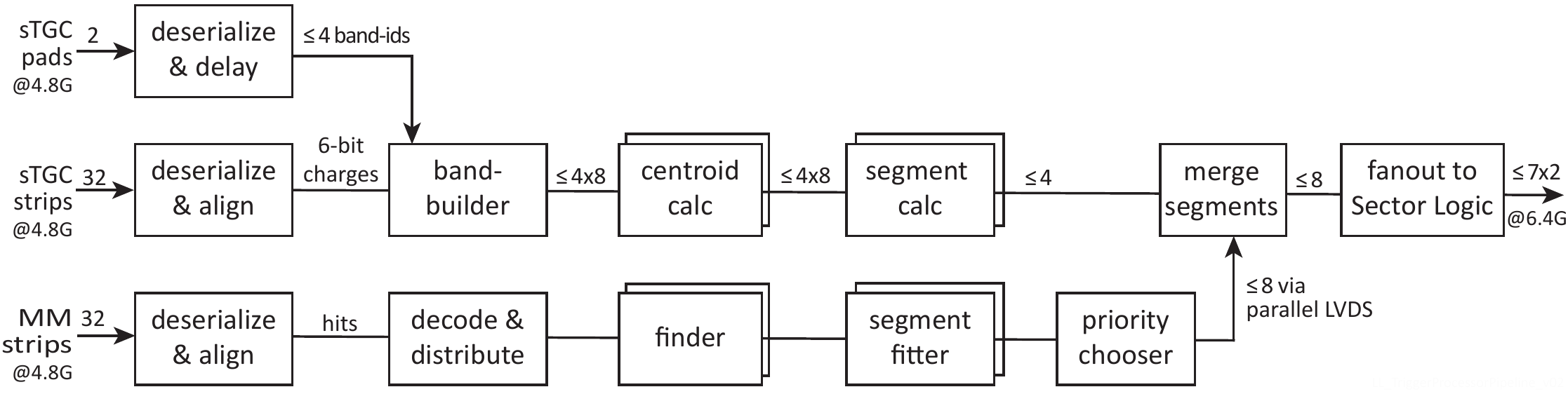}
\caption{The Trigger Processor segment finding pipelines implemented in the mezzanine FPGAs.}
\label{fig:TrigProcPipeline}
\end{figure}

The Trigger Processor segment finding pipeline in the mezzanine FPGAs is shown in Figure\,\ref{fig:TrigProcPipeline}.
A brief overview is given below and further details can be found in\,\cite{TrigProcFWspec}.

\subsubsection{\MM specific trigger processing}

\begin{figure}[ht]
\centering
\includegraphics[width=0.99\textwidth]{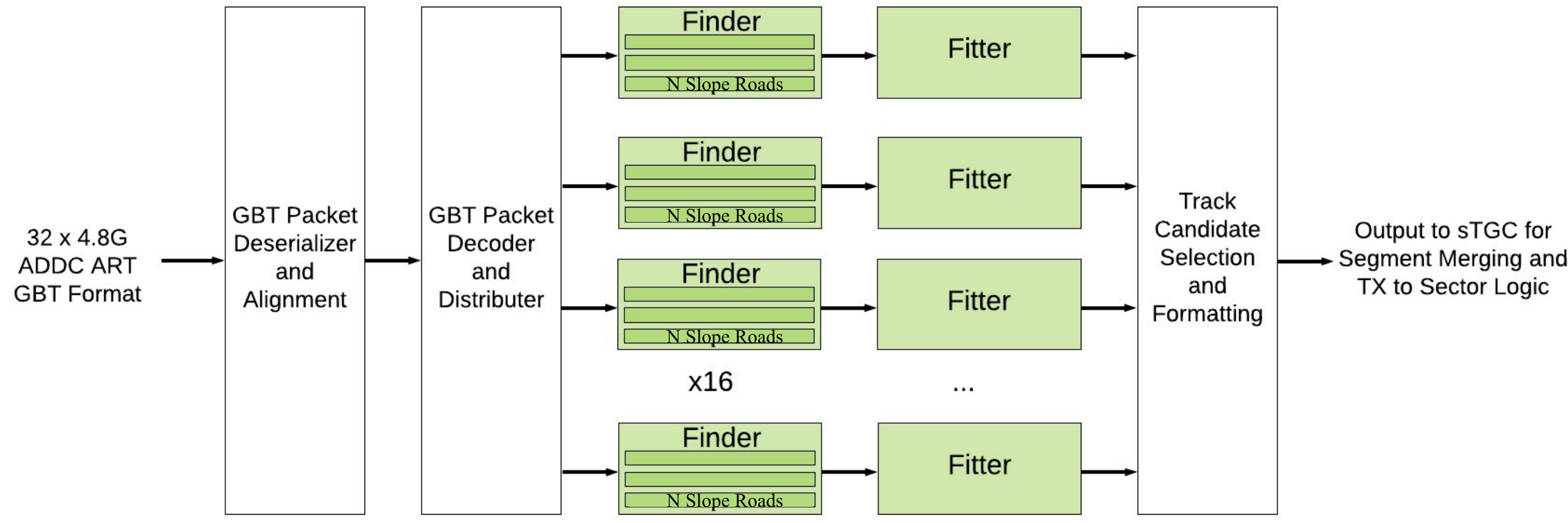}
\caption{The \MM Trigger Processor block diagram.}
\label{fig:GI_MMAlg}
\end{figure}

The main blocks of the \MM Trigger Processor (MMTP) algorithm are shown in Figure\,\ref{fig:GI_MMAlg}.  Its functionality requires sixteen copies of the algorithm operating in parallel. Each copy is sharing information with its neighbour to avoid boundary issues.  The algorithm generates its 320\,MHz clock from the main BC clock.

\para{Input capture and alignment}
The first stage of the firmware receives data from 32~GBT links from the 16 ADDC boards in one sector.  The deserializer uses the 320\,MHz clock derived from the main bunch crossing clock.  The data from the GBTx ASICs on the ADDC boards are captured by fixed latency receiver firmware\,\cite{GBTfixedLatency}.  In order though to account for each fibre length and provide fixed latency to a single BC clock, the data is registered twice using a clock that is phase aligned to the recovered BC clock. The first register is used to account for any individual fibre length differences. The second register provides a configurable phase adjustment but the phase is a single constant set for all fibres. This register aligns the data from all fibres and provides a fixed latency.

\para{Decoder}
Incoming strip hit addresses are decoded into global slope values. 
A strip's stored slope value is defined as the orthogonal distance between a given strip and the beam-line divided by the $z$ location of the relevant detection plane. It is pre-computed taking into account a strip offset and a $z$ position stored for each of the 8 planes and 16 radial segments of each wedge. The slope range is divided into approximately 1000 slope-roads which are used to form a track candidate.

\para{Finder}
The track finding algorithm is slope-road based with the wedge divided from bottom to top into approximately 1000 roads which corresponds to about eight detector strips per road. Hit data from the decoder
is routed to the corresponding road and that road is marked as being hit. Each slope-road can hold a single hit for each plane and this hit
expires after the hit configurable integration time of up to eight BC clocks. The slope-roads will overlap each other to accommodate tracks on the  boundaries.
Each slope-road is checked once per bunch crossing to determine if a coincidence threshold has been met.
Coincidence requires a minimum number of planes to be hit and the oldest hit of the track to be expiring.
Coincidence identification is accomplished using combinatorial logic and a priority encoder. The strip
number and slope for each hit are calculated and passed to the track fitting algorithm.

\begin{figure}[ht]
\centering
\includegraphics[width=0.7\textwidth]{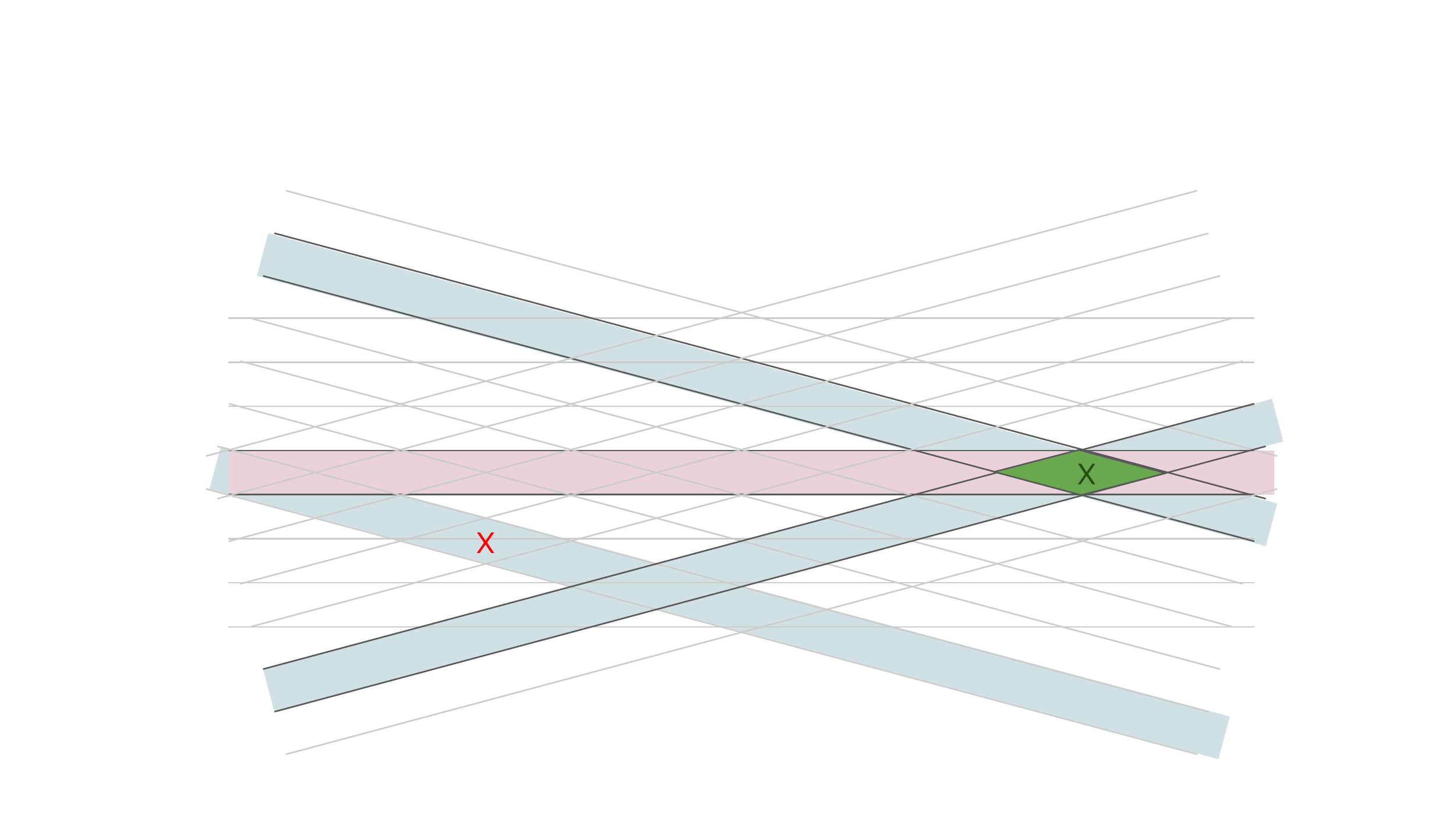}
\caption{Illustration of the \MM XUV trigger candidate. The X-plane candidate is merged with the UV-candidate to form a coincidence on the so-called diamond road which corresponds to the geometrical coincidence of the strips.}
\label{fig:GI_XUVTrigger}
\end{figure}

The Finder algorithm separates the X-plane (perpendicular to the $\eta$ coordinate) and the UV-plane coincidences ($\pm1.5^{\mathrm{o}}$ with respect to the X-plane respectively). The fitter will first identify a coincidence trigger on the X planes
and then scan from left to right all of the UV-roads that overlap the triggered X-road. The overlapping
area of UV-roads are the so-called ``diamonds''. Figure\,\ref{fig:GI_XUVTrigger} shows such a trigger coincidence.  For each X trigger there can be up to 57 UV-roads that are searched. The number of diamonds is configured for each algorithm region to match the length of
the Micromegas detector strips. Every bunch crossing, each of the 16 algorithm regions can process two
independent X-road coincidences. For every X-road coincidence, the Finder can process three UV-road
coincidences.
The amount of resources used by the Finder is proportional to the product of the number of roads and the number of planes.
The Finder implementation focuses on minimizing the resources used in each slope-road.
In the total design, it uses the most resources.

\para{Fitter}
In the fit, individual hit slopes in a slope-road are used to calculate global slopes associated
with each plane type.
From these  slopes, the  expressions for the fit quantities $\theta$ (the zenith),  and $\phi$ (the azimuth) can be derived. The $\Delta\theta$ (the difference in $\theta$ between the direction of the segment extrapolated back to the interaction point
and its direction when entering the detector region) can be derived.

The expected tolerance for installation of the Micromegas chambers is roughly 2\,mm, although it is possible
to have larger deviations. In order to retain the optimal performance of the trigger, alignment corrections
need to be performed. The corrections need to be performed rapidly to minimize latency, and also to have
a small number of constants to avoid large look-up tables.

\para{Candidate selection and output}
The process starts by collecting the segments from the
16 algorithm regions. Each region can generate eight segments every bunch crossing. A priority encoder
is used to select eight total segments each bunch crossing from all regions. The selected segments
are transferred to the sTGC FPGA using sixteen  640\,Mb/s LVDS signals. The receiver in the sTGC FPGA, uses
independently delayed versions of the same signal to set the sampling point to the center of the data ``eye''
and provide segment data to the sTGC logic that will be aligned to the sTGC clock with no additional
clock domain crossing being necessary.

\subsubsection{sTGC specific trigger processing}

\para{Input capture and deskew}
The FPGA multi-gigabit serializer-deserializer pairs do not have fixed latency.
After every power cycle or link reset, the data are written to the deserializer's parallel output bus with an uncertain shift of $\pm$1\,deserializer output bus clock.
The capture block waits for a configurable worst-case delay, including differing cable delays, to ensure all the data has arrived.
This delay has been determined by observation, and all inputs are continuously monitored to confirm that they never exceed the configured worst case.
The strip-TDS band data are transmitted in four 30-bit scrambled packets in one BC.
The packet format is shown in\,\cite{HU2022167504}.)
The Xilinx FPGAs in the Router and Trigger Processors work with 20-bit packets.
The capture block unscrambles the data and rebuilds the 104-bit payload.

\para{Band-builder} The Pad Trigger may find up to four coincidence towers.
For each band of strips passing through a tower, each of eight strip-TDS ASICs, one (or two for split bands) per layer,
will transmit the band's strip charge information via the Router handling its layer to the Trigger Processor.
Each Router has four fibres to be able to transmit up to four bands.
In the case of multiple bands, the bands' strip charge information arrives at the Trigger Processor on one of the four fibres from each Router, but not necessarily with the same fibre id for all of the eight layers.
For each band, the Band-builder routes the data for all eight layers of that band to one of the four instances of the trigger algorithm.
Latency is reduced by using the Band-ids and $\phi$-ids provided by the Pad Trigger a few BC's earlier than the Router data arrive
to address all the Look-up Tables needed to process that data for that Band-id and $\phi$-id.
The 128 channels of the strip-TDS are not necessarily aligned with the boundaries of the bands.
Consequently, about 7\% of the bands in a layer span over two strip-TDS ASICs.
The Band-builder concatenates the two charge vectors received from the two input fibres.
There are 86 bands for which all eight layers are available for use in the trigger.

\para{sTGC algorithm}
There are four identical instances of the algorithm that take the strip charge information of the bands in eight layers and calculates a track segment.
A track segment consists of a radial R-index, an azimuthal $\phi$-id and a $\Delta\theta$ as described in Section\,\ref{sec:sTGCtrigger}.
They are calculated from the centroids of the charge distributions received for each layer.
Each stage of the algorithm is described below:

\vspace{3pt}\noindent\textit{Cluster selection:}
If the charge distribution in a layer of a band is inconsistent with a minimum ionizing particle, then that layer of the band is rejected.
Valid charge distributions, ``clusters'', are defined by a look-up table whose address is a vector with ``ones'' for strips above a threshold
and whose value gives the location of the beginning of the cluster (or -1 for an invalid cluster).
The look-up table requires cluster widths between two and five and with at most one extra isolated noise strip. This rejects bands with ionization from neutrons and $\delta$-rays.

\vspace{3pt}\noindent\textit{Layer centroid:}
The centroid of each cluster is calculated and projected onto the plane of the wires.

\vspace{3pt}\noindent\noindent\textit{Quadruplet centroid:}
The required three or four valid centroids of a quadruplet are projected onto the central plane of the quadruplet,
assuming the track originates from the interaction point, and the average radial position is calculated.
This accounts for the different $z$\,positions of the valid planes.

\vspace{3pt}\noindent\textit{Segment calculation:}
To calculate  $\Delta\theta$, a piecewise linear approximation to $\tan(\theta_{\textrm{IP}} - \theta_{\textrm{local}}$) is made.
Each linear segment corresponds to the region of a Band-id.
The extrapolation of $R$ is done to the sTGC wire plane furthest from the Interaction Point and assigned to an R-index.
The $\phi$-id is provided by the Pad Trigger data.

%
%
%
%

\subsubsection{Segment merging and duplicate removal}
The \MM Trigger Processor finishes finding segments before the sTGC does.
It transfers up to eight track segments found to the sTGC Trigger Processor.
They are transferred via sixteen 640\,Mb/s LVDS signals within one BC clock.
They are held until the sTGC Trigger algorithms complete.
From the up to four sTGC segments and up to eight \MM segments, a maximum of eight segments can be sent to the Sector Logic.
The Merge block removes duplicates and drops segments beyond the eight allowed.
Priority is currently given to sTGC segments.
There are options to ignore one or the other of \MM or sTGC segments and for ``duplicates'', to take the $\phi$-id from \MM and the other variables from the sTGC segment.


\subsubsection{Ancillary functions}

\para{Connection to FELIX:}
The Trigger Processor interfaces to FELIX with the Xilinx\,\gls{GTH} serializer/deserializer and CERN's GBT-FPGA firmware\,\cite{GBT-FPGA, GBT_FPGA2}.  Together they emulate the GBTx ASIC.
This firmware runs on the Carrier Sector FPGAs. Each \MM and sTGC sector have a FELIX link.
The GBT-FPGA recovers the BC clock with fixed latency and decodes/encodes a received/transmitted 120-bit GBT packet every BC clock.
The received and transmitted 120-bit words are mirrored between Carrier and mezzanine FPGAs every BC via 8b/10b encoded serial transmission at 6.4\,Gb/s.
The mezzanines connect to the appropriate fields in the 120-bit word that correspond to specific E-links.
This mirror connection does not have fixed latency, and so is suitable only for the readout, configuration and monitoring paths, but not for the eight TTC signals.
These are transferred to the mezzanine via eight LVDS lines along with the BC clock.
Zero-delay PLL jitter cleaners recondition the BC clock and ensure fixed latency of the FELIX link interface.
A soft processor implemented in the Carrier Sector FPGA manages the PLL and GTH transceiver during the clock acquisition sequence and runtime operation.

\para{Readout of Level-1 Accept data:}
All input data, output segments and some intermediate data are captured every bunch crossing in FIFOs.
When a Level-1 Trigger arrives at the Trigger Processor, data from a configurable window of bunch crossings is formatted and queued in a derandomizer for output to FELIX via E-links.
To increase the total bandwidth, several 320\,Mb/s E-links can be used by sending groups of input channels and segment output via different E-links.
Furthermore, since Level-1 Triggers on consecutive bunch crossings are allowed (Phase-2), the readout block allows the data for a given bunch crossing to be transmitted in more than one output event packet.

\para{Configuration of operating parameters and reporting status:}
The SCAx package\,\cite{SCAxIEEE, SCXug} provides read/write access to operating parameters, configurable look-up tables and status words.
It emulates the \ItwoC channel of the SCA ASIC.
This enables the configuration and status reporting  to be done using the same software based on OPC\,UA,\cite{OPC-UA} as used for the Front-ends.
Unexpected conditions such as \gls{BCID} mismatch, can be reported by sending an exception code and context information via an exception E-link. These exceptions are further handled by software.

\begin{figure}[ht]
\centering
\includegraphics[width=0.9\textwidth]{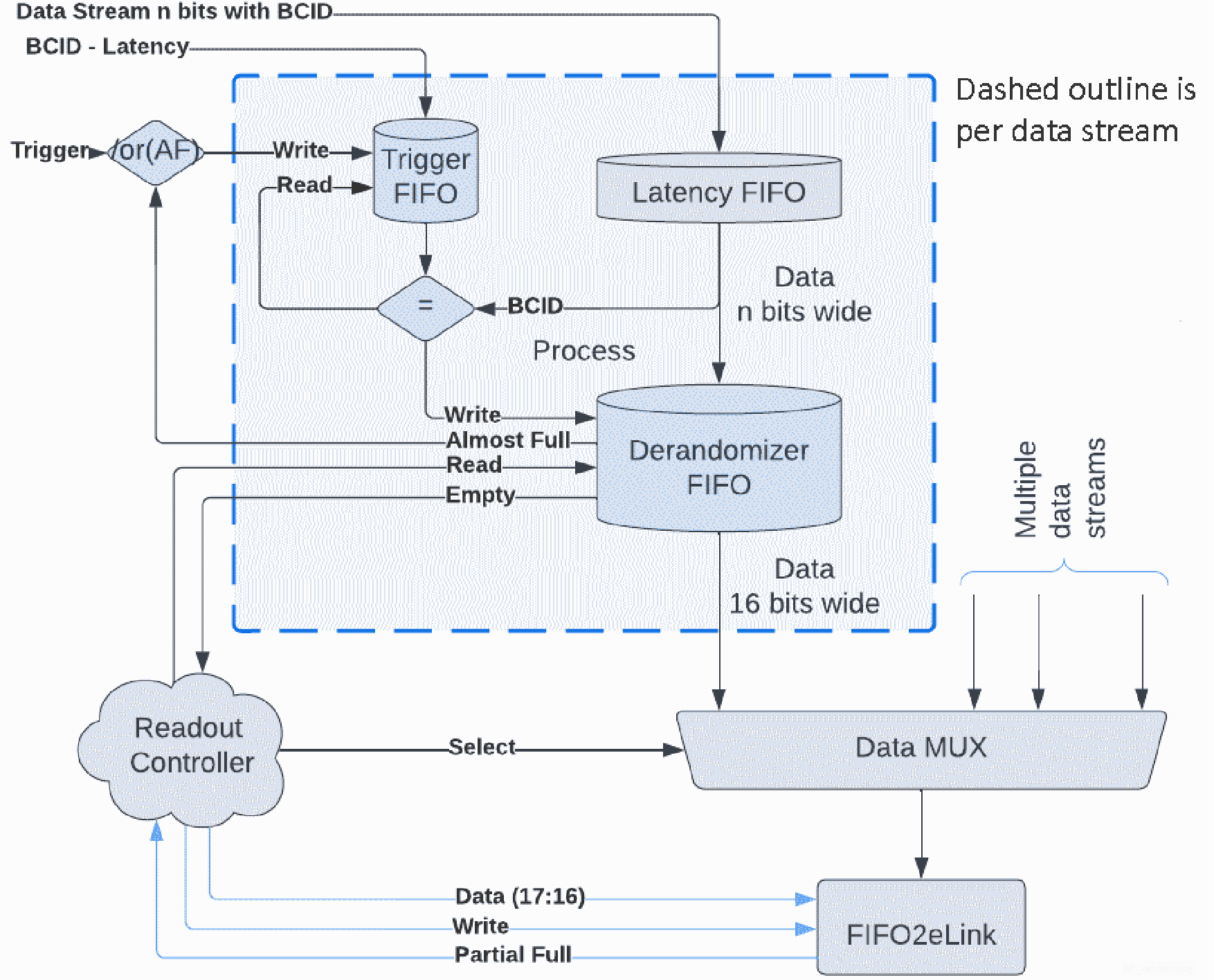}
\caption{Monitoring in the Trigger Processor}
\label{fig:NF_monitoring}
\end{figure}

\para{Monitoring ``interesting'' non-Level-1 Accepted bunch crossings:}
A separate readout E-link exists for monitoring bunch crossings of interest, e.g.\ receiving a 3-out-of-4 Pad Trigger coincidence or corner cases and anomalies.
Figure\,\ref{fig:NF_monitoring} shows the architecture of the monitoring logic.
Intermediate pipeline data such as band-builder output, segments, and merge candidates are buffered in the Latency FIFOs.
Various points along the processing pipeline may trigger the current bunch crossing as one to be monitored.
The data from triggered BC's are transferred to the Derandomizer FIFOs.
For bunch crossings that have been triggered, the Monitoring Readout Controller builds the data  for that bunch crossing from all Derandomizer FIFOs, into a monitoring packet that is sent out on an E-link dedicated to monitoring.
In this way, the FPGA firmware can be debugged and verified.

\para{Level-1 readout of all ATLAS for monitoring ``interesting'' NSW bunch crossings}
If data only from bunch crossings that are accepted by the Level-1 trigger are recorded,
we cannot know if NSW segments did not result in a trigger due to improper functioning of the NSW trigger.
To address this, the Trigger Processor can flag in the  packet sent to the Sector Logic that this bunch crossing is ``interesting'', e.g.\ a 3-out-of-4 coincidence.
The Sector Logic forwards the flag via the MUCTPI to the CTP,
which generates a Level-1 Accept with trigger type ``NSWMON'', unbiased by any other detector or trigger processing.
This allows a monitoring process to correlate NSW trigger data with the more detailed NSW front-end data, the Sector Logic and the MUCTPI data for that bunch crossing.
The NSWMON trigger type is broadcast to all swRODs via TTC; it allows other detectors and NSW sectors that did not generate that NSWMON event to exclude these events from their monitoring sample.
The CTP ensures that the rate of these events does not exceed a few 10's of Hertz.


\subsection{Bunch crossing synchronization of the trigger paths}
\label{sec:BCsync}

Data from the sTGC Pad Trigger, Routers and the \MM Trigger Processor for a given orbit and bunch crossing must arrive at a point in the sTGC Trigger Processor simultaneously.
This alignment is done by defining the first bunch crossing in the first orbit in a run by
broadcasting an Orbit Count Reset Request via the TTC path to the Pad Trigger and \MM Trigger Processor just before the start of the run.
A flag in the data transferred to sTGC from \MM and the Pad Trigger indicates if the data are before the first bunch crossing in the first orbit in the run.
The sTGC Trigger Processor then discards such data but saves unflagged data in a FIFO that holds it until needed in the sTGC pipeline.
For the sTGC path, the Pad Trigger sends the non-existent band-id, 0xFE, on the first bunch crossing of the first orbit in the run.
The reception of a data packet with this band-id by the sTGC Trigger Processor begins the tagging of bunch crossings in its pipeline as being in the run.

%
%


\section{CERN ASICs}
\label{sec:cern_asics}
The ASICs described in this section were developed by the CERN Electronic Systems for Experiments Group for use in several experiments.

\subsection{GBTx -- GigaBit Transceiver}
\label{sec:GBTx}

The GBTx ASIC\,\cite{Moreira:2009pem, Wyllie:2012cua}, aggregates many slow (2,\,4, or 8~bits per bunch crossing, i.e.\ 80, 160 or 320\,Mb/s)
serial data links called \emph{E-links} into a single serial link running at 4.8\,Gb/s.
It provides one such link in each direction; the two directions are completely independent.
The GBTx ASIC is radiation hard and uses forward error correction to assure uncorrupted data transmission. 
Its net throughput is 3.2\,Gb/s with error correction, or 4.48\,Gb/s without error correction in ``Widebus mode'' used by the ADDC (See Section\,\ref{sec:addc}.).
The 120-bit payload frames are transported with fixed latency, synchronous with the LHC bunch crossing clock.
One GBTx can transport event, configuration, control and monitoring streams, and, by virtue of its fixed latency, the bunch crossing clock and TTC (Timing, Trigger and Control\,\cite{TTC, Gallno:1999ws}) signals.
The GBT-FPGA firmware\,\cite{GBT-FPGA, GBT_FPGA2} implements the GBTx protocol into an FPGA.

\noindent\textbf{E-links}
consist of a serial input, a serial output and an output clock, all of whose rates are separately configurable.
\gls{E-links} use the SLVS\,\cite{slvs} standard.
The  properties of different E-links are somewhat constrained (see\,\cite{FelixUserGuide}).
The output data and the output clock are transmitted with a fixed phase with respect to the recovered BC clock. 
To accurately capture the input data, each E-link's internal receive clock phase with respect to data arrival must be calibrated.
Data is transmitted transparently.
In the NSW, all E-links carry 8b/10b\,\cite{8b10b} encoded data, except the SCA E-links which carry High-Level Data Link Control protocol (HDLC)\,\cite{HDLC} encoded data and the TTC links which are not encoded.
The 8b/10b encoding further reduces the throughput to 2.56\,Gb/s.
Control symbols in the 8b/10b standard are used to delineate start and end of packets.
The jitter of the E-link clocks is of the order of 5\,ps which is within the working range of the data acquisition but is marginal for \gls{FPGA} gigabit transceiver reference clocks\,\cite{Elinkjitter}.
To accurately capture the input data, each E-link's internal receive clock phase with respect to data arrival must be calibrated. See Section\,\ref{sec:PhaseAlign}.

\subsection{SCA -- Slow Control Adapter}
\label{sec:SCA}
The \gls{SCA} \gls{ASIC} (Slow Control Adapter)\,\cite{GBT-SCA}, developed by CERN, is used to configure all the ASICs in the NSW.
It is part of the GBT chip-set and communicates to FELIX through a 80\,Mb/s GBTx E-link.
The E-link is encoded in \gls{HDLC}.
It provides communication to other chips using \glslink{I2C}{I$^2$C}, \gls{SPI}, JTAG protocols as well as General Purpose I/O.
The industry-standard OPC\,UA\,\cite{OPC-UA} was chosen as the software interface to the SCA to be easily compatible with the Detector Control System's SCADA program and to take advantage of industry standard software. Although initially the ASIC was featuring a faster ADC with more accurate sampling, in the end,  this could not be realised and a slower implementation based on a Wilkinson architecture was adopted.
The ADC implemented is a 12-bit ADC\,\cite{SCA-ADC} based on a single-slope Wilkinson architecture with a range up to 1.0\,V with a maximum conversion rate of 3.5\,kHz.
Conversions are triggered by software commands.
This implementation underestimates the VMM's measured ENC and scaling should be implemented (See Section\,\ref{sec:SCAconfig}).

\subsection{FEAST -- DC-to-DC converter}
\label{sec:FEAST}
The CERN \gls{FEAST ASIC}\,\cite{FEAST2.1} provides up to 10\,W of Point-of-Load non-isolated DC power from $\sim$10\,V input.
\gls{FEASTMP} pluggable modules\,\cite{FEASTMP} are also used.
The output voltage is set by a resistor.
Several voltages between 1.2\,V and 3.3\,V are required by the various NSW boards.
The device is radiation tolerant for Total Ionization Dose (TID) above 200\,Mrad(Si) and has an adjustable switching frequency between 1\,and 3\,MHz. For the FEAST ASICs in the NSW, the range is between 1.3 and 1.7\,MHz.

\subsection{VTRx, VTTx -- optical interfaces}
\label{sec:VTRxVTTx}
The \gls{VTRx} and \gls{VTTx} are radiation tolerant bi-directional and dual transmitter optical link interfaces respectively.
They target data transmission between the on-detector and off-detector electronics at rates up to 5\,Gb/s with an emphasis on High luminosity LHC level radiation resistance, low power dissipation and low mass components.
They are protocol-agnostic and use a standard \gls{LC}-LC optical connector mounted on a pluggable module.
Their development was a joint ATLAS-CMS project, the Versatile Link project\,\cite{versatile, Xiang:2012vua,Vasey:2012xjw}.
Care must be taken to adequately cool them which was foreseen in the NSW project.

\section{NSW ASICs}
\label{sec:asics_boards}
\label{sec:asics}

\begin{figure}[ht]
\centering
\includegraphics[width=0.45\textwidth]{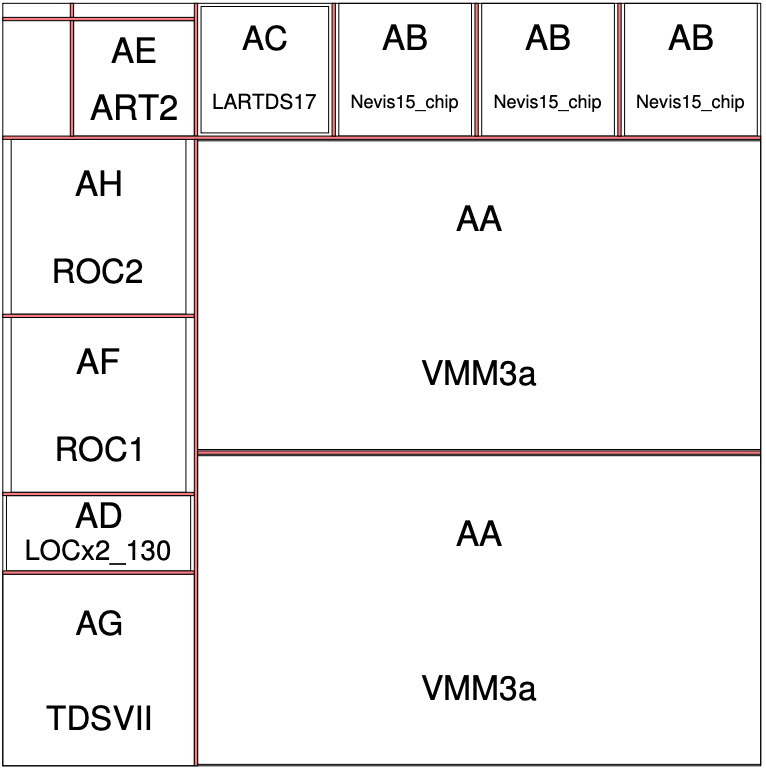}
\caption{The NSW wafer reticle, showing the locations and relative sizes of two copies of the VMM3a, one ROC1 and one ROC2 along with the TDSvII and the ART2 ASIC. There are other ASICs for the ATLAS Calorimeter in the same reticle. As seen, the reticle contains two versions of the ROC ASIC. This was done in case the ROC2 version had an issue and the fall-back solution for scheduling reasons was the ROC1.  There are 86 reticles in the 8-inch wafer. Those on the perimeter, however, could not be used.  Each wafer contains 113 VMM3a, 66 ART2, 62 ROC1, 65 ROC2 and 65 TDSvII chips. The number that can be extracted depends on how the wafer is cut. The reticle size is $20798.32 \times 20795\,$\textmu m.}
\label{fig:reticle}
\end{figure}

The four NSW ASICs (VMM, ROC, TDS, ART) are fabricated in the Global Foundries 130\,nm 8RF-DM CMOS process on a common 8-inch wafer.
The reticle plan is shown in Figure\,\ref{fig:reticle}.

\subsection{VMM -- Mixed-signal front-end ASIC}
\label{sec:VMM}
The VMM\,\cite{9724214, Iakovidis_2020, vmmuserguide} is a custom Application-Specific Integrated Circuit (ASIC). It is designed to be the front-end ASIC of both the Micromegas and sTGC detectors of the New Small Wheels.  For the NSW, it is packaged in a 400-ball $21\times 21\, \textrm{mm}^2$ Ball Grid Array (\gls{BGA}) with 1\,mm ball pitch.

\subsubsection{Requirements}
The 64 channels with highly configurable parameters meet the processing needs of signals from all sources of both detector types:

\paragraph{The Micromegas}\hspace{-0.3cm}signals from the anode strips (negative polarity signals), depending on the chosen gas
gain and shaper integration time, can be up to a maximum 250\,fC, but typically half or even smaller charge is expected. The fast electron current is followed by the positive ion current which typically lasts for $\sim$150\,ns\,\cite{georgePhd}. In addition to the current signal duration and maximum input charge, the other relevant parameter is the electrode (anode strip) capacitance which varies from about 50 to 300\pF depending on the length of the strips. The noise is a critical parameter for the Micromegas determined by the requirement of single primary electron detection with a threshold  five times the RMS noise, a gas gain of  10,000, and the maximum possible electrode capacitance of 300\,pF. These conditions determine the
required noise level to be at 0.5\,fC or about 3,000 electrons RMS.

\paragraph{The sTGC}\hspace{-0.3cm}feature three different types of active elements on a detector: strips, wires, and pads. All three are read out via the VMM. Strips provide the precision radial coordinate measurement for track reconstruction, wires the azimuthal coordinate; pads are used for a ``pre-trigger'' that requires a configurable coincidence performed by the Pad Trigger (See Sections\,\ref{sec:trigpath} and \ref{sec:pad_trigger}) firmware among the signals of pads in consecutive layers. The wire signals have negative polarity, while both the strip and pad signals are positive. Hence the need for the VMM to handle both polarities. The total charge and the long ion tail  impose specific requirements on the processing of the sTGC signals.  The VMM should recover from wire and pad signals of 6\pC and 3\,pC, respectively, within 250\ns while maintaining linearity up to 2\,pC.
For pads, it should provide the Time-Over-Threshold (ToT) and recover within 1\us from high charges up to 50\,pC. The pads impose challenging requirements since their capacitance can be up to 3\,nF. For the strips, an average charge of 1\pC is expected while the input capacitance is $\sim$200\,pF. As mentioned above, the sTGC signals span a very large range from 1\pC on a given
strip to about 50\pC on a pad. The dynamic range for the precision strip measurement is 2\,pC. The need to measure 2.5\% of this charge with a 2\% resolution and a 200\pF electrode capacitance, requires a noise level for a 25\ns integration time to be about 1\,fC RMS.
The noise for the signals from the pads with much larger capacitance (up to 3\nF) is
substantially higher.

Both detectors have similar readout requirements as the architecture is the same. A maximum trigger latency of 10\,$\muup$s must be supported, so the VMM needs deep enough FIFOs to buffer the data for this length of time. Moreover, a hit rate of up to 1\MHz per channel is expected.
On the other hand, the two detectors have different trigger requirements. The Micromegas need to provide the address of the VMM channel that fired first in a bunch crossing.
For the sTGC pad pre-trigger, which requires binary pad hits to select relevant strips, the VMM provides a Time-over-Threshold signal.
For the sTGC trigger, the VMM provides a 6-bit charge measurement of the strips within $\sim$50\,ns.
The sTGC wires do not participate in the trigger formation.

The VMM will  operate in a harsh radiation environment, see Table\,\ref{tab:radEnv} and\,\cite{Ameel, amideiDCDC, ATL-MUON-PUB-2022-001, nswTDR}.  Design techniques are applied to mitigate issues that may affect the operation of the ASIC under the above-targeted conditions.  Although \gls{TID} (Total Ionisation Dose) may degrade the performance of the ASIC,  the VMM3a was tested for TID tolerance in the  $^{60}$Co source irradiation facility at BNL for the expected radiation and no performance degradation was noticed.  Single event upsets (\gls{SEU}), though, become increasingly more serious. To overcome SEU's   in the vulnerable logic blocks (see Table\,\ref{radtablevmm}), Dual Interlocked storage Cells (\gls{DICE})\,\cite{DICE,DICE2} and Triple Module Redundancy (\gls{TMR})\,\cite{TMR} protection techniques are used.
For the large storage elements such as the latency \gls{FIFO}, an upset is just flagged once detected and the FIFO is reset.

\begin{table}[h]
\caption{ Single Event Upset protection schema in the VMM}
\vspace{5pt}
\label{radtablevmm}
\centering
\setlength{\tabcolsep}{3pt}
\begin{tabular}{ p{7cm}p{7cm}  }
\toprule
 Block & Type of protection \\
\midrule
  Global configuration and channel registers & DICE\\
  VMM State Machine & TMR\\
 Bunch Crossing Counter   & TMR\\
 L0 FIFO Control  & TMR\\
 L0 Event Builder & TMR\\
 L0 Accept register, NSkip Circuit & TMR\\
  Latency FIFO & Parity on pointer, FIFO reset on parity error\\

\bottomrule
\end{tabular}
\end{table}

\subsubsection{Architecture}
The analog front-end section of each channel integrates a three-stage low-noise charge amplifier\,(\gls{CA}) followed by a third-order shaper.  The charge amplifier implements a programmable input polarity, a test capacitor connected to the integrated pulse generator, a power-down option, a fast recovery option for very high-charge events, and several programmable bias adjustments to accommodate a broad range of signals. The input \gls{MOSFET} is a \mbox{p-channel}. It is followed by a dual cascode stage and a mirrored rail-to-rail output stage. The shaper features programmable peaking time of 25,\,50,\,100, and 200\,ns.  The gain is adjustable in eight values (0.5,\,1,\,3,\,4.5,\,6,\,9,\,12,\,16\,mV/fC). A low-frequency non-linear feedback baseline holder\,(BLH) stabilizes the output baseline, referenced to an on-chip band-gap reference circuit set at 160\,mV. The BLH has a programmable bandwidth that allows the user to enable either a mild or a strong (effective bipolar shape) compensation introduced to handle the \gls{sTGC} long current due to the long drift of ions.

\begin{figure}[ht]
\centering
\includegraphics[width=0.99\textwidth]{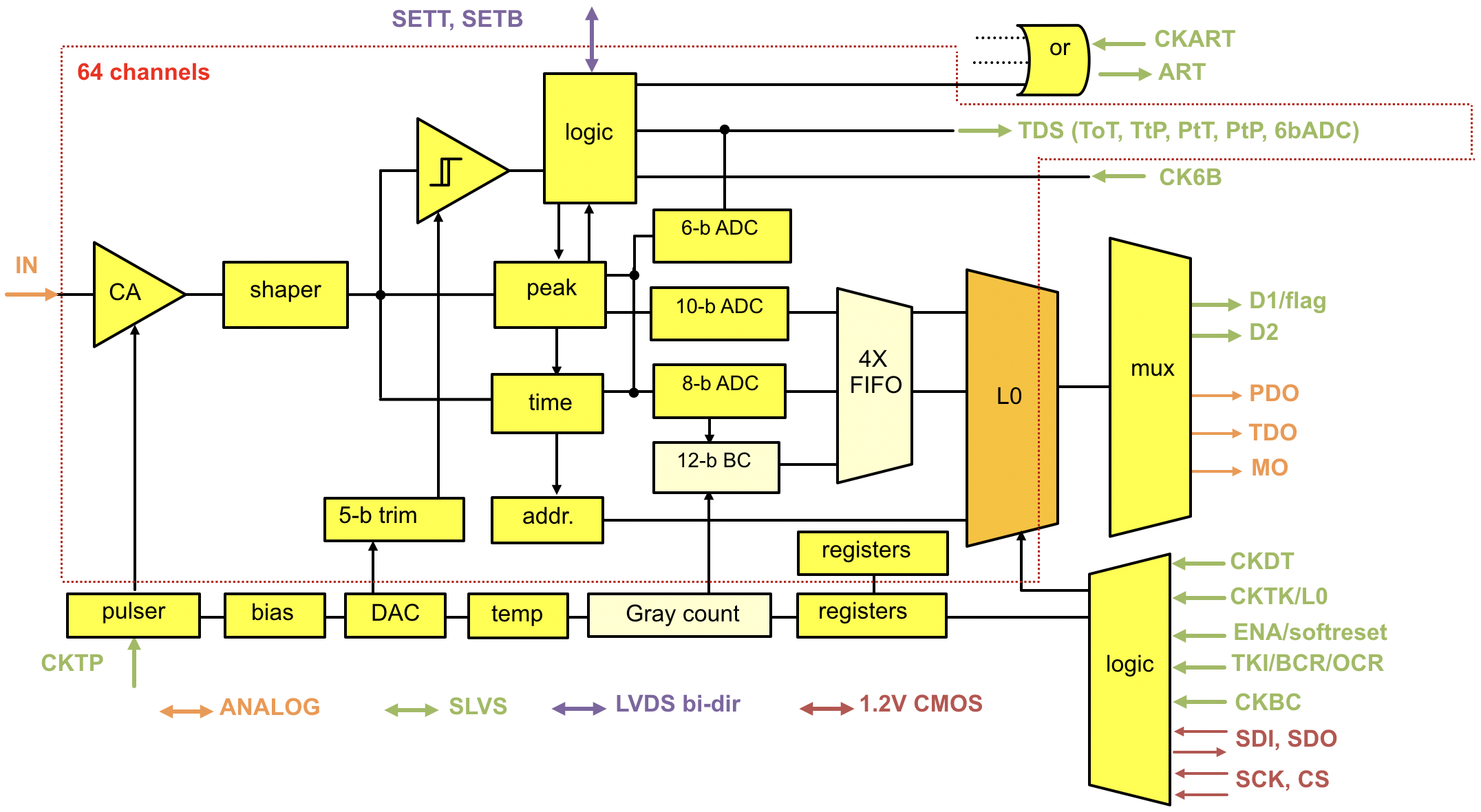}
\caption{Overview of the VMM architecture}
\label{fig:GI_VMM_architecture}
\end{figure}

Following the analog front-end is the mixed-signal section that includes discrimination, peak and timing detection measurements and the corresponding analog-to-digital conversions. The threshold is adjusted by a 10-bit Digital to Analog Converter (DAC) common to all channels plus a local 5-bit trimming \gls{DAC} independently adjustable in each channel in 31 steps of approximately 1\,mV each. The peak detector
measures the peak amplitude and stores its output (PDO) in an analog memory.  Additionally, it provides the timing signal at the time of the peak of the analog pulse. The time detector measures the timing using a time-to-amplitude converter (\gls{TAC}) which arms at the rising edge of the bunch crossing clock (CKBC) and latches at its falling edge. The time detector output (TDO) value is stored in an analog memory. The ramp duration can be configured to 60, 100, 350 or 650\,ns.
For the NSW, the value of 60\ns is used; this is enough to cover the duration of the CKBC while the TAC is in its linear range. The  block diagram of one of the 64 identical channels is shown in Figure\,\ref{fig:GI_VMM_architecture},  delimited with a dashed box, along with the relevant parts shared by all the 64 channel circuits and signals.

The VMM neighbor option triggers the two neighbors of a triggered channel, irrespective of whether they cross threshold.
This allows raising the threshold while still digitizing the edges of a spatial charge distribution. The functionality is applicable across different VMMs through dedicated electrical lines.

The mixed-signal part of the ASIC is followed by three current mode ADCs per channel. A pedestal of 150\mV is subtracted before digitization. This way the range of the ADC's is increased. The 10-bit and 8-bit ADC's are two-stage conversion digitizers providing charge and time measurements, respectively. The per-channel dead-time is driven by the 10-bit \gls{ADC} conversion that is configurable down to $\sim$250\ns, giving an effective rate of $\sim$4\,MHz per channel. The 8-bit ADC and a coarse 12-bit BC counter provide a 20-bit time stamp.  Each channel has a direct dedicated output (\gls{DDO}) where the 6-bit ADC provides the same charge measurement from the \gls{PDO} but in a much faster dedicated path with $\sim$50\ns dead-time.  The channel remains inactive until the 10-bit ADC completes the conversion. The ASIC provides the ability to interrupt the 10-bit conversion once the 6-bit conversion finishes, such that the DDO dead-time is small. In that case, the 10-bit information is unusable. The same DDO can be configured to provide pulses indicating Time-Over-Threshold\,(\gls{ToT}), Time-To-Peak\,(\gls{TtP}), Peak-To-Threshold\,(\gls{PtT}) or a Pulse-at-Peak\,(\gls{PtP}) of 10\ns duration. The address of the channel that registered the first hit per CKBC cycle, is output on a dedicated per-chip serial line. This is called the Address-in-Real-Time\,(\gls{ART}).

\subsubsection{Readout schema}
Although the VMM features more than one readout schema, the so-called ``L0'' mode is designed for operation within the ATLAS experiment. The output of the 10-bit and 8-bit ADCs enter into a 64-deep FIFO per channel called the ``Latency FIFO''.  Given the size of this FIFO and the 250\ns dead-time per channel, a maximum guaranteed latency of 16.0\,$\upmu$s where no data is lost can be achieved. This is larger than the minimum 10\,$\upmu$s required by ATLAS Trigger-DAQ{\,\cite{CERN-LHCC-2017-020}}.
Note that each channel is autonomous and this FIFO is filled asynchronously.

\begin{figure}[ht]
\centering
\includegraphics[width=0.65\textwidth]{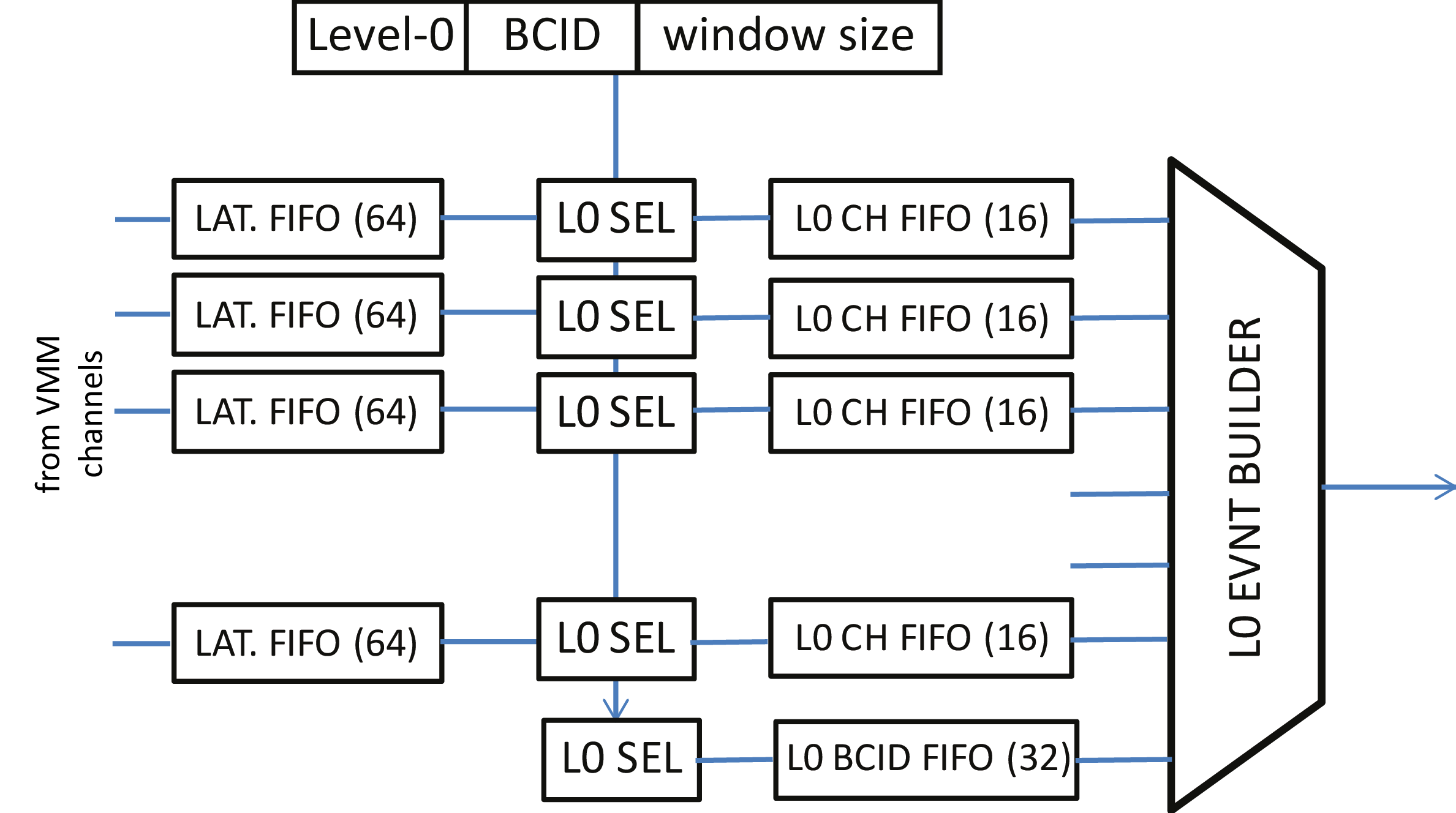}
\caption{Overview of the L0 buffer}
\label{fig:VMM3intBuf_V04}
\end{figure}

Each channel has a Level-0 Selector circuit that is connected to the output of the channel's latency FIFO.
The selector finds events within the BCID window (configurable at a maximum size of 8 BC's and the BC is offset by the latency) of a Level-0 Accept and copies them to the ``L0 channel'' FIFO.  If a channel's L0 selection circuit does not find a hit within the BC window,  a ``no data'' item is passed to the ``L0 channel'' FIFO. In this way the ``L0 channel'' of all 64 channels overflow synchronously. The ``L0 BCID'' FIFO is made deeper, 32-deep, than the ``L0 channel'' FIFO. This way, once the ``L0 channel'' overflows, the VMM data can still indicate on which BC this happened and can skip a configurable number of triggers to recover from the overflow while still maintaining its synchronous data-taking. The buffering scheme of the VMM L0 is shown in Figure\,\ref{fig:VMM3intBuf_V04}.

The data transfer from the VMM is done via two serial lines running at 160\,MHz with Double Data Rate (DDR), giving a total bandwidth of 640\,Mb/s. Two lines are used to reduce the clock rate. The Readout Controller supplies the clock for this transfer. The data is encoded in 8b/10b with one or more comma characters transmitted continuously between Level-0 events. The 8b/10b encoding reduces the effective bandwidth to 512\,Mb/s.

\subsubsection{Trigger outputs}
The VMM must provide different trigger primitives for the four different detector elements connected.
This is achieved through different data outputs and specific configuration. It is possible to turn off the \gls{SLVS} drivers of trigger output lines not in use in order to reduce power consumption.

\paragraph{The \gls{Micromegas} trigger}\hspace{-0.3cm}utilises the fine strip pitch and the ionisation spread across the path of a particle crossing the detector at an angle as used in the $\muup$TPC method\,\cite{ALEXOPOULOS2019125}.
The VMM sends out this first address through a dedicated serial line called Address in Real Time (ART).
This signal is asynchronous to the bunch crossing clock so the VMM can be configured to align it with the BC clock.
The ART clock is provided to the VMM externally from the Readout Controller (ROC); see Section\,\ref{sec:ROC}.
The ART signals can be provided at threshold crossing or at peak found, depending on the configuration.
It can be optionally clocked at both edges of the 160\,MHz clock.

\paragraph{The sTGC trigger}\hspace{-0.3cm}is done in two steps as described in Section\,\ref{sec:trigpath}:
1) A pre-trigger: Overlapping induced cathode pads define a candidate track segment by a configurable coincidence in a projective tower in both quadruplets in a sector made by the Pad Trigger module.
Each VMM connected to pads operates its direct outputs in Time-over-Threshold (ToT) mode.
Each channel self-resets at the end of the timing pulse, thus providing continuous and independent operation of all 64 channels.
2) The projective tower defines the bands of strips (if any) to be read out from each TDS in each layer. The VMM channels connected to the strips digitize the charge through the fast 6-bit ADC and provide them through the direct output.  The channel reset occurs after the last bit has been shifted out. The data are clocked at both edges of the 160\,MHz clock.


\subsection{ROC -- Read out Controller ASIC}
\label{sec:ROC}
\begin{figure}[b]
\centering
\includegraphics[width=0.99\textwidth]{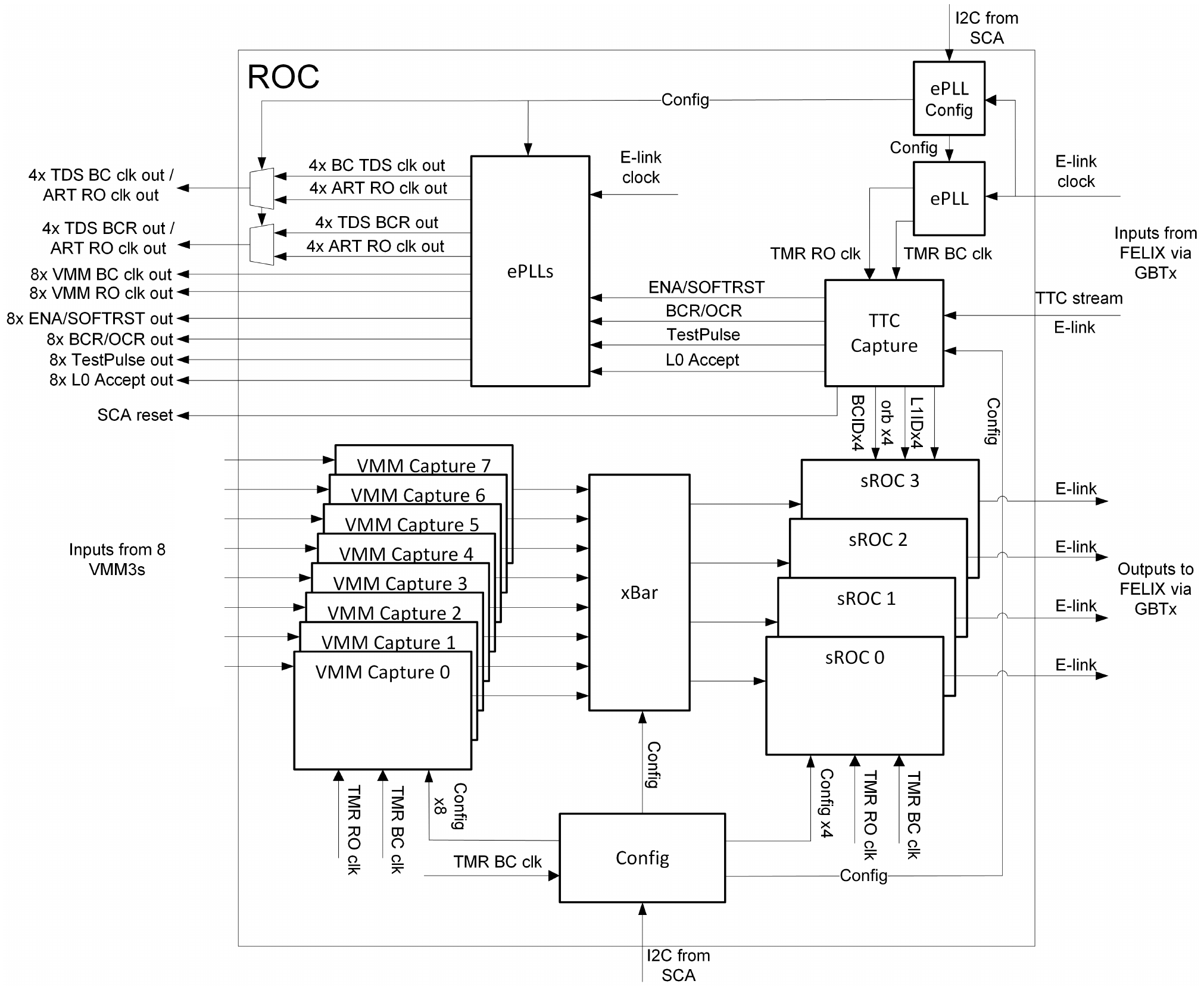}
\caption{Block diagram of the ROC ASIC\,\cite{Popa2022}}
\label{fig:block_ROC}
\end{figure}

The ReadOut Controller (\gls{ROC})\,\cite{Coliban:2016uys, Popa:2019trf, Popa:2020sbm, Popa:2020poz,Popa2022} is a highly configurable data aggregation ASIC designed specifically for the NSW.  The block diagram of the ROC ASIC is shown in Figure\,\ref{fig:block_ROC}.
The ROC must provide readout of the hits buffered from the VMMs in response to a Level-1 trigger at 100\,kHz for Phase\,1 and at 1\,MHz for Phase\,2\,\cite{1MHzReadout}.
Initial specifications for the two-level hardware trigger required handling a latency of 60\,$\upmu$s with a consequent very deep buffer for hit data.
After ATLAS decided on a single-level hardware trigger the latency was reduced to 10\,$\upmu$s, but the ROC ASIC had already been produced.

The ROC receives 8b/10b encoded data from up to eight VMM3a ASICs. There is a dedicated ``Capture'' module for each VMM. The data are first de-serialized and then a dedicated mechanism determines the correct 8b/10b stream alignment and then decodes it. The data parity is also checked and if an error occurs, the relevant counters are incremented. If no data appear then a configurable timeout is asserted.
The decoded L0 packets are enqueued separately for each VMM.
A configurable crossbar allows routing of the data from one to eight VMMs to each of up to four SROC modules.
Each SROC is able to transmit 8b/10b encoded data through a configurable up to 320\,Mb/s E-link to the L1DDC.  Two 320\,Mb/s E-links can be combined to produce a 640\,Mb/s output.
For Phase 1, generally one 320 Mb/s E-link per ROC is sufficient.
For Phase 2, more than one E-link is needed, especially at the inner radius.
VMM occupancy varies strongly with radius.
Front-end boards at inner and outer radii require different numbers of E-links.
The crossbar allows routing VMM outputs to E-links in order to optimize the number of E-links needed.
For load balancing, an SROC can combine high and low occupancy VMMs on the same E-link.
The VMM data integrity is also checked at the SROC level.
If, for example, the ROC receives the so-called ``Magic'' BCID, which is a value outside the range of the expected LHC values, the ROC understands that the VMM FIFOs overflowed and data will not be received from this VMM for the configurable value (on the VMM) of skipped triggers.

The ASIC implements several output formats including dummy hits signalling overflow from the VMM, as well as hits that discard the TDC (TDO) value of the VMM to decrease the bandwidth. Moreover, Busy-On/Busy-Off symbols are injected in the output data stream, if enabled, to signal almost-full buffers in the ROC. The configuration and status of the ASIC is accessed by a dedicated I$^2$C interface through the SCA.

The ROC ASIC receives the TTC stream and BC clock from the GBTx. The alignment of the input stream is determined by detecting the positive edge of the BC clock signal in the readout clock domain. All its internal and externally supplied clock signals are generated by four ePLL blocks\,\cite{Poltorak_2012} driven by the BC clock. It supplies the clock and the TTC commands to all the ART, TDS and VMM ASICs. The three ePLLs supplying the signals and clocks to the other ASICs include phase-shifting circuits. This gives the ability to forward the relevant TTC commands and clocks with a configurable phase. The TTC FIFO (also called BC FIFO) buffers the Level-1 triggers.
In response to the Level-1 trigger, the ASIC checks, aggregates, re-formats and filters the L0 packets from the associated VMM Capture modules, building output packets that are pushed into the SROC FIFO ready to be read out.
The ePLLs are configured and monitored through a separate register bank through a dedicated I$^2$C interface of the SCA.

During the integration of the ROC ASIC and the SCA, it was found that the I$^2$C read-back implementation of the ROC was incompatible with the I$^2$C standard chosen by the SCA. The I$^2$C read-back was therefore emulated using dedicated GPIO lines and a software ``Bit Banger'' application\,\cite{opcuaserver}.

The ROC implements Triple Modular Redundancy\,(TMR)\,\cite{TMR} for all the configuration registers, state machines, control of the bunch crossing FIFO and readout logic to mitigate the effects of SEUs. An SEU counter is accessible by the SCA chip (through the configuration and register bank) to monitor the chip operation in the ATLAS environment.


%


\subsection{TDS -- sTGC Trigger Data Serializer ASIC}
\label{sec:TDS}

The sTGC trigger data serializer (\gls{TDS}) ASIC\,\cite{Wang:2017ols, Wang:2015msa, Wang:2019tay} prepares the trigger data from the VMM for either pads or strips and serializes it for transmission, in the case of pads, to the Pad Trigger, in the case of strips to the Router on the rim of the NSW detector. The data are transferred ultimately to the Trigger Processor.
It is mounted on the Front-end boards and can be configured in either strip or pad mode by connecting a pin to  high or low.

Both the 160\,MHz TDS logic clock and serializer reference clock are generated by an on-chip PLL from the BC clock supplied by the Read Out Controller.
The TDS is configured via the \ItwoC master of the on-board SCA ASIC.
An on-chip pseudo-random binary sequence generator (PRBS-31) is provided for serializer and link testing.

\begin{figure}[t]
\centering
\includegraphics[width=0.999\textwidth]{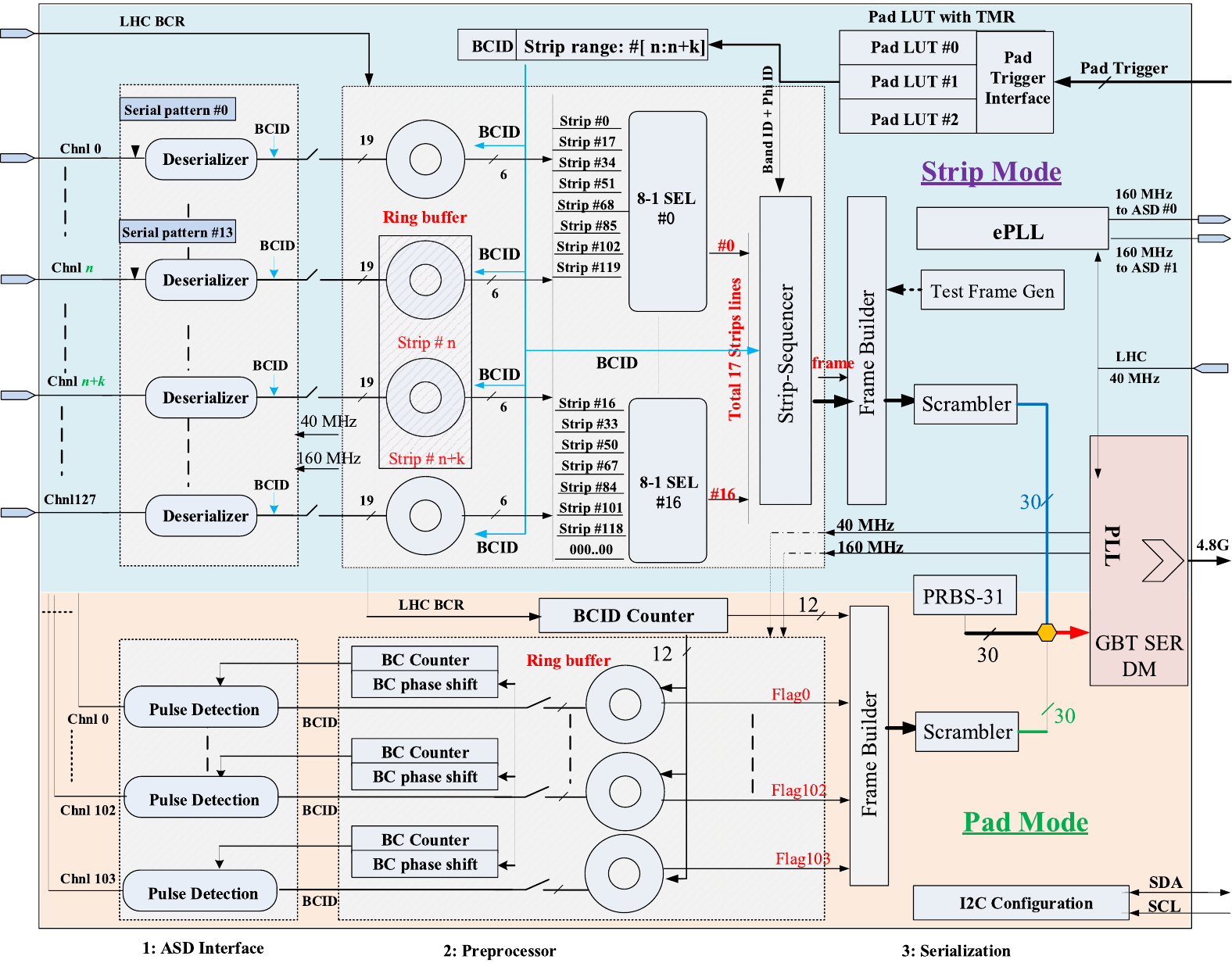}
\caption{Block diagram of the TDS ASIC. Strip mode, top; Pad mode, bottom.}
\label{fig:JW_TDS_blockdiagram}
\end{figure}

The TDS is fabricated in IBM\,130\,nm \gls{CMOS} technology and is packaged as a 400-pin Ball Grid Array (BGA).
It uses a 1.5\,V supply for both the logic part and the serializer and consumes about 0.9\,W.
A block diagram of the TDS ASIC, including both strip and pad parts, is shown in Figure\,\ref{fig:JW_TDS_blockdiagram}.

\subsubsection{Pad TDS mode}  

The pad TDS receives up to 104 pad Time-over-Threshold (ToT) differential signals from two VMM ASICs and transmits the data at 4.8\,Gb/s to the Pad Trigger every 25\,ns.
The rising edge of the ToT signal is used to capture a pad hit and assign its BCID.
Since the 104 pads cover an area of about 3\,m$^2$, the on-detector routing lengths from a pad to the VMM and pad-TDS can differ by up to 2.6\,m, with a propagation time of about 7\,ns/m\,\footnote{Calculated by the PCB layout program and verified approximately by measurements of some pads with a Time Domain Reflectometer.}.
In addition to the differences in time-of-flight from the IP for the pads within a detector,
this difference in propagation time could result in different BCID assignments for pad signals even though they belong to tracks from the same collision.
To compensate for this, each channel has a configurable delay, of up to eight 3.125\,ns steps\,\cite{Wang:2017iuz}.
Assuming that the earliest and latest pad signal arrivals (considering the earliest arrival signal from each pad)
are within 25\,ns of each other,
shifting the phase of the incoming BC clock enables them to be assigned the same BCID.

For serial transmission, 116 bits (104 pad bits plus a 12-bit BCID) are split into four consecutive frames: a 26-bit frame followed by three 30-bit frames.
The four frames are scrambled following a scheme used by the 10\,Gb/s Ethernet physical layer implementation\footnote{IEEE Standard 802.3-2012 with the polynomial function $1+x^{39}+x^{58}$}.
An unscrambled 4-bit header, 0b1010, is prefixed to the 26-bit scrambled frame, marking it as the first of the four 30-bit frames.
Scrambling enables 116 data bits to be transferred in one bunch crossing at 4.8\,Gb/s.
The latency of the pad-TDS from the end of a BC to the first bit of a pad frame exiting the serializer core is 31\,ns.


\subsubsection{Strip TDS mode}  
\label{sec:sTDS}

Each of the two VMM ASICs connected to one strip-TDS, sends 6-bit charge data through 64 independent serial lines, each representing the induced charge on a strip.
The strip-TDS holds the vector of charges, along with its BCID, in a circular buffer.
The Pad Trigger, after finding a coincidence in a tower of pads in eight sTGC layers, sends the ID of the band of strips in each layer that passes through that tower to the strip-TDS that holds the charges of that band.
A configurable look-up table in each TDS contains the channel number of the first strip in each band that it holds.
Strip-TDS ASICs that do not hold charges of any band are sent band-id\,0xff, which is not in its look-up table.
The band-id is the same for all layers, but the 17 strips comprising that band differ from layer-to-layer since the tower points to the interaction point.
When the request arrives from the Pad Trigger, if the charges for strips corresponding to the band-id are in the circular buffer and their BCID matches the requested BCID,
they are transmitted at 4.8\,Gb/s to the Trigger Processor via the Router.
Only the charges of the outer 14 or inner 14 strips are transmitted; a flag indicates which.
The data is sent in four 30-bit packets in one BC.
The packet format is shown in\,\cite{HU2022167504}.
This repeats for every bunch crossing.
The latency of the strip-TDS from arrival of the request from the Pad Trigger until the first bit of data frame exits the serializer core is 75\,ns.

\subsubsection{Test functions}
The TDS has several embedded functional self-tests in case of any system malfunctions, system commissioning and to enable testing without inputs.
The 4.8\,Gb/s serializer can be driven by an on-chip pseudo-random binary sequence generator\,\cite{PRBS}, PRBS-31, for testing and commissioning the link.
Specifically, the test modes are: \\
\textbf{Bypass Trigger mode:} In strip mode, probes inputs from an individual or a programmable number of strip channels bypassing the channel ring buffer and without trigger input from the Pad Trigger.
In pad mode, a similar diagnosis function provides access to an individual or a programmable group of input channels. \\
\textbf{TDS-Router Training Frame:} For strip mode, sends fake test frames to the Router without relying on the input from the Pad Trigger.\\
\textbf{Global-Test mode:} performs a full strip-mode function and output without the input from the Pad Trigger or the Front-end electronics.


%

\subsubsection{The 4.8\,Gb/s serializer}   

The 4.8\,Gb/s serializer for the TDS was developed and prototyped prior to the remainder of the TDS\,\cite{Wang:2015msa}.
Rather than developing such a challenging circuit from scratch, it was adapted from the CERN GBTx serializer, changed from loading 120 bits at 40\,MHz to loading 4\,$\times$\,30 bits at 160\,MHz.
CERN's ePLL design\,\cite{Poltorak_2012} was also used.
The metalization layers were also changed to match the variant of the IBM\,130\,nm CMOS process used by the other NSW ASICs so that all could be fabricated on the same wafer.
The serializer's output into a 100\,$\Omega$ load is about $\pm$\,500\,mV.

A jitter analysis of the transmission showed a total jitter of 49.7\,ps at a bit-error-ratio (\gls{BER}) of $10^{-12}$.
A BER test with an embedded PRBS checker inside a Xilinx\,7 FPGA was also performed.
Error-free running for three days was achieved, which corresponds to a BER less than $10^{-15}$.

\subsubsection{Radiation tolerance}
To mitigate Single Event Upsets, the TDS employs Triple Modular Redundancy\,(TMR)\,\cite{TMR} for the following:
Serializer, Serial protocol logic,
BCID, BC clock generator, BC clock phase shift, size of the matching window of the strip channel,
all configuration registers, Pad Trigger \gls{LUT}.

\subsection{ART -- \MM trigger data aggregator and serializer ASIC}
\label{sec:ART}
The ART ASIC is part of the trigger path of the \MM chambers.
It receives the prompt ``Address in Real-Time'' (ART) data from 32 VMM front-end ASICs which consist of the address of the first arriving hit in each 64-channel VMM for a given bunch crossing.  It then selects up to eight  addresses for each bunch-crossing and sends the data to the \MM Trigger Processor via one GBTx chip configured to operate in ``Wide'' mode, i.e.\ without Forward Error Correction (FEC).

\noindent In particular, the ART ASIC performs the following functions:\vspace{-6pt}\begin{itemize}\itemsep-4pt
\item Deserialize each ART stream and phase-align the hits to the BC clock.
\item Selects the strip addresses of up to a fixed number of hits by means of cascaded priority encoders.
\item Append the 5-bit geographical physical VMM address to the strip address of each hit (defined by the cable connections).
\item Send the ART addresses and the 12-bit BCID to the \MM Trigger Processor via a GBTx.

\end{itemize}

\begin{figure}[ht]
\centering
\includegraphics[width=0.95\textwidth]{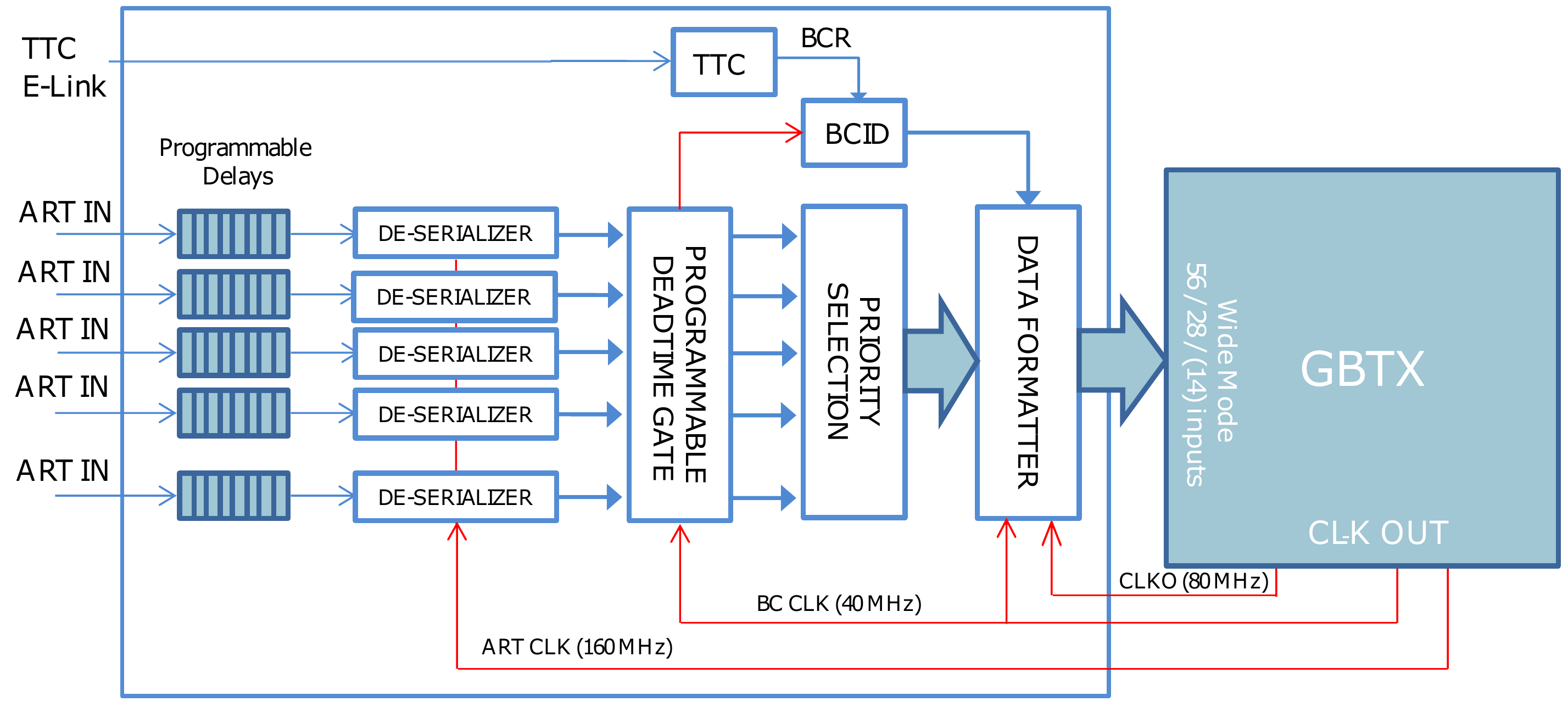}
\caption{Block diagram of the ART ASIC}
\label{fig:block_ART}
\end{figure}

The block diagram of the ART ASIC is shown in Figure\,\ref{fig:block_ART}.  The ASIC receives its TTC stream, configuration and clock signals from the GBTx (which receives them through the Level-1 Data Driver Card (\gls{L1DDC}) downlink, see Section\,\ref{sec:addc}). The ASIC has the option to output the Bunch Crossing Reset (\gls{BCR}) \gls{TTC} signal received to another ART ASIC\footnote{The second ART on the ADDC.}.
The ASIC is configured through the SCA using an I$^2$C bus.  The ASIC is packaged in a 128-pin LQFP package.
The latency of the ART ASIC was measured in several test setups to be $\sim$44\,ns.
Further information can be found in\,\cite{ARTASIC}.

\paragraph{Programmable delays:}\hspace{-8pt}The purpose of the Programmable Delay block is to be able to skew the input signals to the local clock phase and adjust the ART stream to it.  To perform that, the ASIC uses four copies of the 8-channel Phase Aligner core developed at CERN\,\cite{Tavernier_2012}.

\paragraph{Deserializer:}\hspace{-8pt}The first signal that precedes the ART 6-bit address is a flag.
This is generated by the VMM and in DDR mode, the flag pulse raises asynchronously to the ART clock (CKART) and is kept high through the next two falling edges of the 160\MHz clock, being lowered by the second falling edge.
Optionally, the rising edge of the flag can be registered to the CKART clock. A 10\ns reset period is applied after each ART sequence.  The eight flip-flops form an 8-bit DDR shift register which deserializes the incoming data stream.

\paragraph{Programmable dead time:}\hspace{-8pt}This block creates an artificial dead time for each VMM input which is controllable via configuration. Subsequent data on a particular input is ignored for between 0 and 7 BC's. This prevents responding to any subsequent ART signals from the same particle crossing the \MM.
See Figure\,\ref{fig:ART_concept}.

\begin{figure}[ht]
\centering
\includegraphics[width=0.9\textwidth]{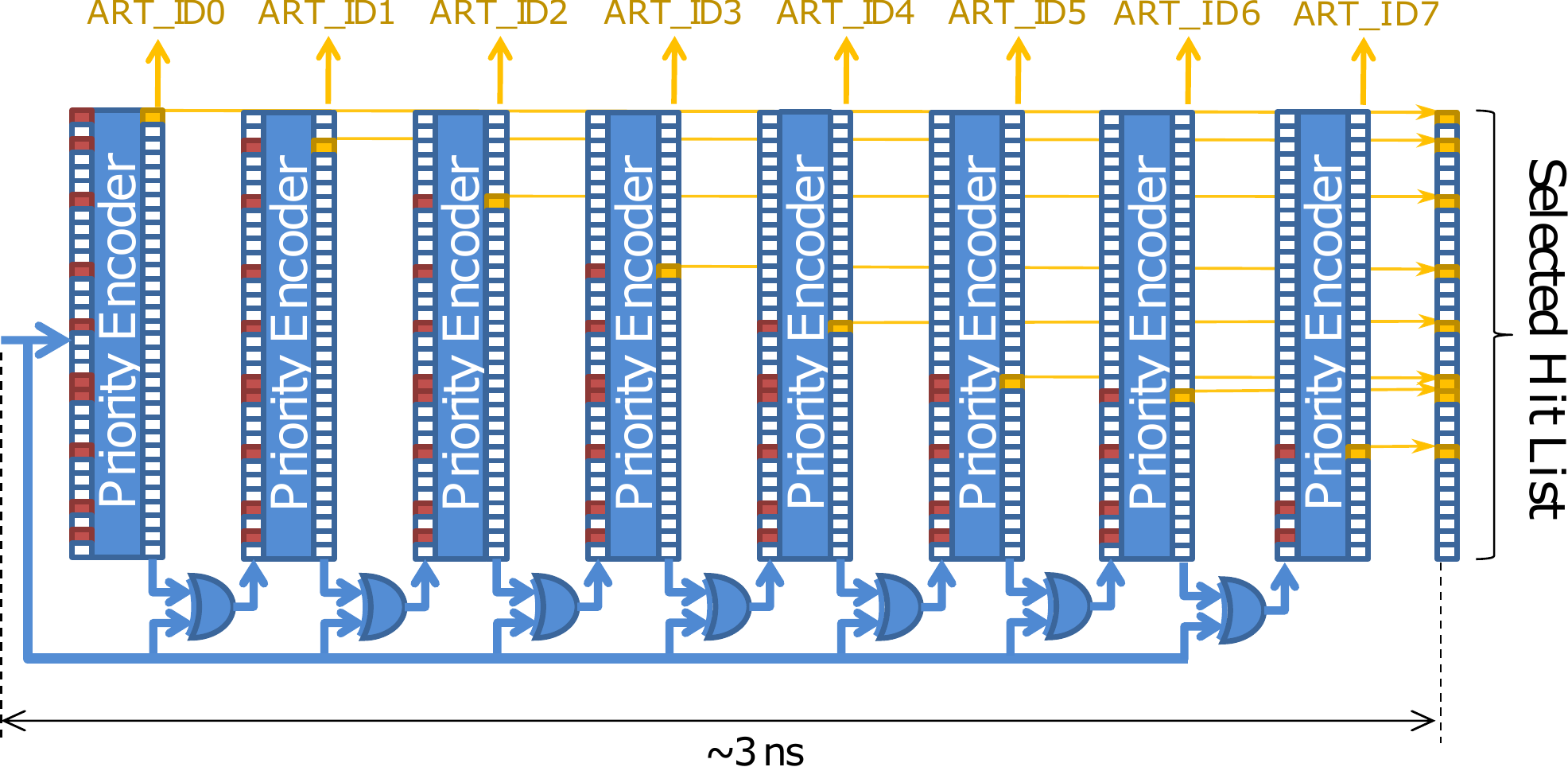}
\caption{Hit map generation circuitry based on priority encoders. The duration of the procedure is $\sim$3\,ns.}
\label{fig:art_priorityenc}
\end{figure}

\paragraph{Priority selection:}\hspace{-8pt}
The Hit Selection circuit is based on eight layers of cascaded priority encoders.
The first priority encoder selects the first ART flag (most significant bit which is not zero from the 32-bit ART flag word).
The 32-bit word is readout as the first ART and then removed from the array. The consequent word (without the first ART) is presented to the following stage. The second priority encoder selects the second non-zero bit from the ART flag word in the same manner as the first stage.
The operation is cascaded eight times to select a maximum of eight non-zero flags.
The operation is illustrated in Figure\,\ref{fig:art_priorityenc}. If there are more than eight ART signals in the same BC window,  then the ART ASIC selects only eight of them, based on the priority scheme explained. The ART can be configured to give priority to different radii of the detector\,\cite{nswTDR}.

\paragraph{Output:}\hspace{-8pt}For each bunch crossing, the ART ASIC transmits to the Trigger Processor through the GBTx, the following data:
\vspace{-6pt}\begin{itemize}\itemsep-4pt
\item{Up to eight  VMM ART signals which had a hit in that bunch crossing}
\item{The 12-bit BCID in which the ART signals occurred}
\item{Other information (error flags, parity bits)}
\end{itemize}

\noindent The Wide Bus mode of the GBTx chip (see Section\,\ref{sec:GBTx}) allows for 112-bits to be transmitted in one bunch crossing on two 80\,MHz clock edges. The ART data are transmitted to the GBTx chip in two batches:\vspace{-6pt}\begin{itemize}\itemsep-4pt
\item{The first 56-bits contain the selected VMM hit list based on the flag bits issued. This can be configured in two different modes: either transmit a ``Hit Map'' 32-bit word where each bit corresponds to one of the 32 VMMs connected to the ART ASIC or, the ``Hit Address'' option containing a 5-bit VMMID for each of the VMMs selected. In both cases, the 12-bit BCID is transmitted along with the Hit Information.}
\item{The second batch of 56-bits contains the eight 6-bit VMM ART addresses of the VMMs selected along with an 8-bit parity bit for each word.}
\end{itemize}

\paragraph{Debugging mode:}\hspace{-8pt}Besides the normal operation modes, the ASIC implements a debugging mode where different parts of the system are bypassed or fixed, or, repeating calibration patterns are transmitted. The following debugging modes are implemented:
\vspace{-6pt}
\begin{itemize}\itemsep-4pt
\item{Full bypass mode where the output of the programmable delays are directly made available to the outputs of the ASIC. This is used to verify and measure the propagation delay during ASIC initial verification and it is not accessible during normal operation in ATLAS.}
\item{Priority encoders bypass mode where the input channels are connected directly to the output logic. This mode is controlled by a configuration register and allows the verification of the transmission between VMM chips and ART ASIC bypassing any selection, while permitting the definition of the propagation delays set in the Programmable Delays. }
\item{Fixed output calibration pattern where a fixed pattern is sent continuously. The pattern is stored in local configuration registers and is accessible via the configuration path which allows the verification of the data transmission.}
\end{itemize}

\paragraph{Radiation protection:}\hspace{-8pt}The ART ASIC is designed with several protection mechanisms in order to ensure protection against SEU events or to flag uncertain conditions.
The SEUs in the data will not seriously affect the trigger. However, simple parity bits are calculated by the input deserializer circuits and can be used downstream to tag possible data corruption.  The circuits protected with triple modular redundancy (TMR)\,\cite{TMR} or parity bits are shown in Table\,\ref{radtable_ART}.

\begin{table}[h]
\centering
\caption{TMR protection scheme in the ART ASIC}
\label{radtable_ART}
\begin{tabular}{lll}
\toprule
       &   Data     & State machine \\
 Block & protection & protection \\
\midrule
  Programmable Delays & No & -- \\
 \gls{DDR} Deserializers & Parity bit & Yes\\
 Programmable dead time   & -- & Yes\\
 Priority Selection & No & --\\
 BCID Counter & Yes & --\\
 Output Logic & No & Yes\\
  Register Matrix & Yes & Yes\\
  \ItwoC  Slave & Yes & Yes\\
\bottomrule
\end{tabular}
\end{table}




\section{NSW boards}
\label{sec:boards}
The electronics boards developed for the NSW upgrade have several commonalities:
\begin{itemize}[topsep=2pt, itemsep=0pt, parsep=1pt]
\item DC (non-isolated) Point-of-Load power from FEAST ASICs\,\cite{FEAST2.1} or FEAST modules\,\cite{FEASTMP}
\item SCA ASIC for configuration and monitoring
\item LVDS (Low Voltage Differential Signalling) or Scalable Low Voltage Signaling (SLVS)\cite{slvs} used for most on-board chip-to-chip communication
\item \gls{MiniSAS} twin-ax ribbon cable\,\cite{twin-ax,8F36} used for board-to-board communication, carrying SLVS, LVDS and 4.8\,Gb/s serial signals.
           In all, there are about 40\,km of MiniSAS twin-ax cables.
\item Active water cooling of all boards (except the Serial Repeaters). To prevent leaks, the system runs with less than atmospheric pressure.
\end{itemize}

\subsection{\MM Front-end board -- MMFE8}
\label{sec:mmfe8}

\begin{figure}[t]
\centering
\includegraphics[width=1.0\textwidth]{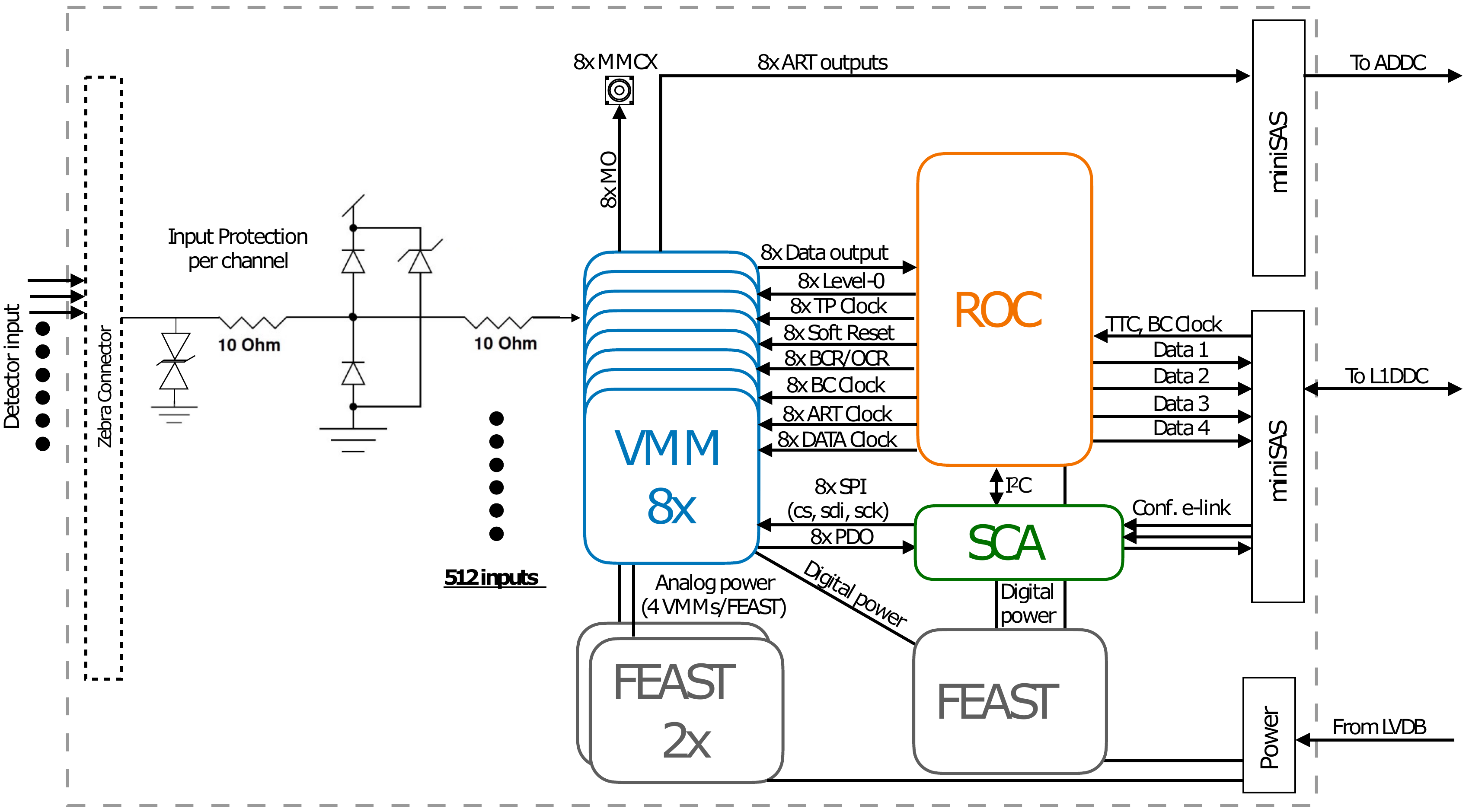}
\caption{The MMFE8 block diagram showing the \gls{ESD} protection,
         the VMM Front-end ASICs, the Readout Controller ASIC, the Slow Control Adapter ASIC, the FEAST DC-to-DC converter ASICs, and the MiniSAS twin-ax connectors for the different interfaces.}
\label{fig:mmfe8_schematics}
\end{figure}

The \gls{MMFE8} board is the front-end electronics for \MM detectors. It is the interface between the detector, the trigger (ADDC), and data acquisition (L1DDC) electronics. Due to the high number of readout channels on the \MM detector ($\sim$2.1\,M), 4096 MMFE8 boards are needed, each handling 512 channels. The board must meet demanding space and electrical requirements and constraints. The board dimensions are 215\mm in length by 60\mm wide. It is comprised of 14 electrical layers and is 2.54\mm thick.  A block diagram of the MMFE8 is shown in Figure\,\ref{fig:mmfe8_schematics}.

The interface to the detector is achieved via two \gls{ZEBRA}\textsuperscript{\textregistered} elastomeric connectors\,\cite{zebra}.  The type and layout of the connectors was carefully studied. The pitch on the board was chosen to be 400\um with a contact width of 200\um which is compatible with the detector pitch.
The connector has six lines of through wires per strip which ensure contact between the detector strips and the pads on the MMFE8 PCB.
A precision machined slot between the two connectors aligns the board to the detector.
The compression of $\sim$150\um is also regulated with mechanical cams on top of the board where a ground strip connects the board analogue ground to the detector ground through six contacts. On either side of the connector, four lines provide geographical information through an on-detector encoding\footnote{It was unfortunately found during integration that due to wrong connectivity of the pins in the MMFE8 board, the geographical decoding could not be achieved.}.

The VMM inputs on the board are protected through a dedicated diode network which was extensively tested and was found to protect the ASIC from any discharges on the detector\,\cite{Iakovidis:2675779}.  On each channel, a Transient Voltage Suppressor (\gls{TVS}), a Semtech \textmu Clamp\,\cite{semtech}, is able to mitigate fast transients.
Following the TVS diodes, a 10\,$\Omega$ resistor connects to one of the four channels of a TVS Diode Array, SP3004\,\cite{sp3004}.  This diode array mitigates slower input transients.
Another 10\,$\Omega$ resistor is placed after the SP3004\footnote{Early in the project, the NUP4114-D was used. It was discovered that the functional details of this protection diode were modified between 2012 and 2014, but without changing the part number. Its new functionality was not adequate to the NSW needs, so the SP3004 was used instead. }.

																																																																																																																																																																																																																														The board contains eight VMM ASICs, one ROC ASIC, one SCA ASIC and three FEAST ASICs. The on-board FEAST ASICs provide power to all the ASICs. Two FEAST ASICs set for 1.3\,V\footnote{Studies showed that the VMM must be provided with at least 1.2\V and therefore the 1.3\V ensures that voltage by compensating for small voltage drops along the lines.} provide power to the eight VMM's analog section (one FEAST per four VMMs).  One FEAST provides 1.2\V to the digital supplies of all the VMMs and also to the ROC and SCA ASICs\footnote{The SCA ASIC is designed to function at 1.5\,V. Discussions and validation tests showed that the SCA can function properly at 1.2\V with negligible impact on the integrated ADC performance. This was done to avoid integrating another FEAST on the board due to space constraints.}.
The SCA provides a unique 32-bit ID for the board.
The input voltage of 11\,V is provided through the Low Voltage Distributor Board (LVDB) described in Section\,\ref{sec:powerDistribution}.
The power consumption of the board was measured to be $\sim$16\,W.

The interface to both ADDC and L1DDC boards is realised through MiniSAS cables and connectors\,\cite{MiniSASconnector,twin-ax,8F36}. The connector to the ADDC carries the eight ART signals. The connection to the L1DDC carries the TTC signals and bunch crossing clock to the MMFE8 as well as the four data lines to the GBTx (one per SROC). The SCA E-link is carried as well in the same connector, removing ambiguity as to which ROC and VMMs are configured and reset by the SCA. The on-board ROC provides dedicated trigger, test pulse, reset and various clock signals to each VMM.

The design of the MMFE8 was iterated such that it almost reached the theoretical noise levels with respect to the ones defined by the VMM for the \MM detector capacitance, which is estimated to be $\sim$150\,pF/m\,\cite{Iakovidis:2675779}. To achieve that, the FEASTMP2.1\cite{FEASTMP} design was adopted.

\subsection{sTGC Front-end boards -- sFEB and pFEB}
\label{sec:sfeb_pfeb}

The strip and pad Front-end boards\,\cite{Miao_2020} interface to detector strips, pads and wire groups, providing their data to the readout path on receiving a Level-1 trigger signal and to the trigger path on every bunch crossing.
There are separate radiation-tolerant Front-end boards for strips and pads+wires.
Both are the result of several demonstrator and prototype boards, including those with earlier versions of the VMM and those with FPGA readout instead of the Readout Controller (ROC) ASIC.
VMM ASICs (see Section\,\ref{sec:VMM}) provide data to both trigger and readout paths.
An SCA ASIC provides a unique 32-bit ID for the board.

For each gas gap, detector signals connect to adapter boards on each of the two radial edges of the detector.
A strip Front-end board (sFEB) reads out strips from one edge, and a pad+wire Front-end board (pFEB) reads out pads and wires from the other.
In total, there are about 30 wire groups, up to 112 pads, and up to 400 strips reading out an sTGC gas gap.
The adapter boards route the signals to high-density (10\,$\times$\,30 contacts) low-profile matrix interposers (Samtec GFZ)\,\cite{GFZ} (one for pads, two for strips) on the adapter board to which the Front-end boards connect.

Careful placement, layout and shielding of the on-board FEAST DC-DC power converters was essential and is described in\,\cite{Miao_2020}.

\begin{figure}[h]
\centering
\includegraphics[width=0.75\textwidth]{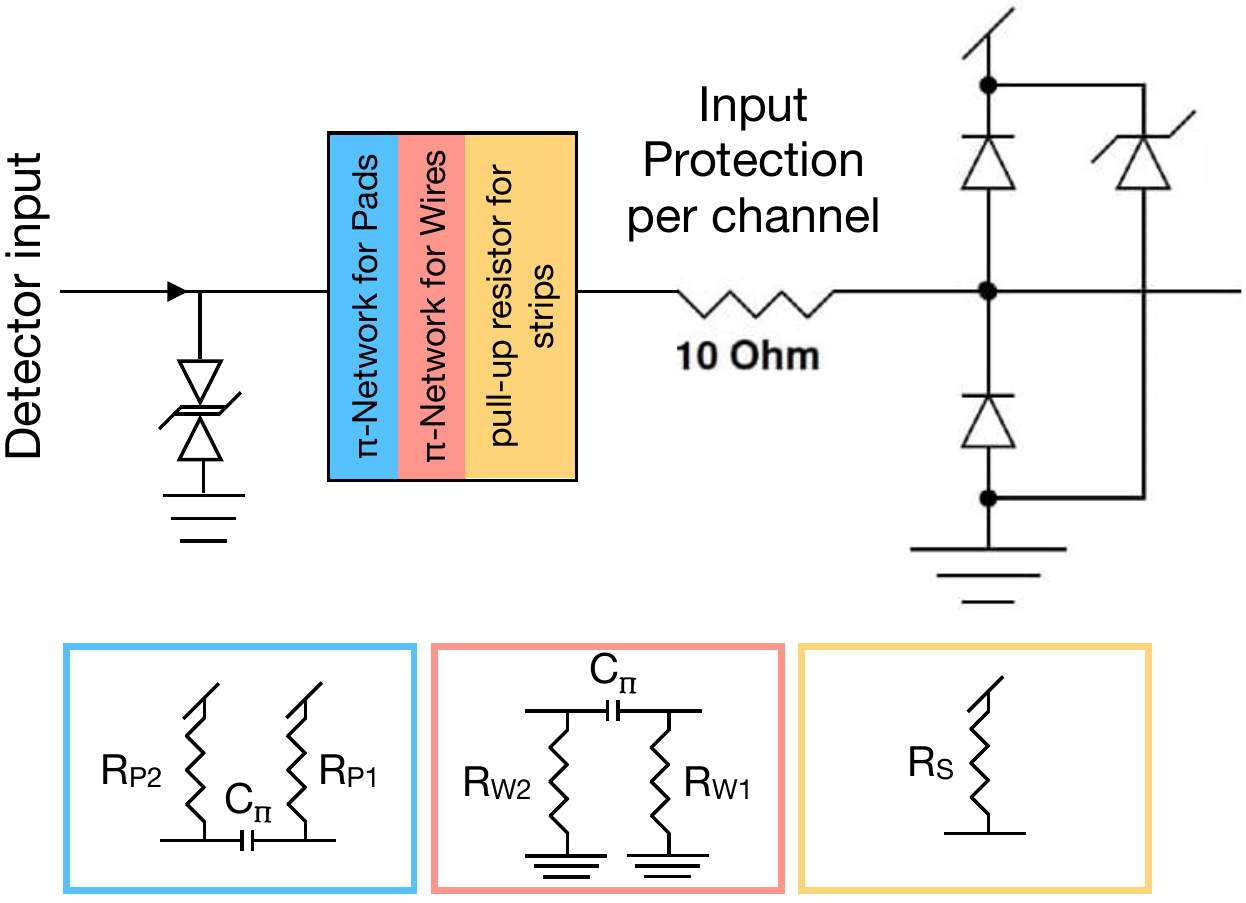}
\caption{The sTGC FEB analog input circuit which protects against transients and compensates for high rate and high charge signals for pads\,(blue, left), wires\,(red, middle) and strips\,(yellow, right).}
\label{fig:TVS_Pi_blockDiagram}
\end{figure}

\para{Input transient voltage suppression and signal conditioning:}
Gas detectors have high voltage discharges and the front-end VMM ASIC inputs must be protected.
Each of the VMM inputs on the board is protected through a dedicated diode network which was extensively tested and was found to protect the ASIC from any discharges on the detector\,\cite{Iakovidis:2675779}.
Figure\,\ref{fig:TVS_Pi_blockDiagram} shows the input circuitry that protects against transients and compensates for the high rate and high charge signals for strips, pads and wires.
The first stage is a Transient Voltage Suppressor (TVS), a Semtech \textmu Clamp\,\cite{semtech}, which can mitigate fast transients.
The last stage is a 10\,$\Omega$ resistor that connects to one of the four channels of a diode array, SP3004\,\cite{sp3004}.
This diode array mitigates slower input transients.
In between, are schemes that compensate for the sTGC high rate and high charge signals, which include a level of uncertainty. Different schemes are needed for wires, pads and strips\,\cite{pi-networks}.

The amplitude and time structure of the sTGC signal require the use of an attenuator circuit in the case of the pads and wires, and a pull-up resistor circuit in the case of the strips, in order for the VMM to operate optimally under the conditions of the HL-LHC.

\vspace{3pt}\noindent\textit{Pad and wire input $\uppi$-networks:}
At the operational voltage of the sTGC, a single cavern background hit can induce a 50\,pC charge into a single sTGC pad, which exceeds the design requirements of the VMM front-end\footnote{At the time of the VMM design the input signal of the sTGC pads was underestimated and the input current would exceed the specifications.}.
This charge is sufficient to saturate the VMM feedback currents (configurable), causing an undesired and overly long recovery time.
To minimize this effect, a $\uppi$-network was implemented\,\cite{pi-networks} and is shown in Figure\,\ref{fig:TVS_Pi_blockDiagram} (opposite polarity for wires and pads).
The network acts as a charge divider, reducing the charge input to the ASIC front end.
The $\uppi$-network capacitor value was optimized to preserve efficiency and minimize recovery time.
Due to the difference in the pad capacitance along the detector radius, a different capacitor is used for the inner, middle, and outer quadruplet.
This circuit is directly implemented on the Front-end boards.
The optimization was based on radiation data (test-beam plus backgrounds), simulation, and charge injection studies.
In these studies, the optimal attenuation factor was found to be approximately 5:1, corresponding to 200\,pF, 330\,pF, and 470\,pF capacitors in the $\uppi$-network for inner, middle, and outer
pads FEBs respectively and 200\,pF for wire FEBs.

\vspace{3pt}\noindent\textit{Input pull-up resistor for strips:}
The signal formation of the sTGC detector is characterized by three components with different timescales:
The first component has a characteristic timescale of $\sim$20\,ns and is due to the electron avalanche drifting towards the wires.
The second component has a characteristic timescale of tens of microseconds and is due to the ion drift towards the cathode planes.
The third component has a characteristic timescale of milliseconds and is due to the charge induction within the resistive layer following the ion arrival to the layer,
and hence has the opposite polarity to the previous two\footnote{Sufficiently high resistance would neglect this component, but in the sTGC detector, in order to dissipate charge from the high background, it is not sufficiently high and a bipolar shaped signal is formed.}.
At the high rates of the HL-LHC, the second component from different signals overlap, causing a constant current into the VMM, which can be compensated by the Front-end, but strong bipolar shaped signals cannot.
Hence, although not necessary, a pull-up resistor was implemented for the strips, as shown in Figure\,\ref{fig:TVS_Pi_blockDiagram}.
In this case, the VMM baseline restoration circuit is protected by providing an additional constant feedback current.
A resistor of 400\,k$\Omega$ connected to supply voltage (1.2\,V nominal) was chosen.

\begin{figure}[ht]
\centering
\includegraphics[width=1.0\textwidth]{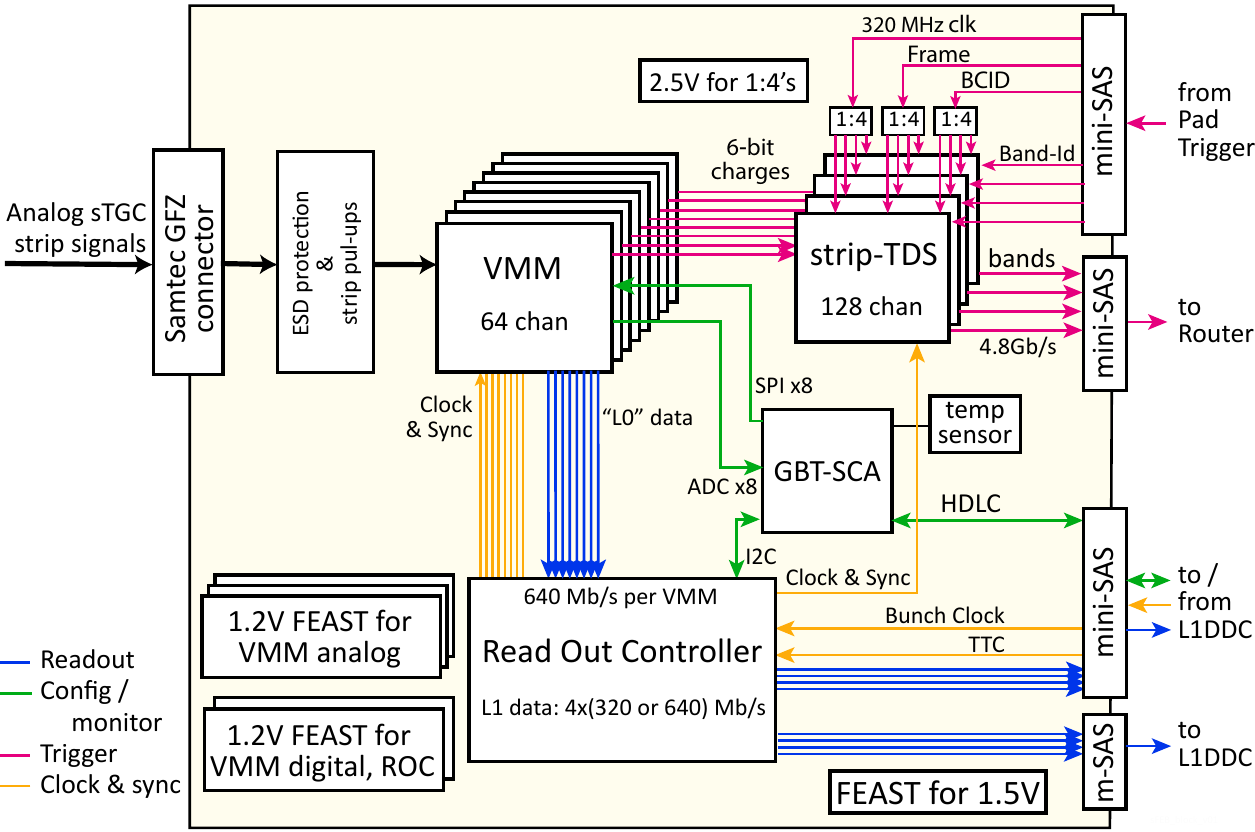}
\caption{Block diagram of the sTGC strip Front-end board that handles the strips in one sTGC gas gap, showing
         its ASICS, connectors and power blocks.
        The second MiniSAS connector is used only for the inner quadruplet.}
\label{fig:sFEB_block_V01}
\end{figure}

\para{Strip front-end boards:}
The sTGC strip Front-end board is a dense 14-layer circuit board (27.5\,cm\,$\times$\,7.6\,cm) comprised of six or eight VMMs, one ROC, one SCA, six FEAST, and three or four strip-TDS ASICs.
The outer two of the three quadruplets have fewer strips and so require only six VMM and three strip-TDS ASICs to be populated on the board.
Boards mounted on one edge of the detector are rotated 180\textdegree\ around its $z$-axis with respect to the other edge.
Therefore the radial coordinates of one are reversed from the other.
Board power consumption is 21\,W.
Figure\,\ref{fig:sFEB_block_V01} shows a block diagram of the board.

The seven serial LVDS lines from the Pad Trigger include four separate Band-id lines for the four strip-TDS ASIC positions on the board, a frame, BCID and a clock.
The BCID, clock and frame are distributed to the four strip-TDS ASICs by three 1-to-4 multiplexers.
Rather than an additional FEAST, three 2.5\,V precision voltage references, TL431AIDBZR\,\cite{TL431}, power them from the 10\,V supply.

\begin{figure}[ht]
\centering
\includegraphics[width=0.92\textwidth]{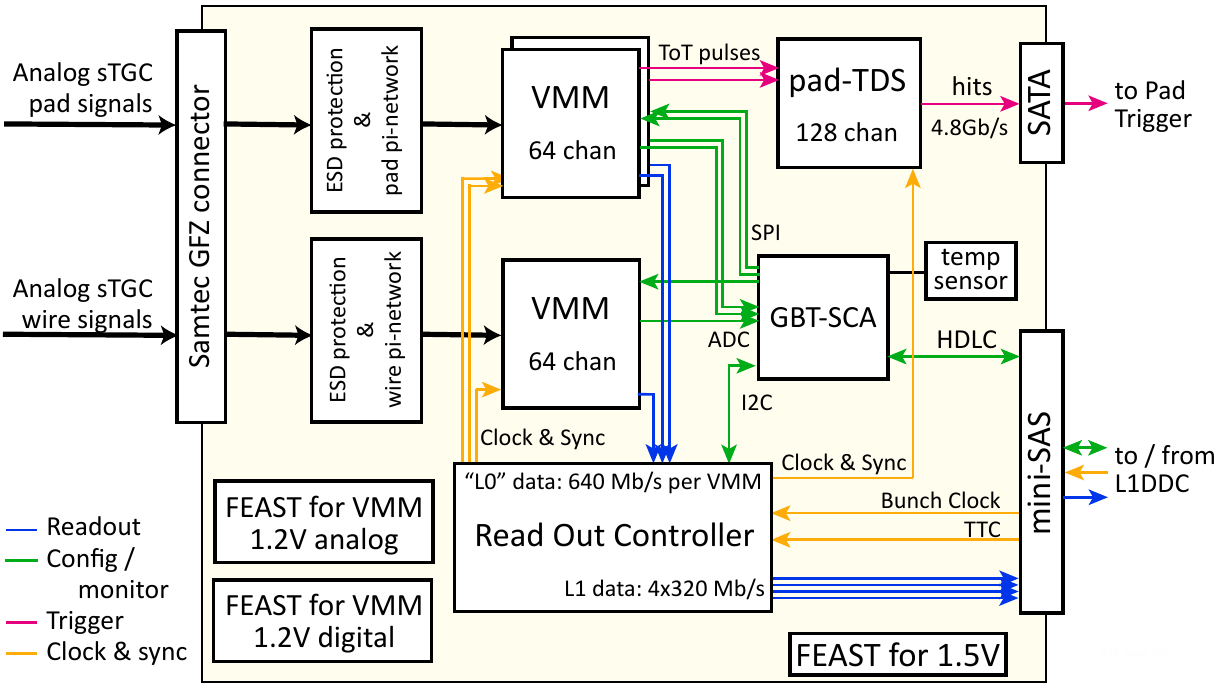}
\caption{Block diagram of the sTGC pad Front-end board that handles the pads and wires in one sTGC gas gap, showing
         its ASICS, connectors and power blocks.}
\label{fig:pFEB_block_V01}
\end{figure}

\para{Pad+wire front-end boards:}
The pad+wire Front-end board is a dense 12-layer circuit board (16.3\,cm\,$\times$\,7.6\,cm) comprised of three VMMs, one ROC, one pad-TDS, one SCA and three FEAST ASICs.
The sTGC pads connect to two of the VMMs, wire groups to the third VMM.
Boards mounted on one edge of the detector are rotated 180\textdegree\ around its $z$-axis with respect to the other edge.
Therefore the wire azimuthal coordinates for half the layers are reversed from the other and the pad numbering differs for different layers.
Board power consumption is 9\,W.
Figure\,\ref{fig:pFEB_block_V01} shows a block diagram of the board.

\subsection{Aggregation of several electrical links to optical links -- L1DDC}
\label{sec:L1DDC}
The Level-1 Data Driver Cards (L1DDC)\,\cite{Gkountoumis:2019ye, Gkountoumis:2779645, Gkountoumis:2016mrs, Gkountoumis:2016dfm} are high-bandwidth radiation-tolerant data aggregator boards based on the GBTx ASIC (Section\,\ref{sec:GBTx}).
On one side they connect through E-links with multiple front-end boards via twin-ax cables; on the other side, they connect through two or more fibres with FELIX.
To accommodate Phase\,2, a Front-end board can provide up to four 320\,Mb/s E-links for L1A readout data; only one is needed for the LHC Run-3.
The L1DDC is completely transparent to the data being transmitted or received.
MiniSAS cables make the connections to the front-end boards.
The optical interfaces are the VTRx and VTTx (see Section\,\ref{sec:VTRxVTTx}).
The boards are powered by FEAST DC-DC converters for 1.5\,V and 2.5\,V.
There are three variants of L1DDC's with the same basic functionality, but with different mechanics and channel counts.
One of them is part of the \MM electronics, and two of them of the sTGC electronics.

\begin{figure}[h]
\centering
\includegraphics[width=0.95\textwidth]{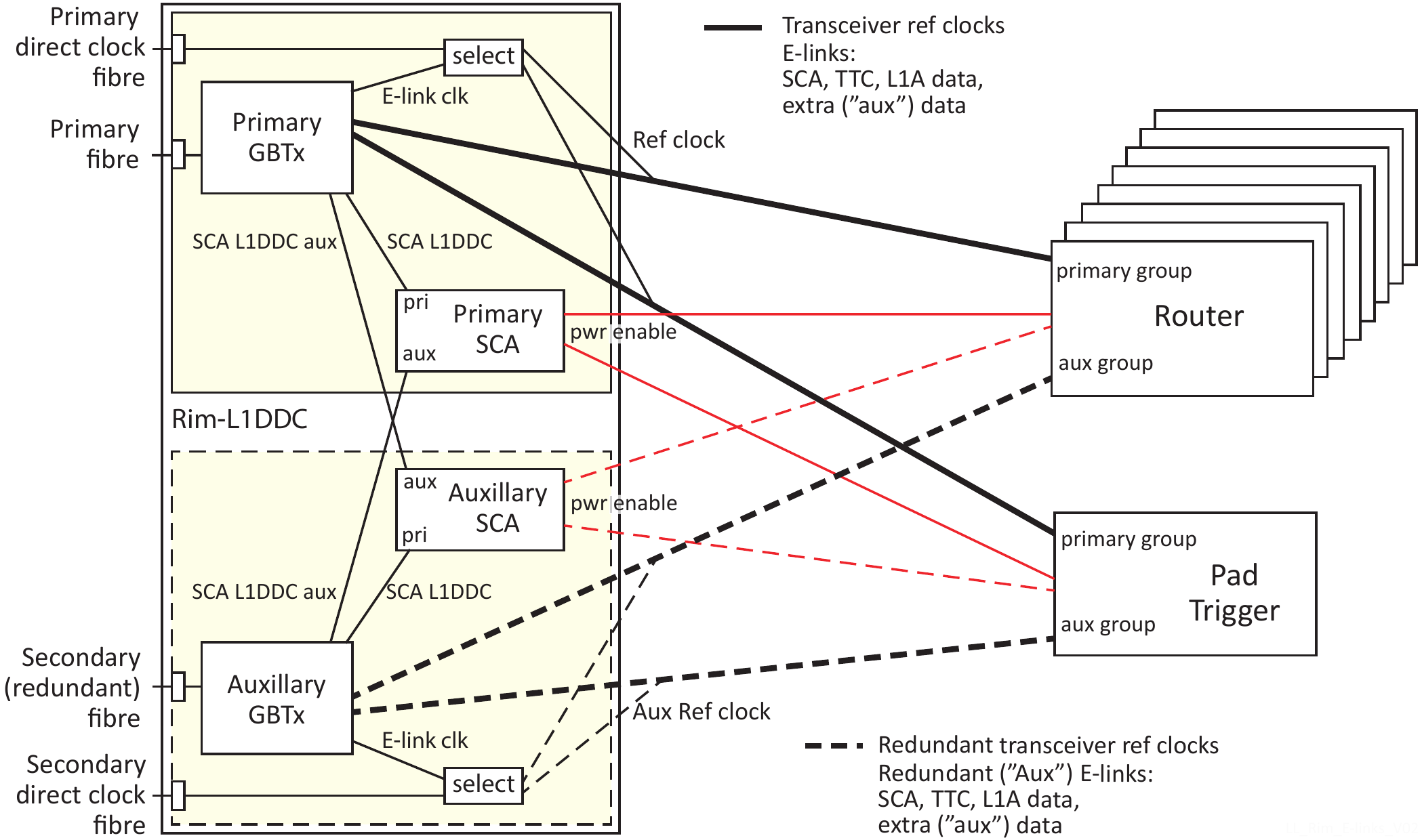}
\caption{The two redundant sections of the Rim-L1DDC and their connections to the Router and Pad Trigger boards}
\label{fig:LL_Rim_E-links}
\end{figure}

\para{The Rim-L1DDC:} Supports the Pad Trigger and eight Router cards in the sTGC trigger path electronics. Those electronics are in a dedicated crate called ``the Rim crate'' since it is located in the rim of the NSW structure.
The Rim-L1DDC consists of two independent boards, one primary and one auxiliary, sharing the same PCB.
Each includes one GBTx, one VTRx and nine MiniSAS connectors. This provides redundancy to the system since a complete sTGC trigger sector depends on it.
The Pad and Router boards are connected to both primary and secondary sections. The on-board SCA's are able to switch on/off the Pad Trigger and Router boards by controlling the enable/disable signal of their FEASTs.
Tests of the Rim-L1DDC showed that its GBTx ASIC's 160\,MHz E-link clock jitter was marginally good as a reference clock for the Xilinx\,7-family FPGA transceivers of the Router and Pad Trigger.
For that reason, a dedicated clock is provided through an additional fibre to the Rim-L1DDC. This direct low jitter 160\,MHz clock (Section\,\ref{sec:DirectClock}) is received through an additional VTRx and is distributed to the Router and Pad Trigger boards via low jitter fan-outs.
The block diagram of the Rim-L1DDC is shown in Figure\,\ref{fig:LL_Rim_E-links}.  The power consumption of the board was measured to be 8\,W.

\vspace{-10pt}
\para{sTGC-L1DDC:} interfaces with three Front-end boards, either strip or pad+wire. Two GBTx ASICs for two independent bi-directional (VTRx) links; only one is used for pad+wire FEBs. The second strip GBTx provides the extra bandwidth needed by Phase\,2.
In all, 21~E-links connect to a sTGC-L1DDC.
The board environment is monitored by an SCA ASIC.
In addition, by means of the SCAs second redundant E-link, both GBTx ASICs can communicate with the SCA.
The second GBTx though is configured through the \ItwoC of the SCA.
The L1DDC's are placed on the upper edge of the sTGC quadruplet.
The power consumption of the sTGC-L1DDC is $\sim$4\,W\footnote{Higher consumption up to 6\,W is measured on the sTGC L1DDC but is due to the consumption of the repeaters which are powered through the L1DDC.}.

\para{\MM -L1DDC:}  interfaces with eight Front-end boards and one ADDC board. It features three GBTx ASICs: one is a full-duplex transceiver (VTRx) and two are simplex transmitters (VTTx) for the extra readout bandwidth needed by Phase\,2.
The GBTx connected to the VTRx is configured directly from its IC link.
The two GBTx ASICs connected to the VTTx  are configured via the on-board SCA ASIC.
The SCA also monitors the temperature and voltage levels of the board and the VTRx's Rx signal strength.
One E-link connects to the SCA on the ADDC board; another distributes the TTC signals (BCR and BC clock) to the ADDC board.
In all, 41~E-links connect to a MM-L1DDC.
One layer of \MM is readout by two L1DDC boards. The L1DDC's are installed along the edges of the outer layers of the \MM quadruplet.
The board power consumption is $\sim$5.5\,W.


\subsection{\MM trigger data serializer -- ADDC}
\label{sec:addc}
\begin{figure}[b]
\centering
\includegraphics[width=0.9\textwidth]{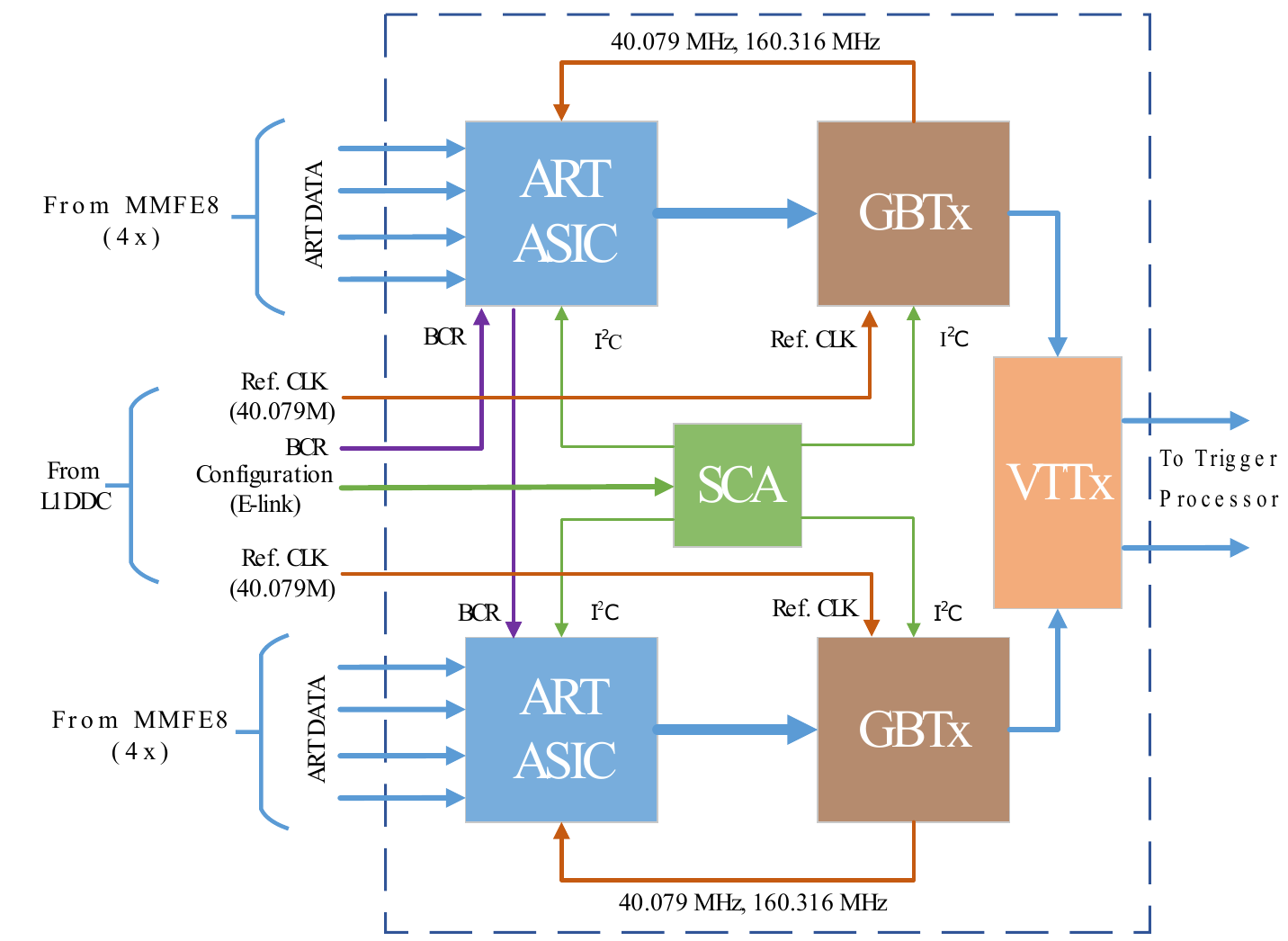}
\caption{The ADDC block diagram. Each ADDC receives 64 inputs of ART data using two ART ASICs. Output to the Trigger Processor is via two GBTx ASICs.}
\label{fig:block_ADDC}
\end{figure}

The Address in Real Time Data Driver Card (ADDC)\,\cite{ADDC, ADDCprod,8376567,9058723} features two ART ASICs, two GBTx ASICs, one GBT-SCA ASIC, one VTTx optical transmitter and two FEAST ASICs.
The ADDC block diagram is shown in Figure\,\ref{fig:block_ADDC}.
Each ART ASIC receives the 320\,Mb/s ART data from 32 VMMs on four MMFE8 front-end boards.
In total 64 VMM ART signals are driven through eight MiniSAS cables to the ADDC. Each ART ASIC then transmits the data to the onboard GBTx.

The hit selection within the two ART ASICs has been explained in Section\,\ref{sec:ART}. Each GBTx ASIC transmits the data then to the Trigger Processor through one transmission channel of the VTTx.  The configuration and control of the ART and GBTx ASIC is achieved by the on-board GBT-SCA ASIC. The ASICs are powered through two on-board FEAST ASICs configured at 1.5\,V and 2.5\,V respectively.

Since the ADDC features only two transmission optical lines to minimize the fibre connections, each ADDC is connected to the GBTx on one L1DDC through a  MiniSAS cable.  Through this connection the following lines are provided to the on-board ASICs:
\vspace{-5pt}
\begin{itemize}\itemsep-4pt
\item{SCA E-link: (three pairs,  BC clock, Tx/Rx data at 80\,Mb/s)}
\item{Two reference BC clocks, one for each of the GBTx ASICs}
\item{The BCR (Bunch Crossing Reset) signal}
\end{itemize}

\vspace{-6pt}\noindent Each of the GBTx ASICs is transmitting by itself the recovered and generated 40.079\MHz and 160.316\MHz clocks to each ART ASIC respectively.
In total 512 ADDC boards communicate with the 4096 front-end boards.

The ADDCs are placed on the edges of \MM detector along with the front-end boards (MMFE8) and the L1DDC boards.			 
The latency of the ADDC was measured to be $\sim$187\,ns, of which around 143\,ns is the latency of the GBTx plus VTTx dual optical transmitter module.


\subsection{sTGC Pad Trigger}
\label{sec:pad_trigger}

\begin{figure}[ht]
\centering
\includegraphics[width=0.72\textwidth]{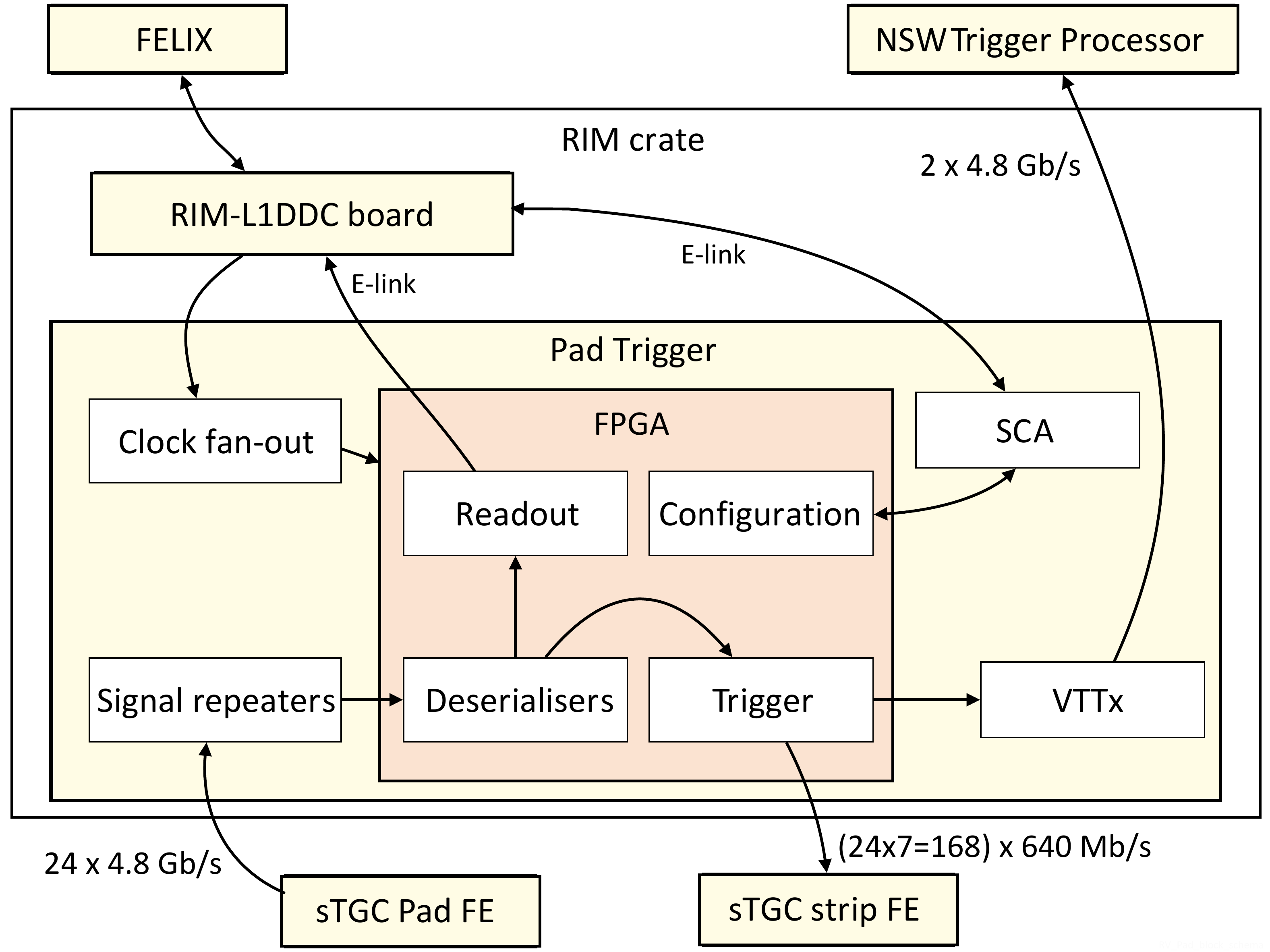}
\caption{Pad Trigger context and block diagram.
         Not shown: The output of the Trigger block is also sent to the Readout block.}
\label{fig:RV_Pad_block_schema}
\end{figure}

\begin{figure}[ht]
\centering
\includegraphics[width=0.99\textwidth]{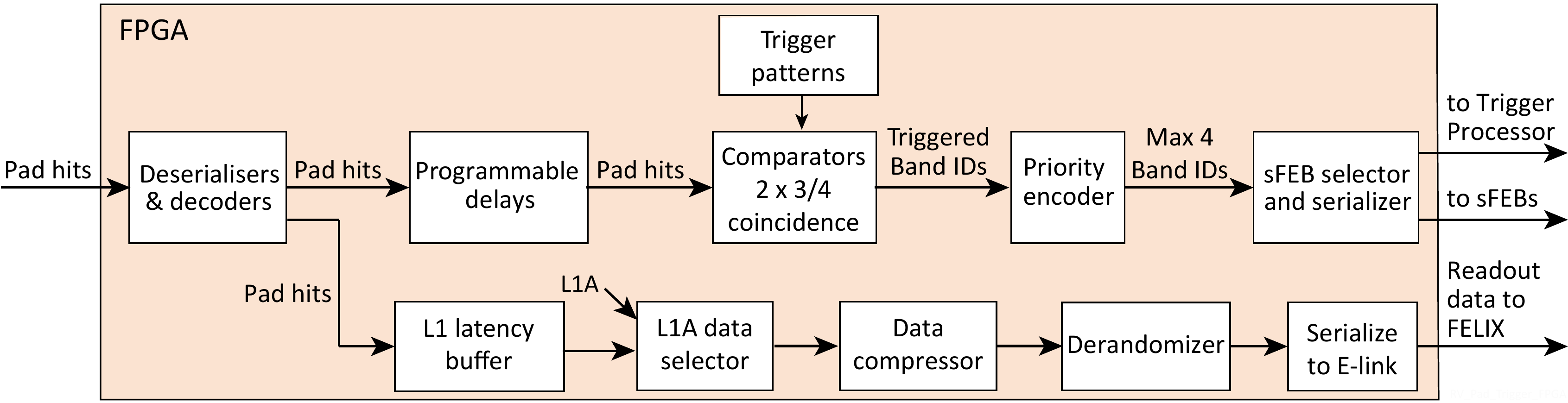}
\caption{Pad Trigger FPGA block diagram.
         Not shown: The output to the Trigger Processor is also read out to FELIX.}
\label{fig:RV_Pad_Trigger_FPGA}
\end{figure}

The sTGC Pad Trigger is a pre-trigger board that finds tracks passing through a pointing tower of logical sTGC pads.
See Figure\,\ref{fig:LL_PadSelect}.
Its purpose is to limit the on-detector trigger electronics to send out only the charge data from those bands of strips that pass through the pad tower coincidences.
Up to four towers can be found.
It receives the binary pad hits from 24 Front-end boards, three per layer, in a sector.
The band of 17 strips in each layer that pass through each tower is identified by a Band-id.
The Band-ids of triggered towers are sent to the relevant strip-TDS ASICs which then send the charge information of each strip in the band to the Trigger Processor via the Router.
One Pad Trigger per sector resides in the Rim Crate for that sector.
Its context and block diagram are shown in Figure\,\ref{fig:RV_Pad_block_schema}.
The internal blocks of the FPGA are shown in Figure\,\ref{fig:RV_Pad_Trigger_FPGA}.


Serial repeater chips\,\cite{ds100br410} condition the 24 4.8\,Gb/s input links from the pad-TDS ASICs.
They are configured via the \ItwoC channels of the on-board SCA ASIC.
After deserialization, each vector of pad hits can be delayed in programmable steps of 4.17\,ns to align them in time.

\para{Band-finding algorithm:} Valid 3-out-of-4 pointing towers for each 4-layer wedge were defined using the ATLAS off-line simulation to track single muons through the sTGC layers of a sector.
The simulation's pad id were mapped to a Pad Trigger input-id.
For each tower, the list of pad input-ids (one per layer), the band-id and $\phi$-id were put in a list of trigger patterns.
There is one comparator for each of about 4700 trigger patterns that compares the input pads to the list of pads in each trigger pattern.
The match is done taking into account the 3-out-of-4 majority logic per each quadruplet.
The comparators are grouped by their corresponding Band-id.
A one or two bunch crossing coincidence window is determined by the choice of a bit file specifically generated for that choice.
The current choice is a two BC window.
The lists of trigger patterns for large and small sectors are defined in separate FPGA bit files.
An output vector of comparator results, one bit per band, is passed to a priority encoder which selects a maximum of four Band-ids to be sent to the strip-TDS ASICs on the sFEBs.
The priority encoder also takes into account that a Band-id may need to be sent to two adjacent strip-TDS ASICs.
In this case the maximum number of Band-ids is reduced by one.
The selected Band-ids are finally directed to their corresponding strip-TDS via cable connections to each sFEB.
Note that each strip-TDS has a configurable lookup table that defines which strips belong to a given Band-id in that layer and that at most three of the four strip-TDS positions on a sFEB are populated.
The Pad Trigger output to each sFEB consists of up to four Band-ids (a maximum of one per TDS ASIC), the BCID, a frame and a clock, and is sent via seven serial LVDS lines (See Section\,\ref{sec:sTDS}.).
The maximum of four Band-ids may be distributed over any of the up to ten strip-TDS ASICs in each layer.
The Band-ids, BCID, and $\phi$-id along with some flags are also sent to the Trigger Processor via two redundant serial links at 4.8\,Gb/s, 8b/10b encoded, via a dual VTTx optical transmitter
(See Section\,\ref{sec:VTRxVTTx}.).

Configuration parameters, partially listed in Table\,\ref{tab:PadTrigConfig}, are set through the SCA server via the Rim-L1DDC and the on-board SCA ASIC. The FPGA configuration of the Pad Trigger board can be realised either by the SCA server via the on-board SCA's JTAG channel or by a Xilinx programmer via a connector. This is configurable through on-board jumpers.


\begin{table}[htbp]
  \centering
  \caption{Partial list of Pad Trigger configuration parameters.}
    \begin{tabular}{l}
    \toprule
    BC counter offset  \\
    Enable readout on L1A  \\
    Enable readout on self-trigger   \\
    Readout window BC offset of the first BC to be read on a L1A   \\
    Readout window size,  in BC units (0\,=\,1\,BC, 1\,=\,2\,BC, ...) \\
    pFEB input delay in steps of 4.17\,ns (0\,=\,4.17\,ns, 1\,=\,8.33\,ns, ...) \,4 bits per input \\
    24-bit mask to force all pad hits of the corresponding pFEB to 0 \\
    24-bit mask to force all pad hits of the corresponding pFEB to 1 \\
    Enable Orbit Count Reset (OCR) mode (where idle state is forced \\ ~~~~~until OCR is received). See Section\,\ref{sec:BCsync}. \\
    \bottomrule
    \end{tabular}%
  \label{tab:PadTrigConfig}%
\end{table}%

In addition to the trigger path output, all inputs and outputs are saved in a latency buffer until a Level-1 Accept is received via the TTC system.
In response to the Level-1 Accept, data for a configurable BC window is sent out on an E-link to the swROD via FELIX.
The input vectors of pad hits are compressed using a sparse binary compression algorithm. See Chapter\,11.5 of\,\cite{salomon2010handbook}.

For redundancy, the Pad Trigger is connected to both of the primary and secondary Rim-L1DDC (See Section\,\ref{sec:L1DDC}.).
The SCA on the Pad Trigger is connected to both as well.
This allows the SCA to choose which Rim-L1DDC provides the 160\,MHz reference clock for the Pad Trigger's gigabit transceivers.
See Section\,\ref{sec:L1DDC}.
The choice of which Rim-L1DDC supplies the E-links is defined by the FPGA design, i.e.\ an FPGA bit file is specific to one Rim-L1DDC or the other.

The Pad Trigger's components are cooled by conduction through heat-conductive gap pads to a copper plate that is water cooled.
Special attention was paid to cool the VTTx optical transmitters which are heat sensitive.
The Pad Trigger board consumes 25\,W from a 9.2\,V supply.

\subsection{sTGC Router}
\label{sec:router}

\begin{figure}[h!]
   \centering
   \includegraphics[width=0.75\textwidth]{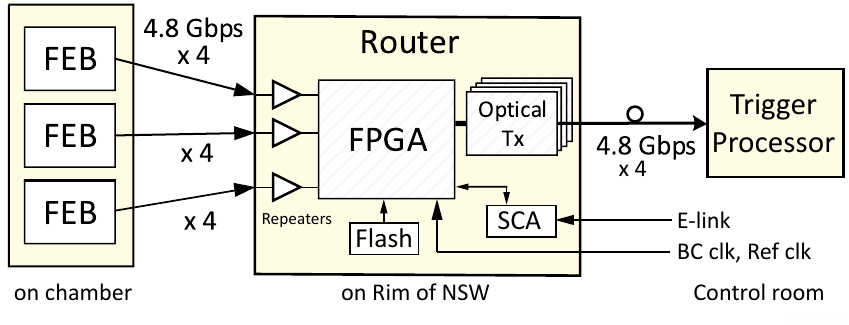}
   \caption{Context diagram of the Router for a layer showing the four electrical inputs from each of the three strip Front-end boards in one layer of a sector and the four fibre outputs.
             }
   \label{fig:RouterContext}
   \end{figure}

The sTGC Router\,\cite{HU2022167504} serves as a packet switch for routing sTGC strip charge information from the strip-TDS to the sTGC Trigger Processor.
It is implemented in a Xilinx FPGA\footnote{XC7A200TFFG1156} from the Artix-7 family.
There is one Router per layer for each sTGC sector.
There are four TDS ASIC positions on a strip Front-end board;
depending on the board location three or four are populated.
A scheme was developed to maintain low and fixed-latency packet multiplexing through the Router\,\cite{RouterFixedLat2, RouterFixedLat1}.
The Router's latency from first bit in to first bit out is 97\,ns.
The Router is unaware of the ATLAS run state and does not send any data in response to a Level-1 trigger.
A context diagram is shown in Figure\,\ref{fig:RouterContext}.

On every bunch crossing the sTGC Pad Trigger chooses up to four bands of strips and sends their band-ids to the strip TDS ASICs that contain those bands.
The selected TDS ASICs transmit the strip data for the band of strips they contain to the Router.
The other strip TDS ASICs transmit a null packet.
The Router inputs are twin-ax electrical serial streams at 4.8\,Gb/s.
The input serial streams are deserialized, unscrambled and aligned to a common clock.
The Router then routes the up to four packets containing the strip charges for a chosen band to the Trigger Processor via the Router's four 4.8\,Gb/s output fibres.
Null packets (containing the sector-id, layer and fibre number) are sent out when there are less than four strip-charge packets.
These are also used in confirming the connectivity of the inputs.

The on-board serial repeater chips\,\cite{ds100br410} condition the 4.8\,Gb/s links from each strip-TDS.
Their parameters were optimized and set with soldered jumpers.
Miniature Transmitter (\gls{MTx}) optical transmitters\,\cite{Zhao:2016czy, Xiao:2016dvu} are used to drive the output fibres.
They are configured via an SCA \ItwoC master.

The on-board SCA  ASIC is used to control the Router.
The sector-id is set in the FPGA via the SCA \gls{GPIO}.
The twin-ax cables from the inner and middle radii front-end boards were made the same length.
The cable from the outer Front-end board is shorter by one or two clock equivalent periods.
In order to align all the inputs, the Router inserts a delay, either one (or two) 160\,MHz clocks, for signals from the outer front-end boards of the large (or small) sectors.

The Router FPGA can be configured from on-board Flash memory or by FELIX via the \gls{JTAG} port of the SCA.
The Router's components are cooled by conduction through heat-conductive gap pads to a copper plate that is water cooled.

\para{Radiation tolerance}

The Router's tolerance to total ionization dose was shown acceptable and reported in\,\cite{RouterRadTol, HU2022167504}.
SEU's in the FPGA fabric or its configuration memory can cause malfunction or halting of the Router operation.
The critical logic sections that handle control and initialization are protected by Triple Modular Redundancy (TMR)\,\cite{TMR}.
The goal is to retain a steady data flow, while allowing bit upsets in the data stream to be mitigated by using the redundancy of multiple detector layers.

Configuration memory bits are protected with Error-Correction Code (\gls{ECC}) and Cyclic Redundancy Check (\gls{CRC}).
However, SEU's may accumulate in essential bits and result in functional failures.
Upsets in configuration memory are handled by the Soft Error Mitigation (SEM) tool from Xilinx\,\cite{XilinxSEM} which monitors the integrity of the configuration
memory and can fix up to two bits upset simultaneously.

If the TMR and SEM protection is corrupted by multiple bit upsets, the Router FPGA can be recovered through reconfiguration, by:
1)~locally reading from the on-board flash memory, which can be updated to the latest version, or
2)~by running JTAG remotely via the SCA JTAG port.
The JTAG programming has the highest priority by default, in case multiple programming paths are available.

Finally, in case of any non-recoverable functional interrupt, e.g.\ single-event functional interrupt (SEFI),
the Router power can be cycled by disabling its FEAST DC-DC converters via a signal from the SCA on the Rim-L1DDC.
Further details of the Router's SEU mitigation strategy may be found in\,\cite{RouterSEU}.


\subsection{Twin-ax serial and LVDS repeaters}
To overcome attenuation due to the up to 6.25\,m twin-ax cable length in the sTGC trigger path (see Table\,\ref{tab:twinaxAttenuation}), repeaters are placed roughly midway along the twin-ax signal paths to guarantee error-free operation.
A 4.8\,Gb/s signalling rate requires good transmission of several odd harmonics of the signalling rate.
The 640\,MHz LVDS transmission is less sensitive to the high frequency attenuation, but does suffer from the attenuation of the fine wires (30\,AWG) in the cable.
The 4.8\,Gb/s serial repeaters restore the signals from the pad-TDS to the Pad Trigger and from the strip-TDS to the Router;
640\,Mb/s parallel LVDS repeaters restore signals from the Pad Trigger to the strip-TDS.
See Figure\,\ref{fig:LL_NSW_ElxOvr} for their locations.
For more details see\,\cite{repeaters}.

\begin{table}[h]
  \centering
  \caption{Attenuation versus frequency for the  ``MiniSAS'' twin-ax ribbon cable (3M~SL8800 Series MiniSAS cables) from\,\cite{twin-ax}. Silver-plated twin-ax cables were finally preferred because they have reduced attenuation at higher frequencies and lower prices. Since a large number of tin-plated cables were already in stock  they were used in the connection from Pad Trigger to sFEB (640\,Mb/s).}
    \begin{tabular}{lrrrrrrr}
    \toprule
    {\textbf{Frequency (GHz)}} & \multicolumn{1}{c}{\textbf{0.5}} & \textbf{1.0} & \multicolumn{1}{l}{\textbf{2.0}} & \multicolumn{1}{l}{\textbf{5.0}} & \multicolumn{1}{l}{\textbf{10.0}} & \multicolumn{1}{l}{\textbf{15.0}} & \multicolumn{1}{l}{\textbf{20.0}} \\
    \midrule
    {Tin plating (dB/m)} & -0.90 & -1.4  & -2.2  & -4.0  & -7.5  & -10.9 & -14.6 \\
    {Silver plating (dB/m)} & -0.85 & -1.2  & -1.7  & -3.2  & -4.9  & -6.8  & -8.8 \\
    \bottomrule
    \end{tabular}%
  \label{tab:twinaxAttenuation}%
\end{table}%

\para{4.8\,Gb/s serial repeaters}
\label{sec:serialRepeaters}
Tests showed that for error-free operation with a safe margin, serial repeaters are necessary for twin-ax cable lengths beyond 4\,m.
A serial repeater for a single 4-pair cable is housed in a small shielded copper box compatible with the width of the cable for easy mounting inline with the cables.
They are powered from a nearby L1DDC.
The quad-channel serial repeater chips\,\cite{ds100br410} are the same as those used as signal conditioners in the Pad Trigger and Router.
Receive equalization, transmit de-emphasis and transmit voltage can all be set by means of pin jumpers.
PRBS test data transmitted from the TDS ASICs showed a bit error rate less than $10^{-14}$.
Different sets of parameters and all cable lengths combinations were used to select the parameters with the lowest transmission error rates.
It was possible to select a common setting for all cable combinations.
A thermal simulation, and subsequent tests using temperature probes, confirmed that the repeater chip was adequately cooled by the copper box.
Power consumption at 2.5\,V is 213\,mW per repeater board. A total of 880 serial Repeater boards were built.

\para{LVDS repeaters}
\label{sec:LVDSRepeaters}
A 6.25\,m twin-ax cable carries the seven serial 640\,Mb/s LVDS lines from the Pad Trigger to the strip-TDS.
The eye-diagram of a test at 640\,Mb/s with a 5\,m cable showed attenuation and an 8\,m cable was unstable with barely an eye.
Consequently, to be safe with a 6.25\,m cable, it was decided to install repeaters.
Single channel Micrel SY58605U 3.2\,Gbps Precision LVDS buffers\,\cite{SY58605U} were used to regenerate the signals.
The high power consumption (a bit less than 1\,W per 7-bit connection) required active cooling.
The LVDS repeaters were located on the spokes of the Wheel, behind the Large Sectors, where cold water for cooling was available.
Boards with six repeaters each, are powered with 2.5\,V from a FEASTMP pluggable module DC-DC converter\,\cite{FEASTMP}.
Two cooling bars are soldered on either side of a copper pipe, carrying cold water.
Two boards are mounted on either side, in thermal contact with the copper bars.
All four boards are enclosed in a common metallic shielding.
A total of 144 LVDS Repeater boards were built.


\subsection{Direct low jitter FPGA transceiver reference clock}

\label{sec:DirectClock}

\begin{figure}[b]
\centering
\includegraphics[width=0.9\textwidth]{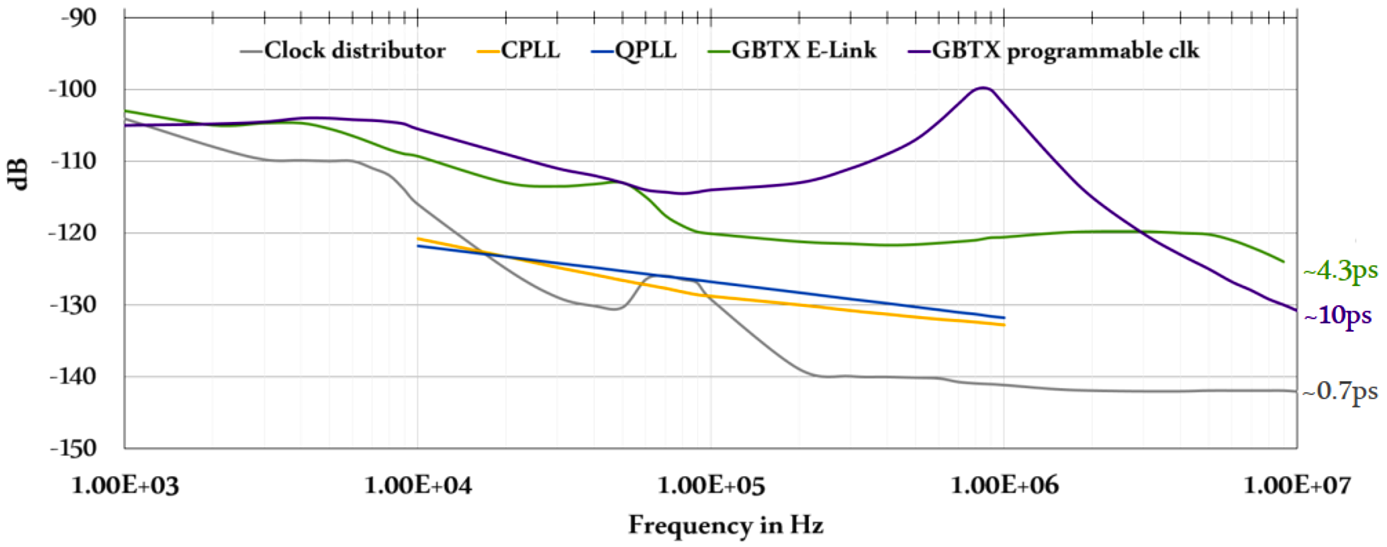}
\caption{Phase noise (random--rms) jitter of the direct and GBTx clocks compared to Xilinx's phase noise mask for the CPLL and QPLL.
         The integrated bandwidth for GBTx E-link and programmable clocks is 1\,kHz to 20\,MHz. The integrated bandwidth for the clock distributor is 1 kHz to 30 MHz\,\cite{directClock}.}
\label{fig:phase noise}
\end{figure}

Initially the E-Link and programmable clocks of the GBTx ASIC were used as 160\,MHz reference clocks for the 4.8\,Gb/s serial receivers in the 288 sTGC Pad Triggers and Routers.
They finally proved to be marginal during the data integrity tests due to their high jitter of about $4.3\,\mathrm{ps}$ and $10\,\mathrm{ps}$ respectively as shown in the phase noise plot in Figure\,\ref{fig:phase noise}.
The jitter cleaner with the required low jitter was tested and found to suffer from fatal single event upsets in the foreseen radiation environment\,\cite{ATL-MUON-PUB-2022-001}.
Consequently, an additional clock distribution scheme\,\cite{directClock} was designed to deliver a low jitter clock directly
from the radiation-protected room outside the ATLAS collision cavern to each Rim L1DDC via $63\,\mathrm{m}$ OM3 fibres.
A clock distributor board per endcap receives the timing, trigger and control (TTC) stream from the ALTI module\,\cite{ALTItwiki} and
recovers a 160\,MHz clock based on the LHC bunch crossing clock, using the Clock and Data Recovery (\gls{CDR}) ADN2814 chip from Analog Devices.
The recovered clock is fanned out to four Si5345 jitter cleaners from Silicon Labs.
Each of their eight outputs is further fanned out by 1:4 fan-outs, Si53306, from Silicon Labs with an ultra-low additive jitter of $50\,\mathrm{fs}$.
The cleaned clocks then drive three 12-channel Avago/Foxconn miniPOD optical transmitters\,({\small AFBR-812FH1Z})
which drive 32 fibres -- one to each of the two redundant Rim L1DDC's per sector.
Each section of the Rim-L1DDC can then be configured by its SCA to forward either the direct clock or an E-link clock to the trigger boards in its Rim crate.
The received clock is chosen and fanned out by means of ultra-low jitter (less than $300\,\mathrm{fs}$ additive jitter) fan-outs.

The clock distributor board is a \gls{VME}\,6U board, designed with several measures\,\cite{directClock} taken to minimize jitter.
The jitter cleaners are configured via \ItwoC using an on-board commercial Ethernet to \ItwoC adapter card.
The VME backplane provides only power to the board.

\subsection{Board and ASIC counts}
\label{sec:CountsOfBoardAndASICs}
Ten types of custom electronics boards were designed for the NSW electronics system; over 8000 of them were installed.
Four custom ASICs were designed; over 49,000 of them were installed.
Table\,\ref{tab:boardsASICs} itemizes the counts of NSW boards and ASICs.

\begin{table}[htbp]
  \centering
  \caption{The numbers of the various boards and ASICs in the NSW system.
  \local{This table summarizes the table in\,\cite{BoardASICcounts}.} }
    \begin{tabular}{lrllrl}
    \toprule
    \textbf{Board} & \textbf{Quantity} & \textbf{Comment} & \textbf{ASIC} & \textbf{Quantity} & \textbf{Comment} \\
    \cmidrule (r){1-3}   \cmidrule (l){4-6}
    MMFE8         &  4096 &                     & VMM   & 40,192 &            \\
    pad-FEB       &   768 &                     & pTDS  &  768   &            \\
    strip-FEB     &   768 &                     & sTDS  & 2304   &            \\
    ADDC          &   512 &                     & ART   & 1024   & 2\,/\,ADDC \\
    MM L1DDC      &   512 &                     & ROC   & 5632   & one\,/\,FEB \\
    sTGC L1DDC    &   512 &		        & GBTx  & 3712   &            \\
    Rim-L1DDC     &    32 &                     & SCA   & 7520   & one\,/\,board  \\
    Pad Trigger   &    32 &                     & VTRx  & 1920   &         \\
    Router        &   256 &                     & VTTx  & 1056   &         \\
    Serial repeater & 768 &                     &       &        &         \\
    LVDS repeater   & 128 &                     &       &        &         \\
    Direct clock  &     2 &  		        & MTx   &  512   &         \\
    FELIX FPGAs   &    60 &                     & FEAST IC & 5760 &        \\
    Trigger       &    16 &  2 sectors each	&       &       & \\
    ~~~~Processor &       &			&       &       & \\
    \bottomrule
    \end{tabular}%
  \label{tab:boardsASICs}%
\end{table}%

\section{Optical fibre interconnects}
\label{sec:fibres}
\begin{figure}[h]
\centering
\includegraphics[width=0.95\textwidth]{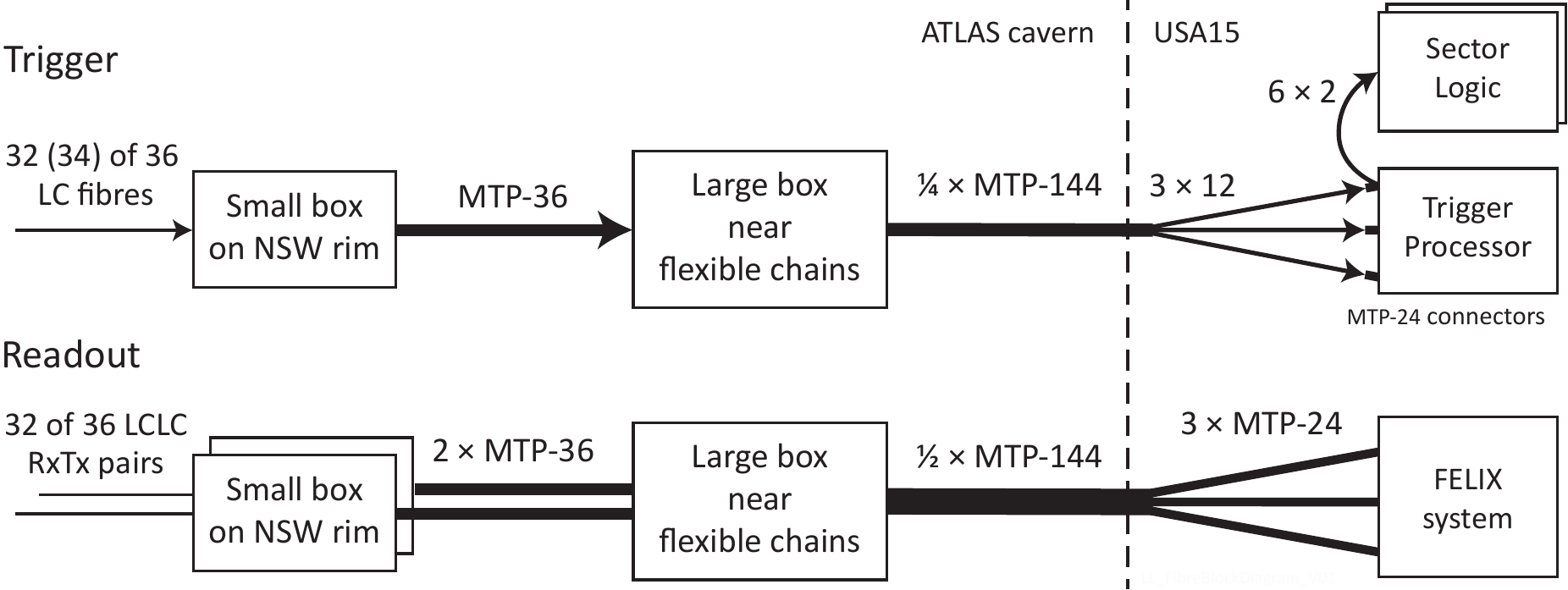}
\caption{Block diagram of the trigger and readout fibre paths for one sector of sTGC \emph{or} \MM.
            \MM has 32 trigger inputs to the Small box, whereas sTGC has two additional inputs from the Pad Trigger for 34 inputs.
            Six output pairs of the sTGC Trigger Processor are used to fan out the segments to up to seven different Sector Logic sectors.
            When seven are required, a second Trigger Processor \gls{MTP}-24 IO connector is used.}
\label{fig:LL_FibreBlockDiagram_V01}
\end{figure}

Fibres are used to connect between electronic boards on the NSW wheel itself in the collision cavern and the radiation-protected room (USA15).
All fibres are multi-mode {\small OM3} (allows transmission up to 10\,Gb/s) and radiation tolerant.
They range from LC-LC pairs to bundles of 36 and 144 fibres.
Figure\,\ref{fig:LL_FibreBlockDiagram_V01} shows a block diagram of the fibre path for one \MM or sTGC sector.

To minimize the trigger latency, the fibres to the Trigger Processors in USA15 pass through a special short path for ATLAS trigger fibres.
The need for several stages of fibre-to-fibre couplers as signals are aggregated to large fibre bundles introduces insertion loss at each coupler.
To have an optical power margin as large as possible despite these insertion losses, direct connections of fibre breakouts from the trunk cables are made to FELIX and the Trigger Processor,
i.e.\ there are no fibre patch panels in USA15 for fibres coming from the cavern.
Altogether, arriving in USA15, there are 2304 trigger fibres and 4608 readout fibres, including spares.
The MTP-36 cables between Small and Large boxes have different lengths depending on their angular position on the wheel.
For connection to FELIX, two \MM sectors share one 144-fibre readout trunk cable and two sTGC sectors another trunk cable.
For the trigger fibres, four sectors share a trunk cable.
The readout trunk cables were made long enough to reach almost anywhere in USA15 since it was not known where in USA15 the Phase\,2 FELIX servers would be located.

\section{Trigger latency}
\label{sec:latency}

The ATLAS Run-3 trigger latency is required to not exceed the limits imposed by those detectors that are unchanged from the previous running period.
Considerable design and implementation effort was made to minimize the NSW latency so as not to be on the critical path.
For example, a low-latency path dedicated to trigger fibres is used between the detector and the radiation-protected room (USA15);
Pad Trigger Band-ids are sent directly to the sTGC Trigger Processor to allow it to prepare the needed look-up table values and
the internal routing of strip data before the strip data arrives.
Table\,\ref{tab:latency} shows the large contributions to latency due to the size of the NSW and the length of the trigger path.
Details of the measured latencies and the signal path may be found in\,\cite{NSWlatency}.

\begin{table}[htbp]
  \centering
  \caption{Contributions to the trigger latency}
    \begin{tabular}{lr}
    \toprule
    \textbf{item} & \textbf{ns} \\
    \midrule
    longest time-of-flight from the IP &  31 \\
    electronic modules       & 405 \\
    serializer-deserializers & 189 \\
    Pad traces \& twin-ax cables &  136 \\
    All fibres               & 329 \\
    ~~~~including from the Large Box to   &      \\
    ~~~~the Trigger Processor (42.5\,m)   &  212 \\
    \bottomrule
    \end{tabular}%
  \label{tab:latency}%
\end{table}%

\section{Configuration}
\label{sec:configuration}
\begin{figure}[h]
\centering
\includegraphics[width=0.8\textwidth]{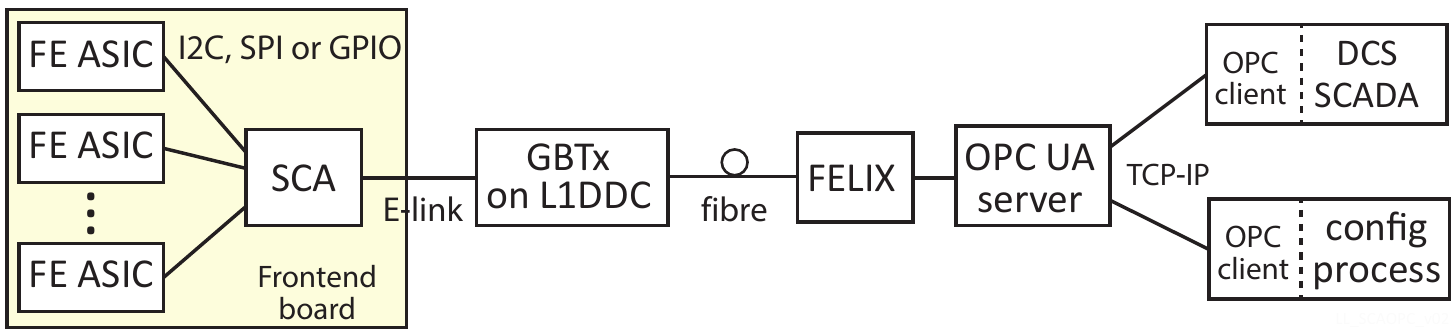}
\caption{The Configuration path with OPC\,UA clients and servers.}
\label{fig:LL_SCAOPC}
\end{figure}

All NSW electronics boards have a Slow Control Adapter ASIC (SCA)\,\cite{GBT-SCA}, see Section\,\ref{sec:SCA} for configuration of the ASICs on the board and readout of board temperatures and voltages.
The FPGA-based Trigger Processor includes firmware, SCAx\,\cite{SCAxIEEE}, that emulates the  \ItwoC sub-device of the SCA ASIC, providing access to registers and memories in the FPGA.
On the software side, a dedicated OPC\,UA server\,\cite{opcuaserver} for the SCA facilitates the SCA communications for the configuration and monitoring of the electronics
The use of OPC\,UA enables the interface to the SCA to be shared by both the Detector Control System and the Data Acquisition System.
It is an industry-standard protocol with an advanced ecosystem of tools.
The details of the configuration path are shown in Figure\,\ref{fig:LL_SCAOPC}.
See\,\cite{Tzanis:2021ttb,Tzanis:2022xje} for additional details.
Many configuration parameters are tuned through the calibration procedures described in Section\,\ref{sec:calibration}.

\section{Calibration}
\label{sec:calibration}


\subsection{Phase alignments}
\label{sec:PhaseAlign}
The primary clock source in the NSW electronics is the Bunch Crossing (BC) clock which is distributed by the ATLAS TTC system\,\cite{TTC}. The NSW electronics generates all the other needed clocks based on the BC clock. Since the data generated in one board may not be aligned with the clock needed to decode these data on another board, a clock-data alignment needs to be performed. This operation is performed by shifting the clock phase with respect to the data such that the decoding is correct.  The following parts need to be aligned in NSW electronics:

\para{ROC TTC reception:}The ROC ASIC receives and distributes the eight TTC bits per BC forwarded by FELIX that are representing the TTC signals specified in Section\,\ref{sec:overall}. Since the eight bits have no begin or end marker, the BC clock needs to be shifted such that the TTC bits are correctly extracted.

\para{VMM-ROC data line:}The VMM transmits data to ROC using a 160\,MHz clock. When not transmitting data, the chip transmits K28.5 idle characters. Although the ROC provides the data clock to the VMM, the signal received is shifted with respect to the transmitted clock. To compensate for that, ROC generates an internal copy of this clock which can be shifted accordingly and hence decode the data correctly.
Errors detected by the ROC while decoding are flagged in a status register that can be read out.

\para{VMM-ART serial stream:}The VMM transmits serially flag and address information to the ART
using a 160\,MHz clock provided by the ROC. The ART shifts the phase of the input serial stream and decodes the received data with its local 160\,MHz clock.

\para{TDS calibration:} Two VMMs transmit serially 6-bit charge information (after a flag) to TDS using a 160\,MHz clock provided by the TDS to which it is connected.
For each of the two VMMs, the phase of the TDS must be calibrated so that the TDS sampling point allows a correct decoding of the data.
The TDS BC clock is provided by the ROC; its phase has to be calibrated to correctly receive the Pad Trigger data.
The relative time offset between the reception of the pad data and the reception of the strip data needs to be calibrated so that the strip data matches the BCID requested by the Pad Trigger.

\para{Trigger Processor - Router, ADDC, Pad Trigger:}The Trigger Processor receives serial streams from the ADDC, the Router and the Pad Trigger boards. In all cases, the streams are deskewed to account for differences in cable lengths and
deserialization. Deskewing requires reading the BCID of each incoming stream independently,
and aligning them with a per-stream delay in units of the local decoding clock. This clock is
240\,MHz for the Pad Trigger and 320\,MHz for the Trigger Processor.

\para {GBTx:}To deserialize the input data stream from, e.g.\ the ROC ASIC, the phase of the GBTx E-link receive clock must be adjusted to sample the incoming data at the correct time.
The GBTx output E-link clock is used by the front-end ASICs to transmit data back to the GBTx. The correct phase of the receiving clock depends on the ASICs internal delays and the cable delay between GBTx and Front-end.
The GBTx features phase aligner circuits, one per E-group each with eight adjustable channels, as all the E-links in an E-group may have different phases.
The phase aligner can operate in three modes: static, automatic phase tracking and initial training with learned static phase selection. The latter two are proposed for the NSW operation but performance is still to be demonstrated.

\subsection{SCA-ADC based calibration}
\label{sec:SCAconfig}
\begin{figure}[ht]
\centering
\includegraphics[width=0.48\textwidth]{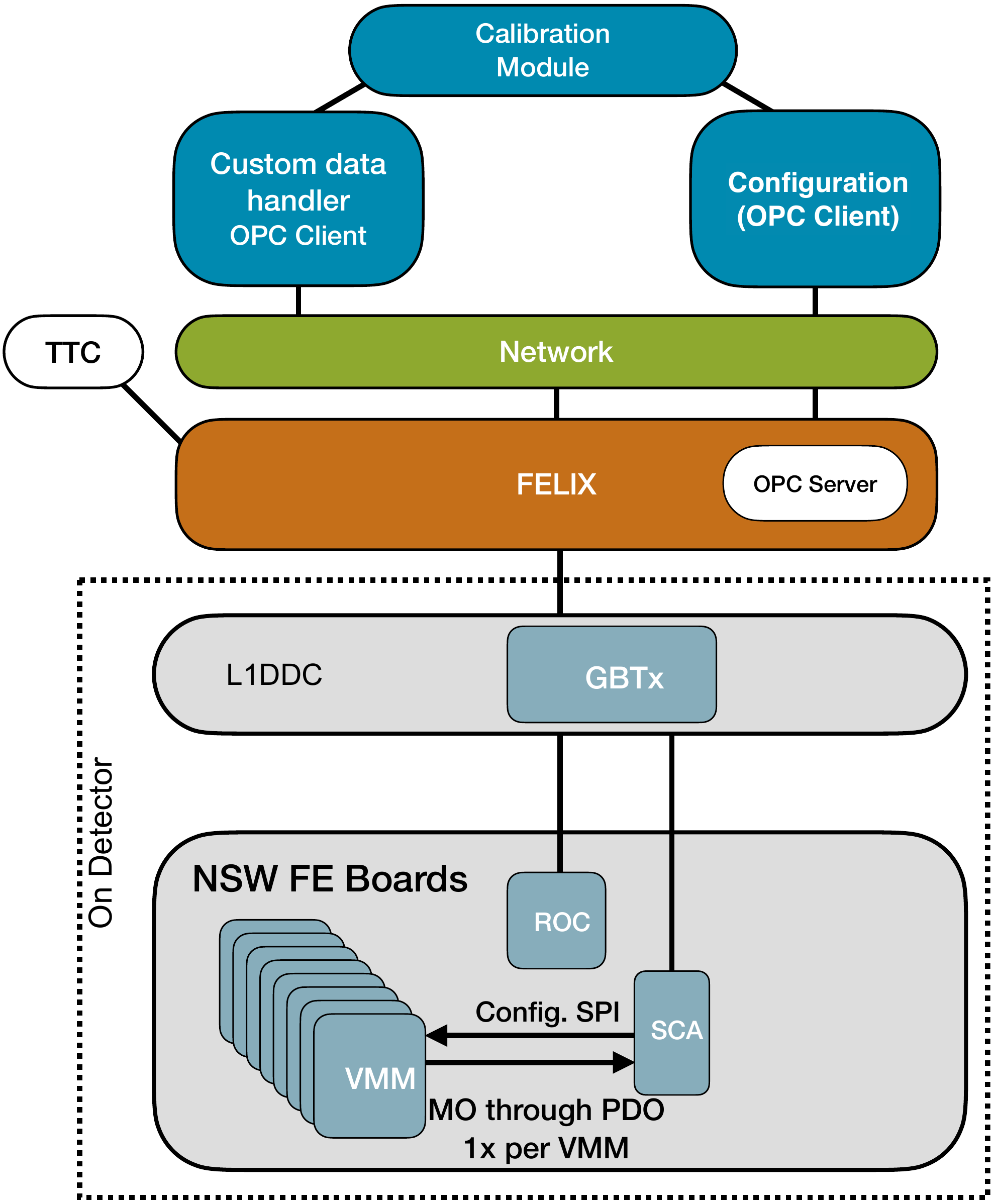}
\includegraphics[width=0.48\textwidth]{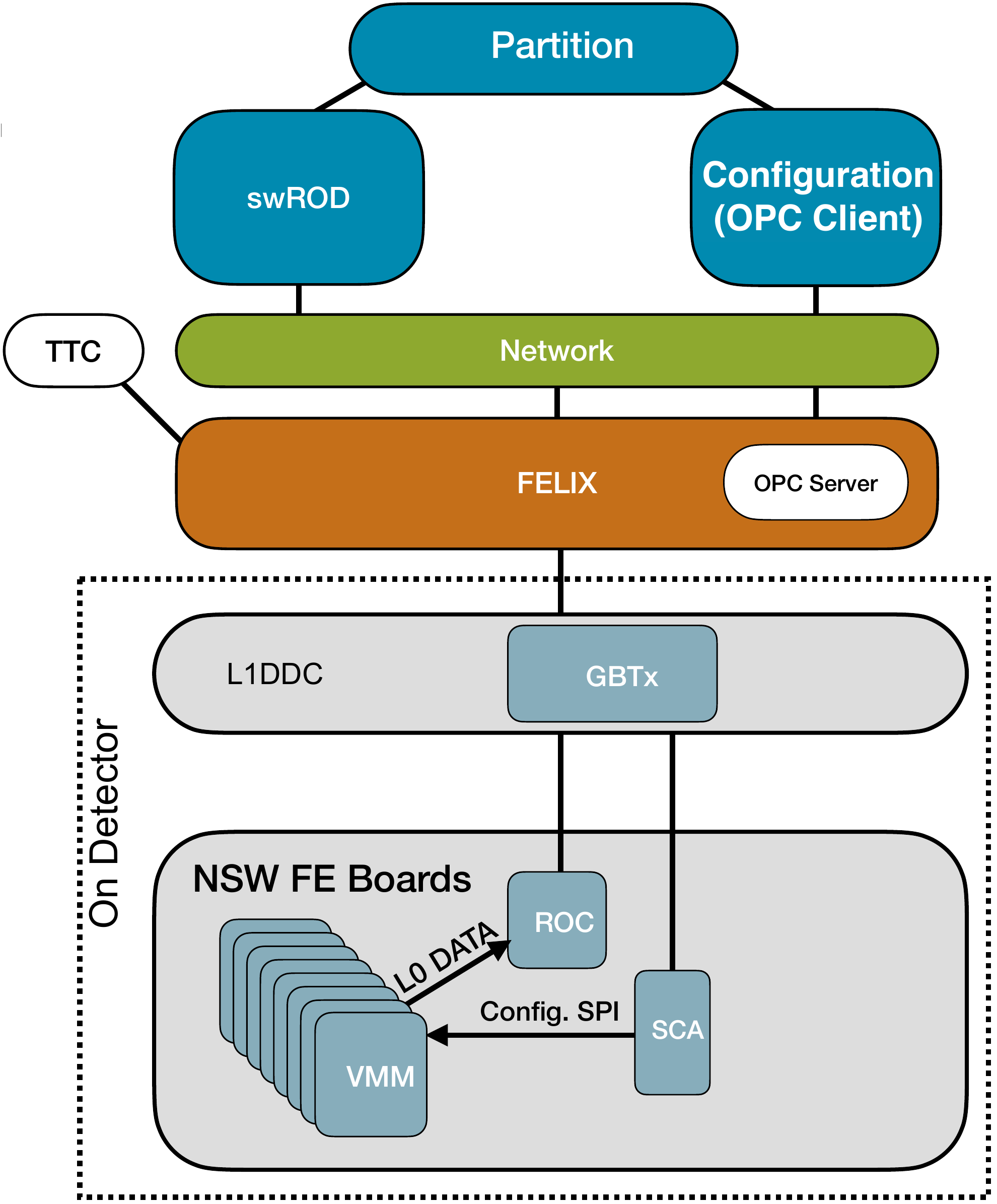}
\caption{Left: Schematic representation of the NSW electronics setup indicating the paths involved in the SCA-ADC based calibration. Commands are transmitted through the OPC client and server via FELIX to the on-detector electronics. The SCA ADC is instructed to sample the analog output of the VMM and data are transmitted back to the custom data handler software. \\
Right: Schematic representation of the data-taking path in the calibration procedure. The dedicated configuration of the DAQ system (partition) configures the system with the OPC client-server through FELIX; the TTC is configured to produce a sequence of test pulses and L1A and L0A signals. Data are captured by the swROD.}
\label{fig:GI_calib}
\end{figure}

This type of calibration does not involve the L1A data. It uses the ADC implemented in the SCA to sample the Monitor Output (MO) of the VMM. Since the MO is routed on a connector physically on the front-end boards, the VMM is configured to copy the MO to the PDO output which is routed through a voltage divider, in the case of MMFE8, to the ADC multiplexer of the SCA. The voltage divider is necessary since the MO output has a range up to 1.2\,V, while the ADC of the SCA has a 1\,V range. The sTGC Front-end board does not implement a divider. All the measured parameters are needed for the optimal configuration of the NSW electronics to acquire data. The procedure is to configure the VMMs to output different parameters, sample the outputs and store the data.
Figure\,\ref{fig:GI_calib}\,(left) shows the schematic of the parts involved. The configuration software transmits commands through the OPC client to the SCA OPC-UA server which, via FELIX, propagates the commands to the Front-end ASICs. Then, through the same path, the SCA is instructed to sample the ADC output and the data are made available through the SCA server to a custom data handler.

\noindent The following types of calibration are foreseen and possible through this procedure:

\para{Baseline and ENC:}The \gls{MO} output of the VMM is configured to be sent out for each channel at a time. In this configuration the analog output of the amplifier after shaping is made available. The channel baseline is sampled in a configurable number of samples from which the mean voltage level can be derived (baseline) and the spread (\gls{ENC} estimation). It is worth mentioning that the SCA ADC has a single slope Wilkinson architecture; hence, the slow ramp is ideal for slowly varying parameters but underestimates the fast ones. Consequently, the noise estimation on the detector is underestimated. To establish the real noise level on the detector, a correction factor needs to be applied.

\para{Pulser and threshold DAC:}Through this measurement, the correct threshold can be applied to the electronics, and the input charge can be converted from DAC counts to mV.

\para{Channel trimmers:}During this calibration, the individual 5-bit trimmer of each VMM channel can be set such that the amount of charge from baseline to the discrimination level is equalised across the channels.

Illustration of the results of these calibrations can be found here\,\cite{9724214}. Through the same procedure, the temperature and the band-gap voltage of the VMM can be measured for monitoring.

\subsection{Data-driven calibration}
Through this type of calibration, several parameters can be extracted that can be used as calibration parameters to reprocess the acquired data. The sequence of this calibration type involves the full data-taking with L1A data. The data are generated through the internal VMM pulser that is driven through a dedicated TTC bit such that they are synchronous along the system.  For this sequence, a dedicated
configuration of the data-acquisition system which involves the configuration procedure through the OPC client-server is used. The TTC system is also part of the calibration since it needs to produce a specific time sequence of signals to drive the VMM pulser and produce a L1A and L0A to read back the data. The data are captured by an instance of the swROD not connected to the ATLAS High Level Trigger. The schematic of this sequence is shown in Figure\,\ref{fig:GI_calib}\,(right).
The following types of calibration are required through this procedure:

\para{VMM channel gain calibration:}By injecting different amounts of charge in the VMM and storing the PDO, the gain of the electronics can be established which allows correction of the charge measurement results. During this procedure, the time-walk can also be measured using the \gls{TDO} data.

\para{Time-to-Amplitude Converter (TAC) calibration:}The VMM records as TDO a voltage level which is the amplitude of the TAC starting either at peak or threshold and stopping at the falling edge of the bunch crossing clock. The command to inject charge arrives via the TTC system and hence is synchronous to the system and the bunch crossing clock.
The ROC has a configurable delay between the bunch crossing clock and the clock used to inject the charge. By stepping this delay, the slope of the TAC ramp can be measured and used to translate ADC counts into meaningful time units.

\subsection{Configurable delays}
As mentioned above, the clock distribution in NSW has to follow paths of different length to reach various front-end boards. In addition, the time-of-flight from particles originating from the ATLAS interaction point (IP) differs along the radius of the detector.  The sTGC pad-generated signals are transmitted through PCB traces with significantly different lengths; hence they must be aligned.
To compensate for the above-mentioned effects, configurable delays have been implemented in different ASICs. Those delays must be tuned to assign the correct time stamp on the recorded event.
\section{Power distribution and grounding}

\subsection{Power}
\label{sec:powerDistribution}
All the NSW on-detector electronics utilise the FEAST DC-DC converter\,\cite{FEAST2.1} for the on-board Point-of-Load DC regulators. The input power to the FEASTs is provided by power supplies developed by CAEN\,\cite{caen}.
The Low Voltage power distribution\,\cite{nswlv} adopts a two step voltage conversion. The New Generation Power Supply (NGPS)\,\cite{ngps} located in the US15 service area of ATLAS provide 280\,V to the on-wheel  Intermediate Conversion Stage (\gls{ICS}) modules which are based on the EASY BRIC system\,\cite{ics}.
The ICS modules convert the 280\,VDC to 11\,V (configurable) and supply the voltage to the on-detector electronics through a Low-Voltage-Distributor-Board (LVDB). One LVDB can supply up to eight Front-end boards (analog section) and up to four digital boards (e.g.\ L1DDC, digital section).

\subsection{Grounding}
\label{sec:grounding}
The grounding for the New Small Wheel tries to follow the general ATLAS guidelines as much as possible\,\cite{Blanchot:1073170} in a ``Star'' topology. Each detector technology sector of the NSW is considered an isolated system attached to its mechanical frame via isolating attachments. The low inductance,  low resistance connection to ground must not carry current. The ``Analog'' ground of the Front-end boards should connect directly to the detector ground with the shortest path possible.  The Low Voltage lines are connected to a floating power supply.  All ground, drain, and shield connections in LVDS cables should be AC coupled to the ground plane of the Front-end Board only. The other end will be DC coupled on the digital electronics. The grounding scheme with principles and rules that were implemented can be found here\,\cite{Levinson:2845656}. The integration of the electronics in the wheel is not described in this manuscript.



\acknowledgments
This research was supported in part by Grant No 2871/19 from the Israeli Science Foundation (ISF).

This work was funded in part by the U. S. Department of Energy, Office of Science, High Energy Physics under Contracts DE-SC0012704.

We acknowledge support of this work by the project ``DeTAnet: Detector Development and Technologies for High Energy Physics'' (MIS 5029538) which is implemented under the action ``Reinforcement of the Research and Innovation Infrastructure'' funded by the Operational Programme ``Competitiveness, Entrepreneurship and Innovation'' (NSRF 2014–2020) and co-financed by Greece and the European Union (European Regional Development Fund).

The authors would like to acknowledge the valuable help and support of the ATLAS TDAQ, FELIX and DCS teams during the development of the electronics described in this manuscript. Moreover the authors would like to acknowledge the CERN ESE department for the support and collaboration.


\clearpage
\section*{Glossary}
\addcontentsline{toc}{section}{Glossary}
\begin{multicols}{2}
\printglossary[type=main]
\end{multicols}

\clearpage
\bibliographystyle{JHEP}
\bibliography{bib/nsw_electronics, bib/nsw_electronics_GI, bib/nsw_electronics_LL}
\end{document}